\documentclass[iop,revtex4,numberedappendix,appendixfloats,twocolappendix]{emulateapj}
\usepackage{times}
\usepackage{appendix}
\usepackage{pdflscape}
\pdfoutput=1

\begin{document}

\newcommand{\oii}{\text{[\ion{O}{2}]}}
\newcommand{\neiii}{\text{[\ion{Ne}{3}]}}
\newcommand{\oiii}{\text{[\ion{O}{3}]}}
\newcommand{\woiii}{\text{$W_\lambda(\oiii)$}}
\newcommand{\nii}{\text{[\ion{N}{2}]}}
\newcommand{\hei}{\text{\ion{He}{1}}}
\newcommand{\heii}{\text{\ion{He}{2}}}
\newcommand{\ha}{\text{H$\alpha$}}
\newcommand{\wha}{\text{$W_\lambda(\ha)$}}
\newcommand{\hb}{\text{H$\beta$}}
\newcommand{\hg}{\text{H$\gamma$}}
\newcommand{\hd}{\text{H$\delta$}}
\newcommand{\he}{\text{H$\epsilon$}}
\newcommand{\hz}{\text{H$\zeta$}}
\newcommand{\hn}{\text{H$\eta$}}
\newcommand{\htheta}{\text{H$\theta$}}
\newcommand{\hiota}{\text{H$\iota$}}
\newcommand{\hi}{\text{\ion{H}{1}}}
\newcommand{\hii}{\text{\ion{H}{2}}}
\newcommand{\hk}{\text{H$\kappa$}}
\newcommand{\caii}{\text{\ion{Ca}{2}}}
\newcommand{\sii}{\text{[\ion{S}{2}]}}
\newcommand{\wlya}{\text{$W_\lambda$({\rm Ly$\alpha$})}}
\newcommand{\wlyaem}{\text{$W_\lambda^{\rm em}$({\rm Ly$\alpha$})}}
\newcommand{\llya}{\text{$L$(Ly$\alpha$)}}
\newcommand{\llyaobs}{\text{$L$(Ly$\alpha$)$_{\rm obs}$}}
\newcommand{\llyaint}{\text{$L$(Ly$\alpha$)$_{\rm int}$}}
\newcommand{\lyafrac}{\text{$f_{\rm esc}^{\rm spec}$(Ly$\alpha$)}}
\newcommand{\lha}{\text{$L$(H$\alpha$)}}
\newcommand{\lhb}{\text{$L$(H$\beta$)}}
\newcommand{\sfrha}{\text{SFR(\ha)}}
\newcommand{\sfrsed}{\text{SFR(SED)}}
\newcommand{\sfruv}{\text{SFR(UV)}}
\newcommand{\ssfrha}{\text{sSFR(\ha)}}
\newcommand{\ssfrsed}{\text{sSFR(SED)}}
\newcommand{\ebmvneb}{E(B-V)_{\rm neb}}
\newcommand{\ebmvcont}{E(B-V)_{\rm cont}}
\newcommand{\ebmvlos}{E(B-V)_{\rm los}}
\newcommand{\nhi}{N(\text{\ion{H}{1}})}
\newcommand{\lognhi}{\log[\nhi/{\rm cm}^{-2}]}
\newcommand{\lognhitable}{\log\left[\frac{\nhi}{{\rm cm}^{-2}}\right]}
\newcommand{\lya}{\text{Ly$\alpha$}}
\newcommand{\lyb}{\text{Ly$\beta$}}
\newcommand{\lyg}{\text{Ly$\gamma$}}
\newcommand{\comment}[1]{}
\newcommand{\wciii}{\text{$W_\lambda$(\ion{C}{3}])}}
\newcommand{\ciii}{\text{\ion{C}{3}]}}
\newcommand{\interoiii}{\text{\ion{O}{3}]}}
\newcommand{\rsiione}{R(\text{\ion{Si}{2}}\lambda 1260)}
\newcommand{\rsiitwo}{R(\text{\ion{Si}{2}}\lambda 1527)}
\newcommand{\nv}{\text{\ion{N}{5}}}
\newcommand{\siii}{\text{\ion{Si}{2}}}
\newcommand{\cii}{\text{\ion{C}{2}}}
\newcommand{\civ}{\text{\ion{C}{4}}}
\newcommand{\siiv}{\text{\ion{Si}{4}}}
\newcommand{\ovi}{\text{\ion{O}{6}}}
\newcommand{\rcii}{R(\text{\ion{C}{2}}\lambda 1334)}
\newcommand{\ralii}{R(\ion{Al}{2}\lambda 1670)}
\newcommand{\qh}{Q(\text{H$^0$})}
\newcommand{\rs}{{\cal R}_{\rm s}}
\newcommand{\fcovhi}{f_{\rm cov}(\hi)}
\newcommand{\fcovmetal}{f_{\rm cov}({\rm metal})}
\newcommand{\fesclya}{f_{\rm esc}^{\rm spec}(\lya)}
\newcommand{\logxi}{\log[\xi_{\rm ion}/{\rm s^{-1}/erg\,s^{-1}\,Hz^{-1}}]}


\title{The Effects of Stellar Population and Gas Covering Fraction on
  the Emergent Ly$\alpha$ Emission of High-Redshift Galaxies\altaffilmark{*}}

\author{\sc Naveen A. Reddy\altaffilmark{1},
Michael W. Topping\altaffilmark{2},
Alice E. Shapley\altaffilmark{3},
Charles C. Steidel\altaffilmark{4},
Ryan L. Sanders\altaffilmark{5,6}
Xinnan Du\altaffilmark{1},
Alison L. Coil\altaffilmark{7},
Bahram Mobasher\altaffilmark{1},
Sedona H. Price\altaffilmark{8},
Irene Shivaei\altaffilmark{2,6}}

\altaffiltext{1}{Department of Physics and Astronomy, University of California, 
Riverside, 900 University Avenue, Riverside, CA 92521, USA; naveenr@ucr.edu}
\altaffiltext{2}{Steward Observatory, University of Arizona, 933 North 
Cherry Avenue, Tucson, AZ 85721, USA}
\altaffiltext{3}{Department of Physics \& Astronomy, University of California,
Los Angeles, 430 Portola Plaza, Los Angeles, CA 90095, USA}
\altaffiltext{4}{Cahill Center for Astronomy and Astrophysics, California Institute of Technology, MC 249-17, Pasadena, CA 91125, USA}
\altaffiltext{5}{Department of Physics, University of California, Davis, One Shields Ave, Davis, CA 95616, USA}
\altaffiltext{6}{NASA Hubble Fellow}
\altaffiltext{7}{Center for Astrophysics and Space Sciences, University of
California, San Diego, 9500 Gilman Drive, La Jolla, CA 92093-0424, USA}
\altaffiltext{8}{Max-Planck-Institut f\"{u}r Extraterrestrische Physik, Postfach 1312, Garching, D-85741, Germany}

\altaffiltext{*}{Based on data obtained at the W.M. Keck Observatory,
  which is operated as a scientific partnership among the California
  Institute of Technology, the University of California, and NASA, and
  was made possible by the generous financial support of the W.M. Keck
  Foundation.}

\slugcomment{DRAFT: \today}

\begin{abstract}

We perform joint modeling of the composite rest-frame far-UV (FUV) and
optical spectra of redshift $1.85\le z\le 3.49$ star-forming galaxies
to deduce key properties of the massive stars, ionized ISM, and
neutral ISM, with the aim of investigating the principal factors
affecting the production and escape of $\lya$ photons.  Our sample
consists of 136 galaxies with deep Keck/LRIS and MOSFIRE spectra
covering, respectively, $\lyb$ through $\ciii$\,$\lambda\lambda 1907,
1909$, and $\oii$, $\neiii$, $\hb$, $\oiii$, $\ha$, $\nii$, and
$\sii$.  Spectral and photoionization modeling indicate that the
galaxies are uniformly consistent with stellar population synthesis
models that include the effects of stellar binarity.  Over the dynamic
range of our sample, there is little variation in stellar and nebular
abundance with $\lya$ equivalent width, $\wlya$, and only a marginal
anti-correlation between age and $\wlya$.  The inferred range of
ionizing spectral shapes is insufficient to solely account for the
variation in $\wlya$.  Rather, the covering fraction of
optically-thick $\hi$ appears to be the principal factor modulating
the escape of $\lya$, with most of the $\lya$ photons in
down-the-barrel observations of galaxies escaping through
low-column-density or ionized channels in the ISM.  Our analysis shows
that a high star-formation-rate surface density, $\Sigma_{\rm SFR}$,
particularly when coupled with a low galaxy potential (i.e., low
stellar mass), can aid in reducing the covering fraction and ease the
escape of $\lya$ photons.  We conclude with a discussion of the
implications of our results for the escape of ionizing radiation at
high redshift.

\end{abstract}


\section{\bf INTRODUCTION}
\label{sec:intro}

Ly$\alpha$ emission is often used as a signpost of ionizing radiation
from star-forming galaxies \citep{partridge67} due to the strength of
the line and the cosmic abundance of hydrogen.  Despite its widespread
use, its large absorption cross-section means that a Ly$\alpha$ photon
may traverse a complicated path through the interstellar medium (ISM)
of a galaxy before being absorbed by dust or escaping
\citep{spitzer78, meier81}.  The intensity of the line will be
severely affected as a result.  Ly$\alpha$ photons will scatter out of
the line of sight by relatively small columns of $\hi$, rendering a
significant fraction of the flux undetectable in typical
slit-spectroscopic observations of galaxies.  The observed Ly$\alpha$
flux will be further diminished in the presence of dust, and its
velocity profile will be shaped by bulk motions that cause the photons
to shift out of resonance (e.g., \citealt{kunth98}).

While the resonant scattering of $\lya$ poses challenges for
interpreting the nature of the ionizing sources and the subsequent
transmission of those photons, it can be exploited to probe the
spatial structure and kinematics of the ISM and circumgalactic medium
(e.g., \citealt{steidel11, momose14, xue17, bacon17, wisotzki18,
  leclercq20}).  Further, because $\lya$ photons are easily scattered,
the residual flux that emerges along the observer's sightline holds
important clues on the porosity of the ISM, a key factor in the escape
of ionizing (``Lyman Continuum'', or LyC) photons \citep{zackrisson13,
  trainor15, dijkstra16, reddy16b, steidel18}.  It is thus not
surprising that many efforts to investigate the sources of cosmic
reionization have focused on galaxies with strong Ly$\alpha$ emission,
or Ly$\alpha$ emitters (LAEs)---see \citet{ouchi20} and references
therein.

Radiative transfer simulations can be used to trace Ly$\alpha$ photons
from the time they are produced to their escape or absorption by dust,
a useful exercise for understanding how the strength and the velocity
profile of the emerging photons is shaped by a clumpy, dusty, or
non-static ISM \citep{dijkstra06, verhamme06, verhamme08, duval14,
  gronke15, gronke16}.  Nevertheless, one can gain significant insight
into the key factors governing the production and escape of Ly$\alpha$
(and LyC) photons---namely the ionizing spectrum and the gas covering
fraction---without resorting to computationally-expensive radiative
transfer simulations.  As discussed below, these two factors appear to
successfully describe the variation in Ly$\alpha$ equivalent widths
and escape fractions measured for the high-redshift galaxies analyzed
here.  The rest-frame far-UV (FUV) and rest-frame optical spectra of
galaxies can be used to constrain the ionizing spectrum and the gas
covering fraction.

The FUV ($\lambda \simeq 1000-2000$\,\AA) galaxy spectrum contains a
high density of emission and absorption lines well suited for
investigating the production and transmission of Ly$\alpha$ photons.
Aside from allowing a direct measurement of the emergent Ly$\alpha$
flux, this region of the spectrum contains numerous stellar
photospheric absorption lines (e.g., \citealt{leitherer01, rix04}) and
stellar wind lines (e.g., \ion{N}{5}\,$\lambda 1240$,
\ion{Si}{4}\,$\lambda 1402$, \ion{C}{4}\,$\lambda 1550$,
\ion{He}{2}\,$\lambda 1640$; \citealt{leitherer01, crowther06,
  brinchmann08}) sensitive to the stellar metallicity, age, and
initial mass function---factors that are among those responsible for
modulating the ionizing radiation field.  Several studies have used
specific features in the FUV, or full-spectrum fitting, to deduce
stellar metallicities and ages of the massive stellar populations in
high-redshift galaxies (e.g., \citealt{halliday08, sommariva12,
  steidel16, cullen19, cullen20, topping20a, topping20b}).

The FUV and optical nebular emission lines provide additional powerful
constraints on the shape of the extreme-UV (EUV; $\lambda < 912$\,\AA)
radiation field, which ultimately sets the intrinsic number of
ionizing photons.  Specifically, these lines allow one to distinguish
between single star population synthesis models and those that include
the effects of stellar multiplicity (binarity).  While both flavors of
models successfully match many of the stellar photospheric absorption
and wind lines in the FUV for various assumptions of the stellar
metallicity and age, only the binary models are able to reproduce the
hard EUV spectrum required to generate the nebular line luminosities
and line ratios measured from composite FUV and optical spectra of
$z\sim 2$ galaxies \citep{steidel16}.

In addition to providing valuable constraints on the ionizing
radiation field, the FUV spectrum includes a number of low- and
high-ionization metal and \ion{H}{1} absorption lines that may be used
to infer gas covering fraction.  Several studies have used the strong
saturated metal transitions occurring in a predominantly neutral
medium (e.g., \ion{Si}{2}\,$\lambda 1260$, \ion{Si}{2}\,$\lambda
1527$, \ion{C}{2}\,$\lambda 1334$, \ion{Al}{2}\,$\lambda 1670$) to
infer the covering fraction of \ion{H}{1} (e.g., \citealt{shapley03,
  heckman11, berry12, jones13, alexandroff15, trainor15, henry15,
  du18, harikane20}).  Many of the same works and others have
suggested a strong connection between the escape of Ly$\alpha$ (and
LyC photons) and gas covering fraction as inferred from these same
lines (e.g., \citealt{kornei10, hayes11, wofford13, borthakur14,
  rivera15, trainor15, reddy16b, steidel18, jaskot19}).

As noted elsewhere, however, the depths of saturated low-ionization
metal absorption lines may underestimate the covering fraction of
\ion{H}{1} \citep{henry15, vasei16, reddy16b, gazagnes18}: these lines
are not sensitive to metal-poor and possibly less dense gas.  The
\ion{H}{1} absorption lines provide the most direct probe of the
covering fraction if the lines are saturated.  Large variations in
Ly$\alpha$ forest blanketing along different sightlines renders it
difficult to robustly model the interstellar \ion{H}{1} lines for all
but the highest-column-density individual (damped Ly$\alpha$) systems
at high redshift.  Thus, one must average the FUV spectra across
typically $N\ga 25$ independent sightlines to reduce uncertainty in
the mean foreground opacity to $\la 10\%$ \citep{reddy16b, steidel18}.

For the most part, analyses of composite spectra of high-redshift
galaxies, as well as the spectra of individual Ly$\alpha$-emitting
(and LyC-leaking) local galaxies, imply a picture where Ly$\alpha$ and
LyC photons that emerge along the line-of-sight (or down the barrel of
the galaxy) do so after escaping the ISM through low-column-density
channels in an otherwise optically-thick (ionization-bounded) medium
\citep{trainor15, dijkstra16, reddy16b, gazagnes18, chisholm18,
  steidel18, trainor19, gazagnes20}.  Other studies of local galaxies
with high-equivalent-width optical emission lines (such as ``Green
Pea'' galaxies) find a high covering fraction of \ion{H}{1} with
$\sim$unity optical depth, where the column density modulates
Ly$\alpha$ escape \citep{henry15, jaskot19}.  There are yet other
studies that have suggested that density-bounded conditions indicated
by very high $\oiii/\oii$ may favor the escape of LyC and Ly$\alpha$
photons (e.g., \citealt{nakajima13, nakajima14, jaskot13, tang19}).

In general, galaxies are likely better-visualized not as a single
\ion{H}{2} region, but as many overlapping regions, with a
complicated ISM structure as a consequence.  Thus, all of the
aforementioned conditions could be present even in an individual
galaxy: e.g., some sightlines through the galaxy may be
ionization-bounded, others may density-bounded, while still others may
be something in between (\ion{H}{1} gas with $\sim$unity optical
depth; see also \citealt{kakiichi21}).  If this is the case, averaging
over many galaxies (and hence many sightlines) will result in an
average (composite) spectrum that appears consistent with an
ionization-bounded and porous ISM; i.e., a non-unity covering fraction
of optically-thick gas \citep{reddy16b}.  One also expects the
visibility of Ly$\alpha$ (and LyC) emission to be highly dependent on
the orientation of the galaxy relative to the line of sight (e.g.,
\citealt{ma16}).  At any rate, modeling of FUV spectra can elucidate
the predominant pathways by which Ly$\alpha$ and LyC photons escape
the ISM of galaxies (e.g., \citealt{henry15, reddy16b, steidel18,
  gazagnes18, gazagnes20}).

Galaxies at redshifts $2.0 \la z\la 2.6$ are uniquely suited for
studying the production and transmission of Ly$\alpha$ photons: the
FUV is shifted to observed wavelengths where the sky background is
extremely low and the throughput of blue-optimized ground-based
spectrographs is high, typical star-forming galaxies are sufficiently
bright to enable detailed studies of individual objects
\citep{reddy09}, the relative transparency of the foreground IGM
allows for constraints on the gas covering fraction for ensembles of
objects \citep{reddy16b}, and the full suite of strong optical
emission lines used to diagnose the state of the ISM are shifted to
the near-IR windows of atmospheric transmission.  

In this paper, we investigate the role of stellar population and gas
covering fraction in the production and escape of Ly$\alpha$ (and LyC)
photons using a sample of 136 typical star-forming galaxies at
redshifts $1.85\le z\le 3.49$ with deep FUV and optical spectra from
the MOSFIRE Deep Evolution Field (MOSDEF) survey \citep{kriek15}.  Our
analysis extends upon previous efforts (e.g., \citealt{steidel16,
  topping20a, cullen20}) by jointly modeling FUV and optical composite
spectra---including the interstellar $\hi$ absorption
lines---constructed in bins of $\lya$ equivalent width and
star-formation-rate surface density.  The galaxies were targeted for
optical and subsequent FUV spectroscopy independent of FUV or optical
emission-line strength and, as such, constitute a sample that is
particularly advantageous in the current context.  While targeted
searches of LAEs and other high-equivalent-width optical line emitters
provide a means of efficiently selecting galaxies that contribute
significantly to cosmic reionization, they lack the dynamic range in
galaxy properties to properly evaluate the physical factors modulating
the emission lines and escape of $\lya$ and ionizing photons.  To
address the question of {\em why} certain galaxies have strong
Ly$\alpha$ emission and others do not, one would clearly want to
obtain spectroscopy in a controlled way for many galaxies spanning a
range of Ly$\alpha$ strength.  Moreover, the availability of deep {\em
  Hubble Space Telescope} ({\em HST}) imaging in the fields targeted
by MOSDEF allow for an exploration of how the compactness of star
formation impacts the porosity of the ISM and the escape of Ly$\alpha$
and ionizing photons (e.g., \citealt{heckman11, verhamme17, marchi19,
  naidu20}).

The outline of this paper is as follows.  The MOSDEF survey, followup
Keck/LRIS FUV spectroscopy and data reduction, individual galaxy
measurements, and construction of composite spectra are described in
Section~\ref{sec:survey}.  Section~\ref{sec:modelfitting} presents the
procedures for fitting the composite FUV spectra and optical emission
line ratios with stellar population synthesis (SPS) and
photoionization models, respectively; and the correlations between
$\lya$ equivalent width and parameters relating to the ionizing
spectrum (e.g., stellar metallicity, age), state of the ionized ISM
(e.g., ionization parameter, gas-phase oxygen abundance), and the
configuration of the neutral ISM (e.g., $\hi$ column density and
covering fraction).  The role of the shape of the ionizing spectrum
and gas covering fraction on $\lya$ escape, as well as the impact of
the star-formation-rate surface density and galaxy potential well on
the escape of $\lya$ and the ionization parameter are discussed in
Section~\ref{sec:discussion}.  We summarize the analysis and our
conclusions in Section~\ref{sec:conclusions}.  Unless indicated
otherwise, quoted wavelength ranges (e.g., EUV, FUV, and optical)
refer to the rest frame.  All equivalent widths are expressed in the
rest frame.  A \citet{chabrier03} initial mass function (IMF) is
considered throughout the paper.  Magnitudes are on the AB system
\citep{oke83}.  We adopt a cosmology with
$H_{0}=70$\,km\,s$^{-1}$\,Mpc$^{-1}$, $\Omega_{\Lambda}=0.7$, and
$\Omega_{\rm m}=0.3$.

\section{\bf SAMPLE SELECTION AND OBSERVATIONS}
\label{sec:survey}

\subsection{The MOSDEF Survey}
\label{sec:mosdefsurvey} 

Galaxies analyzed here were drawn from the MOSDEF survey.  This survey
obtained moderate resolution ($R\sim 3000-3600$) rest-frame optical
spectroscopy of $\approx 1500$ $H$-band-selected galaxies and AGNs at
redshifts $1.4\la z\la 3.8$ in the CANDELS fields \citep{grogin11,
  koekemoer11} using the MOSFIRE spectrometer \citep{mclean12} on the
Keck telescope.  MOSDEF galaxies were selected for spectroscopy based
on pre-existing photometric, grism, or spectroscopic redshifts
($z_{\rm spec}$)---$z=1.37-1.70$, $z=2.09-2.70$, and
$z=2.95-3.80$---that place the strong rest-frame optical emission
lines (e.g., $\oii$, $\hb$, $\oiii$, $\ha$, $\nii$) in the $YJHK$
atmospheric transmission windows.  Details of the spectroscopic data
reduction are provided in \citet{kriek15}.  The final spectroscopic
sample spans ranges of star-formation rate ($1\la {\rm SFR}\la
200$\,$M_\odot$\,yr$^{-1}$) and stellar mass ($10^9 \la M_\ast \la
10^{11}$\,$M_\odot$) typical for galaxies at $z\sim 1.4-3.8$, with the
majority of galaxies in the sample having detections of multiple
rest-frame optical emission lines.  Optical line luminosities were
calculated using the methodology presented in \citet{kriek15}.

\subsection{MOSDEF-LRIS FUV Spectroscopy}
\label{sec:mosdeflrissurvey}

Optical spectra from MOSDEF were complemented with FUV spectra
obtained using the Low Resolution Imaging Spectrometer (LRIS;
\citealt{oke95, steidel04}) on the Keck telescope.  Objects from the
parent MOSDEF spectroscopic sample in the AEGIS, COSMOS, GOODS-N, and
GOODS-S fields were prioritized for LRIS spectroscopy in the following
manner.  Objects with detections of \hb, \oiii, \ha, and \nii\, were
given the highest priority, and those with detections of the first
three of these lines and an upper limit on \nii\, were given the
next-highest priority.  We then included the following in order of
priority: objects with MOSDEF redshifts $2.0\le z_{\rm spec}\le 2.7$,
$1.4\le z_{\rm spec}\la 1.7$, or $3.0\le z_{\rm spec}\le 3.8$; those
for which we were unable to obtain a redshift with MOSFIRE; and
finally objects not observed with MOSFIRE, but which have photometric
or grism redshifts from the 3D-HST survey that place them in the
redshift ranges---and which meet the limit in {\em H}-band
magnitude---targeted by MOSDEF.

Slit masks were milled with $1\farcs 2$-width slits and, to ensure
proper background subtraction, a minimum slit length of $9\arcsec$.  A
total of 259 objects (217 of which have $z_{\rm spec}$ from MOSFIRE)
were observed using 9 slit masks in the AEGIS, COSMOS, GOODS-N, and
GOODS-S fields.  The observations were obtained over 10 nights in five
separate observing runs during the period 2017 January to 2018 June.
Spectra were obtained with the blue and red channels of LRIS (LRIS-B
and LRIS-R, respectively), with the incoming beam split at $\simeq
5000$\,\AA\, using the d500 dichroic.  LRIS-B and LRIS-R were
configured with the 400\,line/mm grism and the 600\,line/mm grating
blazed at 5000\,\AA, respectively.  The red-side grating was tilted so
that the combined LRIS-B and LRIS-R spectrum for each galaxy has
continuous wavelength coverage from the atmospheric cut-off at
$\lambda \simeq 3100$\,\AA\, to $\ga 7000$\,\AA, with the reddest
covered wavelength depending on the position of the slit within the
LRIS spectroscopic field of view.  This setup yielded blue and red
spectral resolutions of $R\sim 800$ and $R\sim 1400$, respectively,
for a typical seeing of around $0\farcs 8$.  Total exposure times
varied from 6 to 11\,hours, with a median of 7.5\,hours.  In addition
to internal flats on the red side, we obtained sky flats taken in
twilight on the blue side as they provide better illumination than the
internal flats at $\lambda \la 4000$\,\AA.  Arc spectra from Cd, Ar,
Zn, He, and Hg internal lamps were acquired for wavelength
calibration, and spectroscopic flux standards were observed at
different airmasses on all observing runs to aid in flux calibration.
LRIS data were reduced using the procedures described in
\citet{steidel18} and \citet{topping20a}.  

\subsection{Individual Galaxy Measurements}
\label{sec:individualmeasurements}

\subsubsection{Redshifts}
\label{sec:redshifts}

The procedure used to derive redshifts for Ly$\alpha$ emission and
each of the strongest interstellar low-ionization metal absorption
lines in the LRIS spectra---\ion{Si}{2}\,$\lambda 1260$,
\ion{O}{1}\,$\lambda 1302$+\ion{Si}{2}\,$\lambda 1304$,
\ion{C}{2}\,$\lambda 1334$, \ion{S}{2}\,$\lambda 1526$,
\ion{Fe}{2}\,$\lambda 1608$, and \ion{Al}{2}\,$\lambda 1670$---is
described in \citet{topping20a}.  Briefly, the redshift of each line
was calculated by fitting the local continuum and the line with a
quadratic and Gaussian function, respectively, and taking the centroid
of the Gaussian function as the observed wavelength of the line.
These fits were manually inspected and only those lines with satisfactory
fits were used to derive the final Ly$\alpha$ emission line and
interstellar absorption line redshifts, $z_{\rm Ly\alpha}$ and $z_{\rm
  IS}$, respectively.  

\begin{figure}
\epsscale{1.2}
\plotone{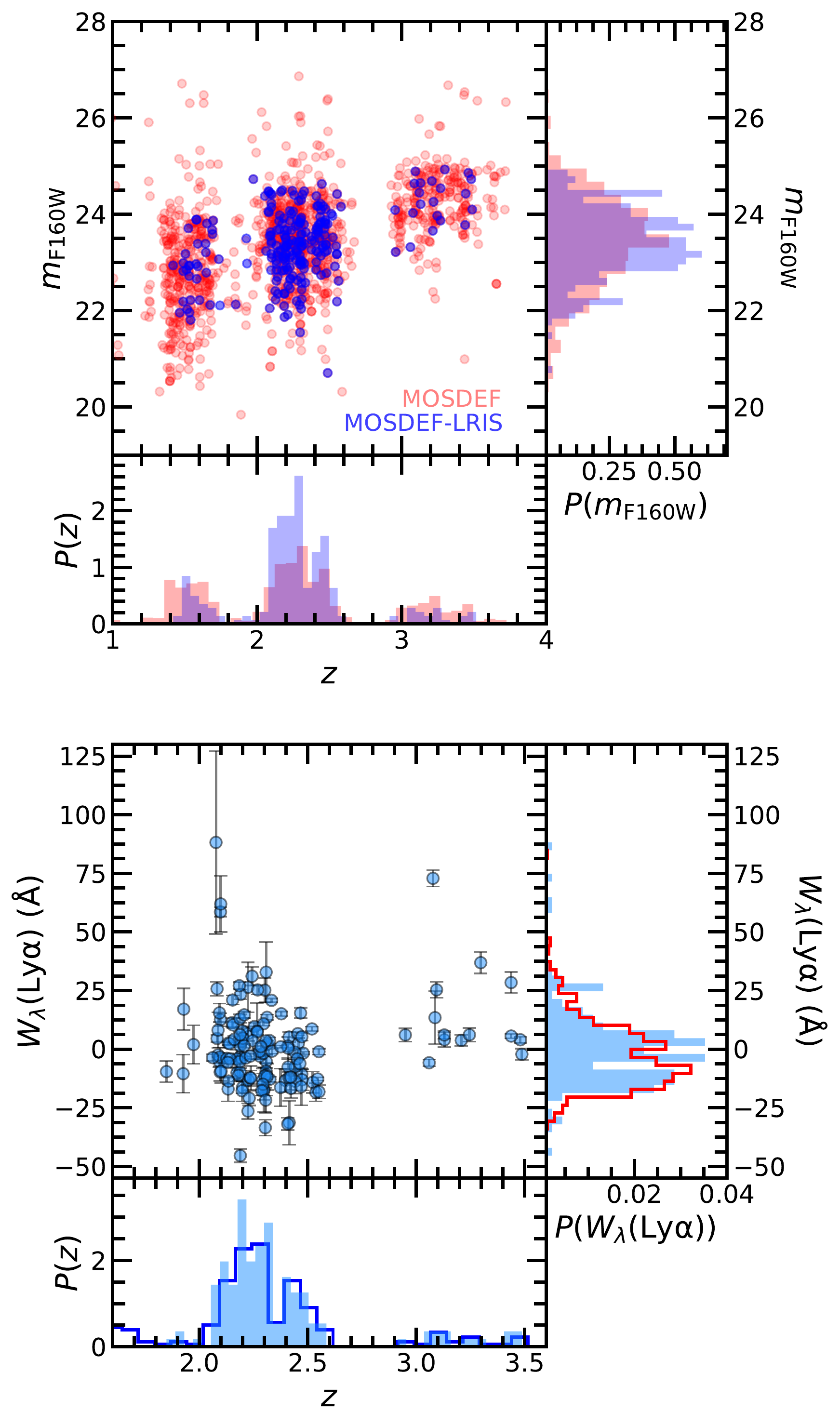}
\caption{{\em Top:} Joint magnitude and redshift distributions of the
  MOSDEF-LRIS sample (blue circles) and the parent MOSDEF sample (red
  circles).  The histograms show the probability density, normalized
  so that the total area below the histograms is equal to unity, as a
  function of redshift (bottom subpanel) and magnitude (right
  subpanel).  {\em Bottom:} Joint $\wlya$ and redshift distributions
  for the final sample of 136 objects used in this work (blue
  circles).  The bottom subpanel shows the redshift probability
  density histogram for the final sample (shaded blue) and that of the
  entire MOSDEF-LRIS sample (blue outline).  The right subpanel shows
  the $\wlya$ probability density distribution for the final sample
  (shaded blue) and that of 978 objects from the UV-selected KBSS
  survey of star-forming galaxies at similar redshifts (red outline).}
\label{fig:zhistwlyahist}
\end{figure}

The spectra were shifted to the rest frame using systemic redshifts,
$z_{\rm sys}$, which were measured from the highest signal-to-noise
(S/N) optical nebular emission lines (i.e., typically
[\ion{O}{3}]\,$\lambda 5008$ or $\ha$) in the MOSFIRE spectra, as
described in \citet{kriek15}.  For the 21 (of 259) objects in the
sample that did not have a spectroscopic redshift from
MOSFIRE---either because a robust redshift identification was not
possible or because the object had not been previously targeted in the
MOSDEF survey---$z_{\rm sys}$ was estimated from $z_{\rm Ly\alpha}$
and/or $z_{\rm IS}$ using the relationships between $z_{\rm sys}$,
$z_{\rm Ly\alpha}$, and $z_{\rm IS}$ for those objects where both
$z_{\rm sys}$ and at least one of $z_{\rm Ly\alpha}$ or $z_{\rm IS}$
were available (see \citealt{topping20a}).  There were an additional
three objects in the sample that did not have redshifts from MOSFIRE
and for which we were unable to derive redshifts from the LRIS
spectra, usually because the LRIS spectrum was either too noisy or
contained irreparable artifacts that prevented a robust redshift
identification.  The redshift and near-IR magnitude distributions of
the MOSDEF-LRIS sample, relative to those of the parent MOSDEF sample,
are presented in the top panel of Figure~\ref{fig:zhistwlyahist}.  The
few objects fainter than the $H=24.0$, $24.5$, and $25.0$ magnitude
limits imposed for MOSDEF target selection in the $z=1.37-1.70$,
$z=2.09-2.70$, and $z=2.95-3.80$ redshift windows, respectively, are
serendipitous objects that happened to fall on MOSFIRE slits and for
which we were able to derive robust redshifts.

\subsubsection{Ly$\alpha$ Equivalent Widths, $\wlya$}
\label{sec:wlyameas}

One of the basic parameters in this study is the equivalent width of
Ly$\alpha$, $\wlya$.  This quantity scales with the escape fraction of
Ly$\alpha$ photons, and thus depends on the intrinsic ionizing photon
production rate and gas covering fraction.  $\wlya$ was measured for
each galaxy following the procedures given in \citet{kornei10} and
\citet{du18}.  Errors in $\wlya$ were calculated by remeasuring
$\wlya$ for many realizations of each spectrum based on the
corresponding error spectrum.  Note that $\wlya$ only includes the
emission measured within the spectroscopic aperture, adjusted for slit
loss by assuming the emission is spatially coincident with the
non-ionizing FUV continuum (Section~\ref{sec:mosdeflrissurvey}).
$\wlya$ does not include the fraction of $\lya$ scattered out of the
spectroscopic aperture (e.g., \citealt{steidel11}).  As discussed in
Section~\ref{sec:wlyavsfcov}, the spectroscopically-determined $\wlya$
is the one most relevant for identifying ISM configurations that are
amenable for the escape of $\lya$ (and ionizing) photons.  Throughout
this work, $\wlya$ refers to the rest-frame value and is positive for
$\lya$ in emission.  In Section~\ref{sec:fesclya}, we consider an
alternative method for computing $\lya$ equivalent widths that
accounts for underlying interstellar and stellar absorption.

\subsubsection{Broadband SED Modeling}
\label{sec:sedmodeling}

To aid in accounting for stellar Balmer absorption when measuring line
luminosities, as well as determining SFRs ($\sfrsed$), stellar masses
($M_\ast$), ages, and continuum reddening ($\ebmvcont$), we fit the
broadband photometry of galaxies in the sample with the Binary
Population and Spectral Synthesis (BPASS) version 2.2.1 models
\citep{eldridge17, stanway18}.  Our choice of these models is
motivated by recent studies (e.g., \citealt{steidel16}) that jointly
model FUV and optical spectra and find that the FUV and optical line
luminosity ratios of typical star-forming galaxies at $z\sim 2$ are
best reproduced when incorporating the physics of stellar binarity.
The BPASS distribution conveniently includes model predictions both
with and without the effects of stellar binarity, as well as for two
different cutoffs of the high-mass end of the IMF, 100 and
300\,$M_\odot$.  The resulting four variations of the BPASS models are
considered in the present analysis: binarity with an upper mass cutoff
of $100$\,$M_\odot$ (``100bin''), binarity with an upper mass cutoff
of $300$\,$M_\odot$ (``300bin''), single star evolution with an upper
mass cutoff of $100$\,$M_\odot$ (``100sin''), and single star
evolution with an upper mass cutoff of $300$\,$M_\odot$ (``300sin'').
We refer to these four flavors of models as the BPASS model ``types.''

The instantaneous burst models provided in the BPASS distribution were
summed to produce constant star formation models for $\log[{\rm
    Age}/{\rm yr}] = 7.0 - 10.0$ in $0.1$\,dex increments.  The
$Z_{\ast}=0.001$ metallicity models were considered in fitting the
broadband photometry, based on the results obtained when directly
fitting the BPASS models to the FUV spectra
(Section~\ref{sec:fuvfitting}).  Nebular continuum emission was added
to the SPS models as described in
Section~\ref{sec:fuvfittingprocedure}.  Based on the analysis
presented in Appendix~\ref{sec:sfrcomparison} (see also
Section~\ref{sec:fuvfitting}), the SMC \citep{gordon03} dust
extinction curve was assumed for the reddening of the stellar
continuum.

The broadband photometry for each galaxy (as compiled in
\citealt{skelton14}) was first corrected for contributions from the
optical emissions lines and Ly$\alpha$.  The corrected photometry was
then fit with the aforementioned models, limiting the age to be less
than the age of the Universe at the redshift of each galaxy and
greater than the minimum dynamical timescale of $\simeq 10$\,Myr
inferred for the most compact galaxies in the sample (e.g.,
\citealt{reddy12b, price16, price20}).  The model parameters that
yielded the lowest $\chi^{2}$ relative to the photometry were taken to
be the best-fit values.  Errors in the parameters were determined from
the standard deviation of parameter values obtained when fitting many
perturbed realizations of the photometry based on the photometric
errors.  Unless indicated otherwise, the SED-fitting results obtained
with the 100bin models and the SMC extinction curve are adopted by
default (see Section~\ref{sec:fuvfittingprocedure} and
Appendix~\ref{sec:sfrcomparison} for further discussion on the choice
of the stellar reddening curve).

\subsection{Final Sample}
\label{sec:finalsample}

Several criteria were applied to the targeted set of 259 objects in
the MOSDEF-LRIS survey to arrive at the final sample for analysis.
Any objects for which the LRIS spectra contained irreparable artifacts
that prevented a robust redshift identification, or that were too noisy
to yield a redshift ($z_{\rm Ly\alpha}$ or $z_{\rm IS}$), were
removed.  AGNs identified using the IR, X-ray, and optical line flux
criteria described in \citet{coil15}, \citet{azadi17,azadi18}, and
\citet{leung19} were removed.  Examination of the LRIS spectra of the
remaining objects did not reveal the presence of any additional
luminous AGN (e.g., such as those identified by broad Ly$\alpha$
emission or significant \ion{N}{5} or \ion{C}{4} emission).  Any
objects for which the MOSFIRE or LRIS spectra indicate that the target
may be blended with a foreground object were removed.  The combined
telescope and instrumental sensitivity of LRIS-B with the 400 line/mm
grism falls below $\simeq 20\%$ at $\lambda \la 3300$\,\AA.  As such,
any objects for which $\lya$ falls at bluer wavelengths (i.e., $z_{\rm
  sys}<1.7$) were removed.  These criteria result in a
final sample of 136 galaxies whose redshift and $\wlya$ distributions
are indicated in the bottom panel of
Figure~\ref{fig:zhistwlyahist}---of these, 79 galaxies have complete
coverage of the strong optical nebular emission lines.  The redshift
distribution is statistically similar to that of the full MOSDEF-LRIS
sample, while the $\wlya$ distribution is statistically similar to
that of the Keck Baryonic Structure Survey (KBSS) sample distribution
for UV-selected galaxies at similar redshifts \citep{reddy08,
  steidel18}.  The percentage of galaxies with $\wlya>20$\,\AA, the
criterion that defines a $\lya$ emitter (LAE), is $12.5\%$ (17 of 136
galaxies).  This percentage agrees well with the $12\%$ found for
star-forming galaxies selected by their FUV colors to lie at similar
redshifts, i.e., ``BX''-selected galaxies, with $\rs \le 25.5$
\citep{reddy08}.

\subsection{FUV and Optical Composite Spectra}
\label{sec:compspec}

\begin{figure*}
\epsscale{1.2}
\plotone{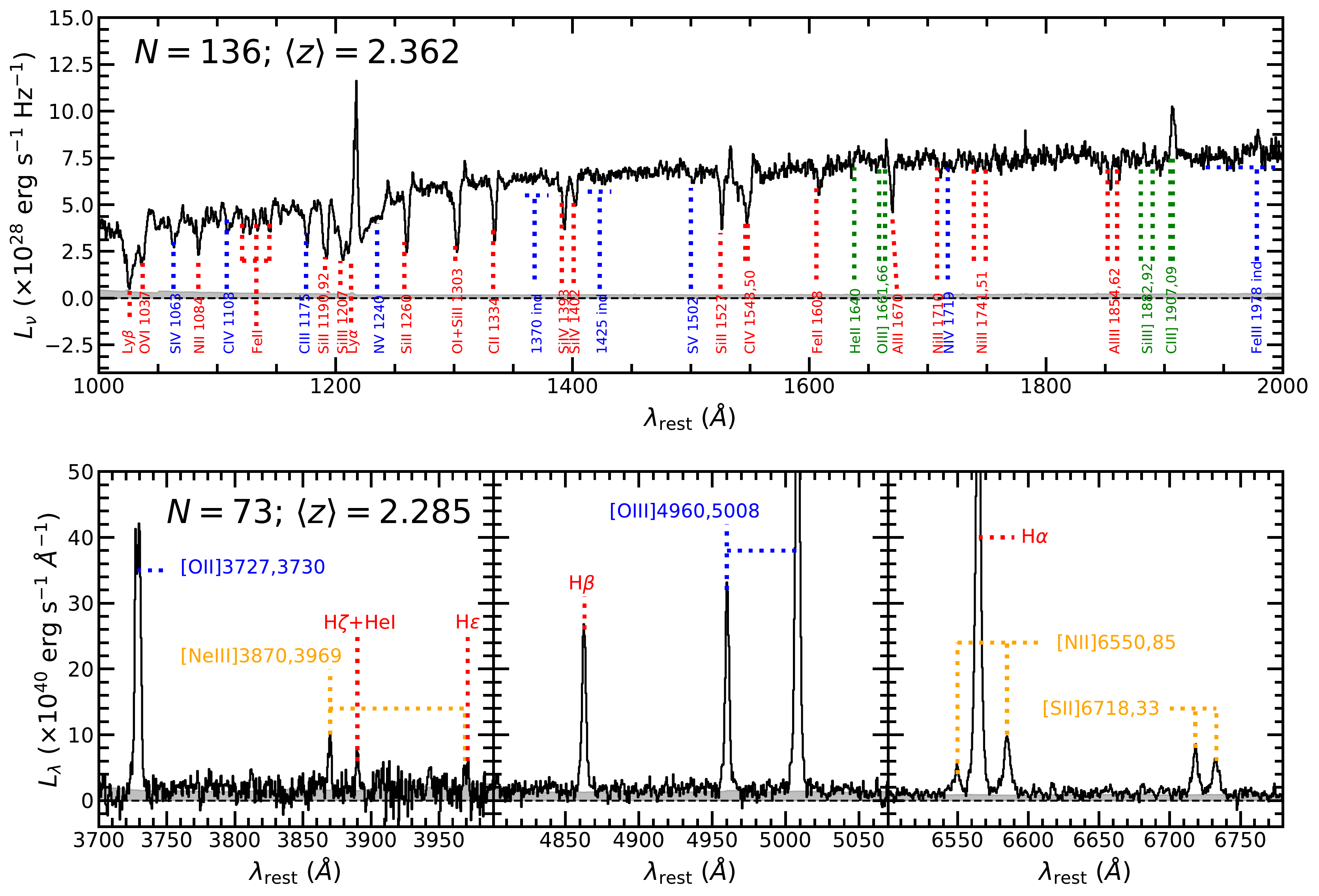}
\caption{{\em Top:} Composite FUV spectrum (based on the LRIS
  spectroscopic observations) for the $N=136$ galaxies in the sample
  with a mean redshift of $\langle z\rangle = 2.362$.  The spectrum
  has been corrected for the combined IGM and CGM opacity appropriate
  for the mean redshift of galaxies contributing to the composite
  spectrum.  Labeled are some of the prominent features in the
  spectrum, including stellar absorption (blue), nebular emission
  (green), and interstellar absorption (red).  $\civ$, $\siiv$, and
  $\ovi$ also include stellar wind absorption.  {\em Bottom:}
  Composite optical spectrum (based on the MOSFIRE spectroscopic
  observations) for the $N=73$ galaxies ($\langle z\rangle = 2.285$)
  in the subsample with complete wavelength coverage of the strong
  optical emission lines, including $\oii$\,$\lambda\lambda 3727,
  3730$, $\neiii$\,$\lambda\lambda 3870,3969$, $\hb$,
  $\oiii$\,$\lambda\lambda 4960, 5008$, $\ha$, $\nii$\,$\lambda\lambda
  6550, 6585$, and $\sii$\,$\lambda\lambda 6718, 6733$.  The prominent
  oxygen emission lines (blue), $\hi$ Balmer recombination emission
  lines (red), and other nebular emission lines (orange) are shown. In
  both panels, the light gray region indicates the composite error
  spectrum.}
\label{fig:compspec}
\end{figure*}

\begin{deluxetable}{llccr}
\tabletypesize{\footnotesize}
\tablewidth{0pc}
\tablecaption{Subsample Construction and Statistics}
\tablehead{
\colhead{Subsample} &
\colhead{Criteria\tablenotemark{a}} &
\colhead{$N$\tablenotemark{b}} &
\colhead{$\langle z_{\rm sys}\rangle$\tablenotemark{c}} &
\colhead{$\langle\wlya\rangle$  (\AA) \tablenotemark{d}}}
\startdata
A & All & $136$ & $2.362$ & $  1.09 \pm 0.46$ \\
AL & All, $\lyb$ & $118$ & $2.410$ & $ -0.79 \pm 0.37$ \\
ALN & All, $\lyb$, neb & $ 73$ & $2.285$ & $ -3.42 \pm 0.47$ \\
   \hline
   WT1 & $\wlya$, T1 & $ 45$ & $2.298$ & $-15.88 \pm 0.71$ \\
   WT1L & $\wlya$, T1, $\lyb$ & $ 39$ & $2.326$ & $-16.89 \pm 0.76$ \\
   WT1LN & $\wlya$, T1, $\lyb$, neb & $ 24$ & $2.311$ & $-19.13 \pm 0.97$ \\
\\
   WT2 & $\wlya$, T2 & $ 45$ & $2.371$ & $ -0.98 \pm 0.38$ \\
   WT2L & $\wlya$, T2, $\lyb$ & $ 39$ & $2.368$ & $ -1.56 \pm 0.38$ \\
   WT2LN & $\wlya$, T2, $\lyb$, neb & $ 24$ & $2.284$ & $ -4.72 \pm 0.63$ \\
\\
   WT3 & $\wlya$, T3 & $ 46$ & $2.417$ & $ 19.72 \pm 1.12$ \\
   WT3L & $\wlya$, T3, $\lyb$ & $ 40$ & $2.533$ & $ 15.67 \pm 0.72$ \\
   WT3LN & $\wlya$, T3, $\lyb$, neb & $ 25$ & $2.260$ & $ 12.92 \pm 0.82$ \\
   \hline
   ST1 & $\Sigma_{\rm SFR(\ha)}$, T1 & $ 26$ & $2.305$ & $ -2.20 \pm 0.99$ \\
   ST1L & $\Sigma_{\rm SFR(\ha)}$, T1, $\lyb$ & $ 24$ & $2.325$ & $ -5.32 \pm 0.90$ \\
\\
   ST2 & $\Sigma_{\rm SFR(\ha)}$, T2 & $ 26$ & $2.266$ & $ -2.24 \pm 0.68$ \\
   ST2L & $\Sigma_{\rm SFR(\ha)}$, T2, $\lyb$ & $ 24$ & $2.292$ & $ -2.84 \pm 0.76$ \\
\\
   ST3 & $\Sigma_{\rm SFR(\ha)}$, T3 & $ 28$ & $2.274$ & $  0.76 \pm 0.59$ \\
   ST3L & $\Sigma_{\rm SFR(\ha)}$, T3, $\lyb$ & $ 25$ & $2.283$ & $ -1.58 \pm 0.60$ \\
   \hline
   sST1 & $\Sigma_{\rm sSFR(\ha)}$, T1 & $ 26$ & $2.281$ & $ -0.88 \pm 1.04$ \\
   sST1L & $\Sigma_{\rm sSFR(\ha)}$, T1, $\lyb$ & $ 24$ & $2.297$ & $ -3.88 \pm 0.96$ \\
\\
   sST2 & $\Sigma_{\rm sSFR(\ha)}$, T2 & $ 26$ & $2.296$ & $ -6.93 \pm 0.69$ \\
   sST2L & $\Sigma_{\rm sSFR(\ha)}$, T2, $\lyb$ & $ 24$ & $2.314$ & $ -7.19 \pm 0.74$ \\
\\
   sST3 & $\Sigma_{\rm sSFR(\ha)}$, T3 & $ 28$ & $2.268$ & $  3.89 \pm 0.51$ \\
   sST3L & $\Sigma_{\rm sSFR(\ha)}$, T3, $\lyb$ & $ 25$ & $2.289$ & $  1.22 \pm 0.53$
\enddata 
\tablenotetext{a}{Criteria used to construct the subsamples: those
  with ``$\lyb$'' include only galaxies with coverage of $\lyb$
  absorption (i.e., $z_{\rm sys}>2.12$), while those with ``$\lyb$,
  neb'' include only galaxies with coverage of $\lyb$ absorption and
  complete coverage of the optical nebular emission lines.  Subsamples resulting from dividing a sample
  into thirds are denoted by T1, T2, and T3, ordered by increasing
  value of the parameter used to divide galaxies into each subsample.}
\tablenotetext{b}{Number of galaxies in the subsample.}
\tablenotetext{c}{Average systematic redshift of galaxies in the subsample.}
\tablenotetext{d}{Average of individually-measured $\wlya$ for 
    galaxies contributing to the subsample.}
\label{tab:compositestats}
\end{deluxetable}

Higher S/N measurements and inferences of mean spectral properties
were obtained by averaging individual galaxy spectra to produce
composite spectra.  The procedures used for constructing composite FUV
and optical spectra are specified in \citet{reddy16a} and
\citet{reddy20}, respectively, and are summarized here.  Galaxies were
first grouped together based on a certain property of interest (e.g.,
$\wlya$, star-formation-rate surface density).  The FUV and optical
spectra for all galaxies in a given grouping were shifted to the rest
frame based on $z_{\rm sys}$, converted to units of luminosity
density, and spline-interpolated to a linear wavelength grid with a
spacing $\delta\lambda_{0} = 0.5$\,\AA.  At each grid wavelength
point, the luminosity densities were averaged after rejecting values
that differ from the median luminosity density by more than $3\sigma$.

No weighting was applied in the averaging to ensure that every galaxy
contributes equally to the composite spectrum and that the predicted
mean IGM+CGM opacity---which is computed assuming all sightlines
contribute equally to the mean decrement shortward of $\lya$---can be
confidently used to correct the composite spectrum
\citep{steidel18}.\footnote{The method of normalizing each individual
  galaxy spectrum by the luminosity of the most frequently detected
  (or highest $S/N$) line of a line ratio can be used to ensure that a
  line ratio measured from a composite spectrum is mathematically
  identical to the average of individual galaxy line ratios.  However,
  normalizing individual galaxy spectra in this manner prevents the
  inclusion of galaxies where the normalizing line may be undetected
  and complicates the IGM+CGM opacity correction to the composite FUV
  spectrum.  As a result, we do not normalize the spectra in this
  manner, opting instead to check that the line ratios measured from
  the composite spectra are consistent with the average of individual
  galaxy line ratios within the measurement uncertainties of those
  ratios.  This is the case for all of the line ratios measured from
  composite spectra of galaxies where all of the relevant lines are
  detected individually.}  The foreground IGM+CGM transmission curve
appropriate for the mean redshift of the galaxies forming the
composite spectrum was derived from the $\nhi$ distribution of
intervening absorbers as a function of redshift given in
\citet{rudie13} (e.g., \citealt{shapley06, reddy16b, steidel18}).  The
composite FUV spectrum was divided by this transmission curve to
correct for the mean foreground IGM+CGM opacity.

The error in the mean luminosity density at each dispersion point
(i.e., the composite error spectrum) reflects both measurement
uncertainties (from the error spectra for individual objects) and the
variance in luminosity densities of objects contributing to each
dispersion point.  The composite FUV and optical spectra for the
entire sample are presented in Figure~\ref{fig:compspec}.

Table~\ref{tab:compositestats} lists the subsets of the sample for
which composite spectra were constructed, along with the number of
objects in each subset and their mean redshifts and $\wlya$.  For
reference, the latter were calculated by simply averaging the
individual $\wlya$ measurements of the galaxies comprising the subsets
(Section~\ref{sec:wlyameas}).  The $\langle\wlya\rangle$ measured
directly from the composite spectra using the procedure described in
\citealt{kornei10}) are systematically lower by $\simeq 7$\,\AA\, than
those computed from the individual measurements owing to the presence
of significantly-detected interstellar $\hi$ absorption underlying the
$\lya$ emission line in the composite spectra (e.g., as seen in
Figure~\ref{fig:compspec}).  Consequently, while $\langle
\wlya\rangle$ is relatively straightforward to compute, differences in
exactly how it is calculated can lead to some ambiguity in its
relation to the production and escape of $\lya$ photons.  Thus, the
$\wlya$ of the emission line itself, $\wlyaem$, and the escape
fraction of $\lya$ photons, $\lyafrac$, are also considered below
(Section~\ref{sec:fesclya}).

\section{\bf MODELING OF THE COMPOSITE FUV AND OPTICAL SPECTRA}
\label{sec:modelfitting}

A central focus of this analysis is to identify the primary factors in
the production and transmission of Ly$\alpha$ (and LyC) photons.  For
the most part, these factors are constrained by fitting SPS and
simplified ISM models to composite FUV spectra, and through
photoionization modeling of nebular emission lines in the composite
optical spectra.  The fitting proceeded in three steps.  In the first
step, the BPASS models were fit to the composite FUV spectra to
constrain stellar metallicities, ages, and continuum reddening of the
stellar population dominating the FUV light.  In the second step,
photoionization modeling of the emission line ratios measured from the
composite optical spectra was used to deduce ionization parameters,
gas-phase oxygen abundances, and nebular reddening.  Additionally, the
residual $\heii$\,$\lambda 1640$ nebular emission in the composite FUV
spectra was used to constrain the BPASS model type (i.e., the
high-mass cutoff of the IMF and stellar binarity).  The parameters
obtained in the first two fitting steps determine the shape of the
ionizing spectrum.  In the third step, models of the neutral
ISM were fit to the composite FUV spectra to infer the line-of-sight
reddening, $\hi$ column densities, and $\hi$ covering fractions.
These three steps---and the correlations between $\wlya$ and the
aforementioned parameters---are described, respectively, in
Sections~\ref{sec:fuvfitting}, \ref{sec:optfitting}, and
\ref{sec:ismfitting}.  The fitting results are summarized in
Section~\ref{sec:fittingsummary}.  Readers who wish to skip the
details of the fitting procedures can proceed to
Section~\ref{sec:fittingsummary}.

\subsection{SPS Modeling and Results}
\label{sec:fuvfitting}

\subsubsection{SPS Modeling Procedure}
\label{sec:fuvfittingprocedure}

Composite FUV spectra were fit with the BPASS 100bin, 300bin, 100sin,
and 300sin SPS models (see Section~\ref{sec:sedmodeling}) using a
methodology similar to that presented in \citet{steidel16} and
\citet{topping20a}, with a few modifications described below.  Similar
to the aforementioned studies, we adopted a constant star-formation
history (SFH).  While there exist several different
physically-relevant models for the SFHs of star-forming galaxies
(e.g., \citealt{chisholm19}), we chose a constant SFH based on
findings that such a SFH (or slowly-rising SFHs) provide a reasonable
approximation of the mean SFH of ensembles of typical star-forming
galaxies at $z\ga 1.5$ \citep{papovich11, reddy12b}.  We also
considered ages of $\log[{\rm Age/yr}] = 7.0$, $7.3$, $7.5$, $7.6$,
$7.8$, $8.0$, $8.5$, and $9.0$.  The publicly available BPASS version
2.2.1 distribution includes model spectra computed at stellar
metallicities $Z_\ast = 10^{-5}$ to $0.04$, expressed in terms of the
mass fraction of metals where, for reference, the solar value is
$Z_{\odot} = 0.0142$ \citep{asplund09}.  The provided spectra were
interpolated to construct a new grid of models with $Z_\ast$ that more
finely sample (and bracket) the values expected for $z\sim 2$ galaxies
\citep{steidel16}.  Specifically, we considered stellar metallicities
of $Z_\ast = 10^{-4}$, $3\times 10^{-4}$, $5\times 10^{-4}$, $7\times
10^{-4}$, $9\times 10^{-4}$, 0.0010, 0.0012, 0.0014, 0.0016, 0.0018,
0.0020, 0.0022, 0.0024, 0.0026, 0.0028, and 0.0030.  Thus, we obtain a
grid of SPS models with different stellar metallicities and ages.
Nebular continuum emission was added to each SPS model in the grid (as
described in \citealt{topping20a}) to produce an ``SPSneb'' model.

Finally, because of the lower resolution of the blue-side LRIS spectra
relative to the red-side (Section~\ref{sec:mosdeflrissurvey}), the
SPSneb models were smoothed to match the resolution of the former at
$\lambda_0 < 1500$\,\AA, corresponding to the rest-frame wavelength of
the dichroic cutoff for the mean redshift of the sample, $\langle
z\rangle \sim 2.3$.  No smoothing was applied at wavelengths redder
than $1500$\,\AA.

\begin{figure*}
\epsscale{1.0}
\plotone{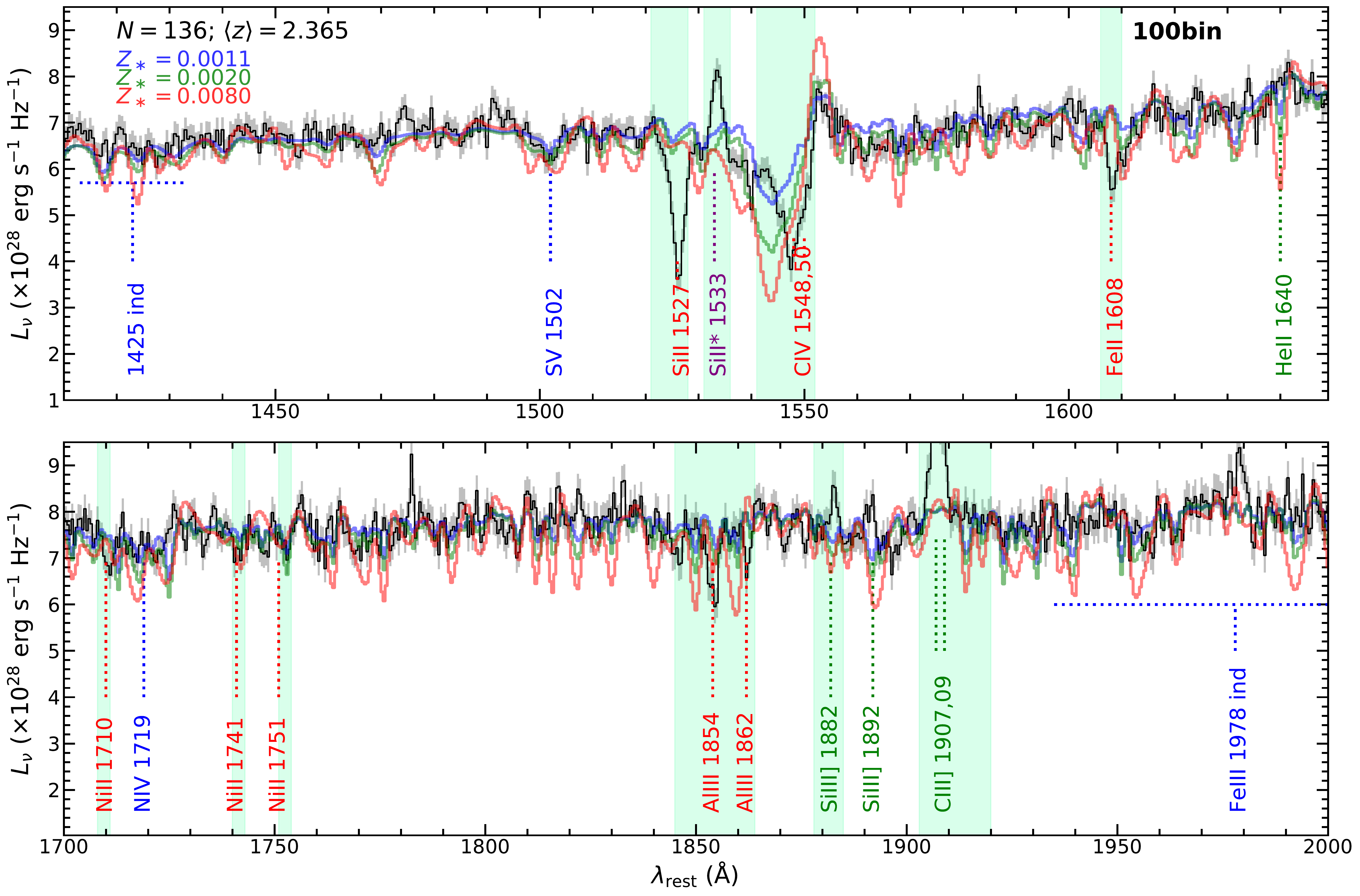}
\caption{Comparison of the composite FUV spectrum of the entire sample
  (black, with the $1\sigma$ error in gray) and the 100bin SPSneb
  models for $\log[{\rm Age/yr}] = 8.0$ and different metallicities,
  focused on the wavelength regions around \ion{C}{4}\,$\lambda\lambda
  1548, 1550$ (top) and $\lambda_{\rm rest} = 1700-2000$\,\AA\,
  (bottom).  Line labeling is similar to Figure~\ref{fig:compspec},
  where we include the fine structure emission line
  \ion{Si}{2}$^\ast$\,$\lambda 1533$.  Regions not included in the
  fitting are indicated by the light green shaded regions.}
\label{fig:zstarageissue1}
\end{figure*}

\begin{figure}
\epsscale{1.0}
\plotone{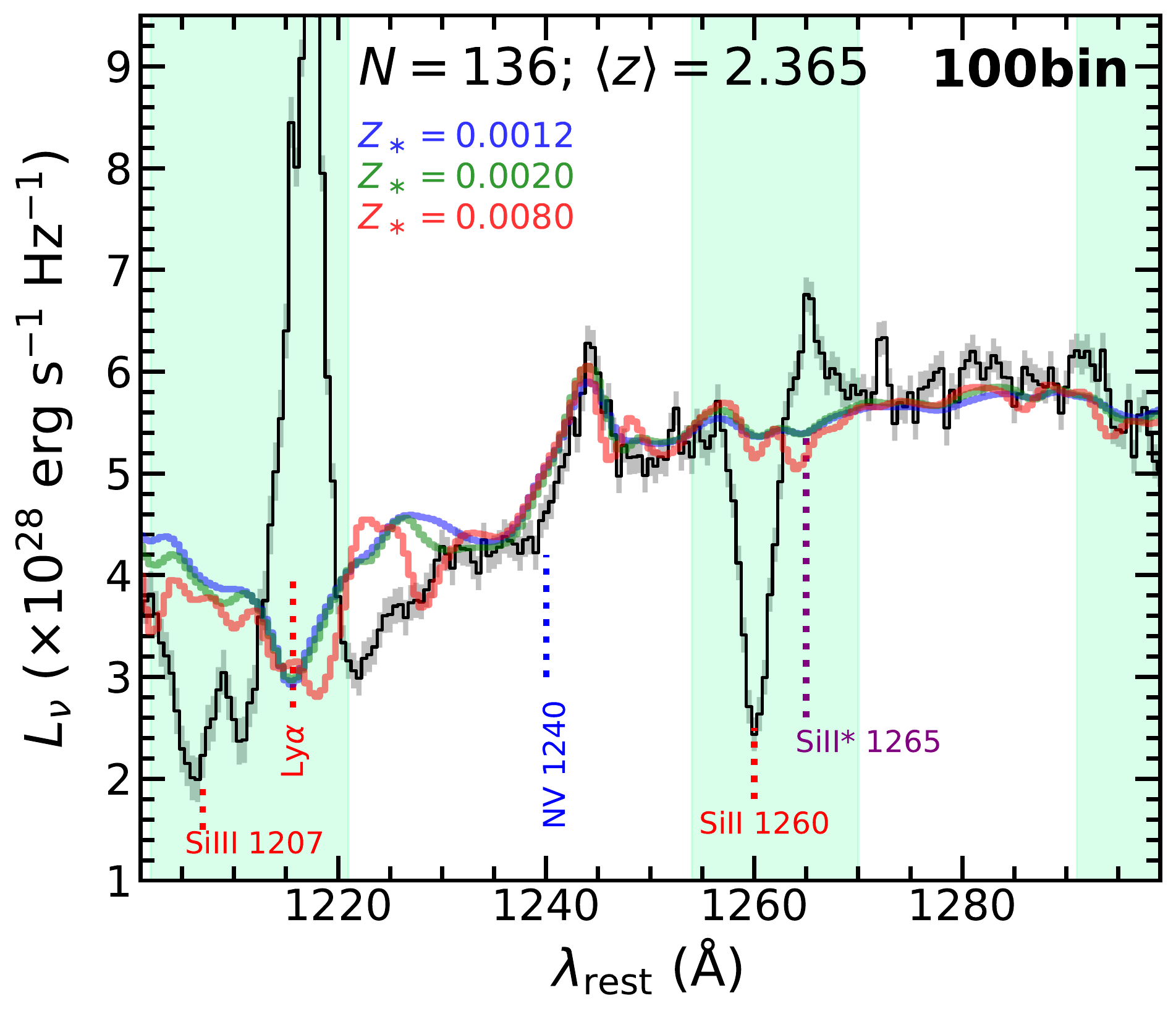}
\caption{Comparison of the composite FUV spectrum of the entire sample
  (black, with the $1\sigma$ error in gray) and the 100bin SPSneb
  models for $\log[{\rm Age/yr}] = 8.0$ and different metallicities,
  focused on the wavelength region around \ion{N}{5}\,$\lambda 1240$.
  The models shown here do not include interstellar $\hi$ absorption.
  Line labeling is similar to Figure~\ref{fig:compspec}, where we
  include the fine structure emission line
  \ion{Si}{2}$^\ast$\,$\lambda 1265$.  Regions not included in the
  fitting are indicated by the light green shaded regions.}
\label{fig:nv}
\end{figure}

Appendix~\ref{sec:sfrcomparison} presents a comparison of the SFRs
obtained with the different continuum attenuation curves, $\sfrsed$,
and those derived from H$\alpha$, $\sfrha$.  This comparison indicates
that only the SMC extinction curve results in UV-based SFRs fully
consistent with the $\ha$-based ones for the galaxy ensembles
considered in this study, irrespective of the BPASS model type
(100bin, 300bin, 100sin, 300sin models; see also \citealt{theios19}).
Thus, the SMC extinction curve---updated in the wavelength range
$\lambda = 950 - 1250$\,\AA\, based on the analysis of
\citet{reddy16a}---was assumed in fitting the SPSneb models to the
composite FUV spectra.  Each SPSneb model was reddened assuming a
range of stellar continuum reddening $\ebmvcont = 0.00-0.40$.  The
$\chi^2$ of each model of a given $Z_{\ast}$, age, and reddening is
\begin{equation}
\chi^2 = \sum_i \left[\frac{l(i) - m(i)}{\sigma(i)}\right]^{2}
\end{equation}
where $l(i)$ is the luminosity density of the composite FUV spectrum,
$m(i)$ is the luminosity density of the model spectrum, and
$\sigma(i)$ is the error in luminosity density, at wavelength point
$i$.  Wavelength points in the ``Mask 1'' windows given in
\citet{steidel16} were used in computing the $\chi^2$ statistic, and
only for those windows for which all galaxies in the composite
spectrum have coverage.  Furthermore, the model spectrum was
normalized to have the same median luminosity density as the composite
FUV spectrum in the Mask 1 windows---the normalization of the model
yields an estimate of the SFR.  The best-fit values of $Z_\ast$, age,
reddening, and SFR were set equal to the mean values obtained when
fitting many realizations of the composite spectrum generated through
random sampling of the galaxy spectra with replacement.  Uncertainties
in parameters were conservatively set equal to the standard deviation
of the values obtained by fitting these realizations---which are
larger than the standard errors in the mean values---and are thus
effectively marginalized over the uncertainties of other fitted
parameters.  This method of deriving the uncertainties applies to all
of the line measurements and model parameters obtained in this work.

\subsubsection{Modeling Constraints on $Z_\ast$ and Age}
\label{sec:zstarandageissues}

Figures~\ref{fig:zstarageissue1} and \ref{fig:nv} show the 100bin SPSneb model fits to
the composite FUV spectrum of all the galaxies in our
sample.\footnote{The fitting to the composite FUV spectrum of all
  galaxies was performed by considering the \citet{steidel16} Mask 1
  windows in the range $\lambda \simeq 1100 - 1600$\,\AA, for which
  all galaxies have wavelength coverage.  For illustrative purposes,
  however, we show the comparison of these fits to the composite
  spectrum in longer wavelength windows (right panel of
  Figure~\ref{fig:zstarageissue1}) to which galaxies at $z\la 2.6$
  primarily contribute.  In practice, the fitting performed using all
  windows is identical to that performed using the subset of windows
  for which all galaxies have wavelength coverage since the sample is
  dominated by galaxies at $2.0\la z\la 2.6$.}  The best-fit continuum
reddening, stellar metallicity, and age obtained with the 100bin model
are $\langle\ebmvcont\rangle = 0.099\pm0.005$, $\langle Z_\ast\rangle
= 0.0011\pm 0.0003$ ($0.076\pm0.018$\,$Z_\odot$), and $\langle
\log[{\rm Age/yr}]\rangle = 8.0\pm 0.2$.  This best-fit model is shown
by the blue spectrum.  The best-fit models of the same age, but with
$Z_\ast = 0.0020$ and $Z_\ast = 0.0080$ are also shown.

While all the models provide an adequate fit to the $\nv$\,$\lambda
1240$ P-Cygni emission feature (Figure~\ref{fig:nv}), comparison of
these models clearly demonstrates the strong preference for lower
stellar metallicities when fitting the entire composite FUV spectrum
(see also \citealt{topping20a}).  For instance, the blend of Si, C,
and Fe photospheric lines centered at $\simeq 1425$\,\AA\, (i.e., the
1425 index; \citealt{rix04}); the depths of \ion{S}{5}\,$\lambda 1502$
and \ion{N}{5}\,$\lambda 1719$; and the \ion{Fe}{3} blend centered at
$\simeq 1978$\,\AA\, (i.e., the 1978 index; \citealt{rix04}) are all
significantly weaker than the predictions of models with $Z_\ast\ga
0.002$.  Moreover, the weak photospheric absorption seen across the
full FUV wavelength range (excluding regions affected by interstellar
absorption or nebular emission) is most consistent with the $Z_\ast
\la 0.002$ models.  Formally, the $Z_\ast = 0.008$ ($\approx
0.56$\,$Z_\odot$) model is excluded with $\simeq 8\sigma$
significance.

Similar fitting of the composite FUV spectrum of $30$ star-forming
galaxies at $z\sim 2.3$ from the Keck Baryonic Structure Survey (KBSS)
also indicates a low stellar metallicity, $Z_\ast \simeq 0.001-0.002$
(\citealt{steidel16}; see also \citealt{cullen19} and
\citealt{topping20a, topping20b}).  Interestingly, the KBSS composite
FUV spectrum exhibits very weak \ion{C}{4} P-Cygni {\em emission},
possibly because of blending with interstellar \ion{C}{4}
absorption \citep{steidel16}.  The strong \ion{C}{4} P-Cygni
emission observed in the composite FUV spectrum of our
sample---similar in strength to the predictions of the $Z_\ast \la
0.002$ models which best reproduce the overall level of photospheric
absorption---may suggest a narrower velocity distribution and/or lower
covering fraction of the \ion{C}{4}-bearing gas in the ISM.

\begin{turnpage}
\begin{deluxetable*}{p{1.2cm}ccccccccccc}
\tabletypesize{\footnotesize}
\tablecaption{SPSneb and ISM Model Fit Results}
\tablehead{
\colhead{Subsample} &
\colhead{$\langle\ebmvcont\rangle$\tablenotemark{a}} &
\colhead{$\langle Z_\ast/Z_\odot\rangle$\tablenotemark{b}} &
\colhead{$\left\langle\log\left[\frac{\rm Age}{\rm yr}\right]\right\rangle$\tablenotemark{c}} &
\colhead{$\langle \ebmvneb\rangle$\tablenotemark{d}} &
\colhead{$\langle Z_{\rm neb}/Z_\odot\rangle$\tablenotemark{e}} &
\colhead{$\langle\log U\rangle$\tablenotemark{f}} &
\colhead{$\langle\ebmvlos\rangle$\tablenotemark{g}} &
\colhead{$\langle \fcovhi\rangle$\tablenotemark{h}} &
\colhead{$\left\langle \lognhitable\right\rangle$\tablenotemark{i}} &
\colhead{$\langle\wlyaem\rangle$ (\AA)\tablenotemark{j}} &
\colhead{$\langle \lyafrac\rangle$\tablenotemark{k}}
}
\startdata
A & $0.099\pm0.005$ & $0.076\pm0.018$ & $8.0\pm0.2$ \comment{& $0.35\pm0.06$ & $0.38\pm0.08$ & $-2.99\pm 0.03$ & $0.096\pm0.014$ & $0.97\pm0.02$ & $20.7\pm 0.3$ & $9.75\pm1.82$ & $ 0.014\pm 0.003$ & $ 260\pm  79$ & $ 383\pm 283$} \\
AL & $0.099\pm0.005$ & $0.075\pm0.017$ & $8.1\pm0.2$ \comment{& $0.36\pm0.06$ & $0.39\pm0.07$ & $-2.98\pm 0.04$} & & & & $0.096\pm0.013$ & $0.97\pm0.02$ & $20.7\pm 0.2$ & $9.11\pm1.67$ \comment{& $ 0.014\pm 0.003$ & $ 212\pm  84$ & $ 354\pm 305$} \\
ALN & $0.101\pm0.006$ & $0.082\pm0.017$ & $8.2\pm0.3$ & $0.33\pm0.04$ & $0.38\pm0.09$ & $-2.93\pm 0.05$ & $0.089\pm0.012$ & $0.99\pm0.01$ & $20.7\pm 0.2$ & $8.65\pm1.36$ & $0.014\pm 0.003$ \comment{& $ 267\pm 105$ & $ 279\pm 223$} \\
\hline
WT1 & $0.116\pm0.009$ & $0.082\pm0.019$ & $8.5\pm0.2$ \comment{& $0.30\pm0.09$ & $0.32\pm0.10$ & $-2.99\pm 0.07$ & $0.114\pm0.014$ & $0.98\pm0.02$ & $21.1\pm 0.2$ & $0.00\pm0.83$ & $ 0.001\pm 0.001$ & $ 221\pm 160$ & $ 102\pm 555$} \\
WT1L & $0.113\pm0.011$ & $0.097\pm0.019$ & $8.4\pm0.3$ \comment{& $0.32\pm0.11$ & $0.35\pm0.11$ & $-2.96\pm 0.08$} & & & & $0.104\pm0.010$ & $0.98\pm0.02$ & $21.0\pm 0.2$ & $0.00\pm0.89$ \comment{& $ 0.001\pm 0.001$ & $ 257\pm 152$ & $ 138\pm 380$} \\
WT1LN & $0.115\pm0.011$ & $0.080\pm0.021$ & $8.4\pm0.3$ & $0.29\pm0.07$ & $0.37\pm0.11$ & $-3.01\pm 0.07$ & $0.102\pm0.015$ & $0.99\pm0.02$ & $21.1\pm 0.2$ & $0.00\pm0.97$ & $ 0.001\pm 0.002$ \comment{& $ 256\pm 205$ & $ 281\pm1213$} \\
\\
WT2 & $0.095\pm0.007$ & $0.126\pm0.022$ & $8.2\pm0.2$ \comment{& $0.44\pm0.10$ & $0.40\pm0.02$ & $-3.09\pm 0.06$ & $0.098\pm0.010$ & $0.97\pm0.03$ & $20.1\pm 0.6$ & $6.50\pm1.51$ & $ 0.009\pm 0.003$ & $ 289\pm 129$ & $ 434\pm 481$} \\
WT2L & $0.093\pm0.007$ & $0.119\pm0.023$ & $8.3\pm0.2$ \comment{& $0.45\pm0.10$ & $0.41\pm0.03$ & $-3.11\pm 0.07$} & & & & $0.100\pm0.012$ & $0.96\pm0.02$ & $20.0\pm 0.7$ & $5.78\pm1.69$ \comment{& $ 0.008\pm 0.003$ & $ 283\pm 150$ & $ 322\pm 740$} \\
WT2LN & $0.099\pm0.011$ & $0.123\pm0.032$ & $8.3\pm0.2$ & $0.48\pm0.11$ & $0.41\pm0.04$ & $-3.15\pm 0.08$ & $0.096\pm0.014$ & $0.97\pm0.03$ & $20.7\pm 0.2$ & $4.89\pm1.37$ & $ 0.006\pm 0.003$ \comment{& $ 386\pm 194$ & $ 247\pm 313$} \\
\\
WT3 & $0.073\pm0.007$ & $0.079\pm0.015$ & $7.7\pm0.2$ \comment{& $0.30\pm0.09$ & $0.34\pm0.15$ & $-2.74\pm 0.07$ & $0.095\pm0.013$ & $0.90\pm0.04$ & $20.3\pm 0.7$ & $24.40\pm3.37$ & $ 0.041\pm 0.012$ & $ 211\pm 200$ & $ 711\pm2566$} \\
WT3L & $0.078\pm0.009$ & $0.078\pm0.017$ & $7.7\pm0.2$ \comment{& $0.34\pm0.10$ & $0.40\pm0.13$ & $-2.77\pm 0.07$} & & & & $0.093\pm0.017$ & $0.90\pm0.03$ & $20.2\pm 0.4$ & $22.50\pm2.64$ & $ 0.035\pm 0.009$ \comment{& $ 104\pm 175$ & $ 999\pm3049$} \\
WT3LN & $0.091\pm0.010$ & $0.087\pm0.033$ & $7.8\pm0.2$ & $0.33\pm0.07$ & $0.44\pm0.08$ & $-2.74\pm 0.07$ & $0.093\pm0.017$ & $0.90\pm0.03$ & $19.7\pm 0.5$ & $17.99\pm3.06$ & $ 0.031\pm 0.006$ \comment{& $ 175\pm  94$ & $ 617\pm 621$} \\
\hline
ST1 & $0.100\pm0.010$ & $0.115\pm0.042$ & $8.2\pm0.2$ \comment{& $0.24\pm0.10$ & $0.31\pm0.15$ & $-3.05\pm 0.06$ & $0.100\pm0.020$ & $0.97\pm0.03$ & $20.7\pm 0.3$ & $7.02\pm2.44$ & $ 0.019\pm 0.009$ & $ 270\pm 195$ & $ 615\pm1647$} \\
ST1L & $0.095\pm0.009$ & $0.091\pm0.032$ & $8.2\pm0.2$ & $0.24\pm0.13$ & $0.28\pm0.13$ & $-3.06\pm 0.07$ & $0.100\pm0.022$ & $0.95\pm0.03$ & $20.8\pm 0.3$ & $5.49\pm1.98$ & $ 0.015\pm 0.005$ \comment{& $ 298\pm 176$ & $ 289\pm2408$} \\
\\
ST2 & $0.113\pm0.011$ & $0.117\pm0.034$ & $8.1\pm0.3$ \comment{& $0.44\pm0.06$ & $0.41\pm0.04$ & $-3.09\pm 0.07$ & $0.115\pm0.013$ & $0.97\pm0.01$ & $20.1\pm 0.7$ & $6.60\pm 2.55$ & $ 0.007\pm 0.003$ & $ 237\pm 147$ & $ 410\pm 426$} \\
ST2L & $0.122\pm0.012$ & $0.146\pm0.059$ & $8.2\pm0.3$ & $0.45\pm0.07$ & $0.40\pm0.03$ & $-3.10\pm 0.06$ & $0.117\pm0.006$ & $0.98\pm0.01$ & $20.3\pm 0.7$ & $6.36\pm2.45$ & $ 0.007\pm 0.003$ \comment{& $ 218\pm 126$ & $ 253\pm 326$} \\
\\
ST3 & $0.083\pm0.007$ & $0.111\pm0.020$ & $7.8\pm0.2$ \comment{& $0.39\pm0.06$ & $0.45\pm0.06$ & $-2.70\pm 0.06$ & $0.077\pm0.020$ & $0.95\pm0.03$ & $20.6\pm 0.4$ & $12.19\pm4.46$ & $ 0.015\pm 0.007$ & $ 348\pm 164$ & $ 337\pm 311$} \\
ST3L & $0.086\pm0.009$ & $0.118\pm0.028$ & $8.0\pm0.2$ & $0.38\pm0.07$ & $0.43\pm0.08$ & $-2.70\pm 0.07$ & $0.082\pm0.017$ & $0.92\pm0.02$ & $20.8\pm 0.2$ & $10.55\pm3.62$ & $ 0.013\pm 0.003$ \comment{& $ 287\pm 142$ & $ 343\pm 352$} \\
\hline
sST1 & $0.127\pm0.012$ & $0.079\pm0.022$ & $7.9\pm0.2$ \comment{& $0.35\pm0.08$ & $0.31\pm0.15$ & $-3.16\pm 0.06$ & $0.123\pm0.004$ & $0.99\pm0.02$ & $20.3\pm 0.6$ & $7.17\pm2.66$ & $ 0.012\pm 0.006$ & $ 193\pm 193$ & $ 463\pm1243$} \\
sST1L & $0.123\pm0.011$ & $0.063\pm0.014$ & $7.8\pm0.3$ & $0.39\pm0.08$ & $0.34\pm0.14$ & $-3.16\pm 0.07$ & $0.122\pm0.005$ & $0.99\pm0.02$ & $20.4\pm 0.2$ & $6.28\pm2.43$ & $ 0.009\pm 0.003$ \comment{& $ 237\pm 193$ & $ 340\pm 598$} \\
\\
sST2 & $0.095\pm0.010$ & $0.163\pm0.052$ & $8.4\pm0.3$ \comment{& $0.40\pm0.07$ & $0.41\pm0.04$ & $-2.96\pm 0.07$ & $0.101\pm0.021$ & $0.95\pm0.02$ & $20.9\pm 0.3$ & $5.20\pm2.23$ & $ 0.007\pm 0.003$ & $ 268\pm 104$ & $ 484\pm 704$} \\
sST2L & $0.099\pm0.012$ & $0.167\pm0.055$ & $8.4\pm0.3$ & $0.38\pm0.08$ & $0.40\pm0.02$ & $-2.95\pm 0.07$ & $0.101\pm0.008$ & $0.95\pm0.02$ & $20.9\pm 0.2$ & $5.17\pm2.34$ & $ 0.007\pm 0.003$ \comment{& $ 247\pm 119$ & $ 324\pm 748$} \\
\\
sST3 & $0.076\pm0.006$ & $0.108\pm0.018$ & $8.1\pm0.2$ \comment{& $0.29\pm0.06$ & $0.42\pm0.11$ & $-2.67\pm 0.06$ & $0.067\pm0.005$ & $0.91\pm0.03$ & $20.3\pm 0.6$ & $12.66\pm3.97$ & $ 0.023\pm 0.008$ & $ 361\pm 200$ & $ 434\pm 785$} \\
sST3L & $0.080\pm0.007$ & $0.115\pm0.014$ & $8.1\pm0.2$ & $0.30\pm0.06$ & $0.41\pm0.08$ & $-2.70\pm 0.04$ & $0.067\pm0.005$ & $0.91\pm0.02$ & $20.4\pm 0.3$ & $11.30\pm3.90$ & $ 0.020\pm 0.003$ \comment{& $ 276\pm 126$ & $ 326\pm 941$} 
\enddata
\tablenotetext{a}{Continuum reddening assuming the SMC extinction curve \citep{gordon03}.}
\tablenotetext{b}{Stellar metallicity in solar units, assuming $Z_\odot = 0.0142$ \citep{asplund09}.}
\tablenotetext{c}{Age.}
\tablenotetext{d}{Nebular reddening assuming the \citet{cardelli89} extinction curve.}
\tablenotetext{e}{Nebular oxygen abundance in solar units.}
\tablenotetext{f}{Ionization parameter.}
\tablenotetext{g}{Line-of-sight reddening of the gas covering the continuum in the clumpy ISM model, assuming the SMC extinction curve.}
\tablenotetext{h}{$\hi$ covering fraction in the clumpy ISM model.}
\tablenotetext{i}{$\hi$ column density in the clumpy ISM model.}
\tablenotetext{j}{$\lya$ emission-line equivalent width.}
\tablenotetext{k}{Line-of-sight escape fraction of $\lya$ photons.}
\label{tab:params}
\end{deluxetable*}
\end{turnpage}

\begin{figure}
\epsscale{1.2}
\plotone{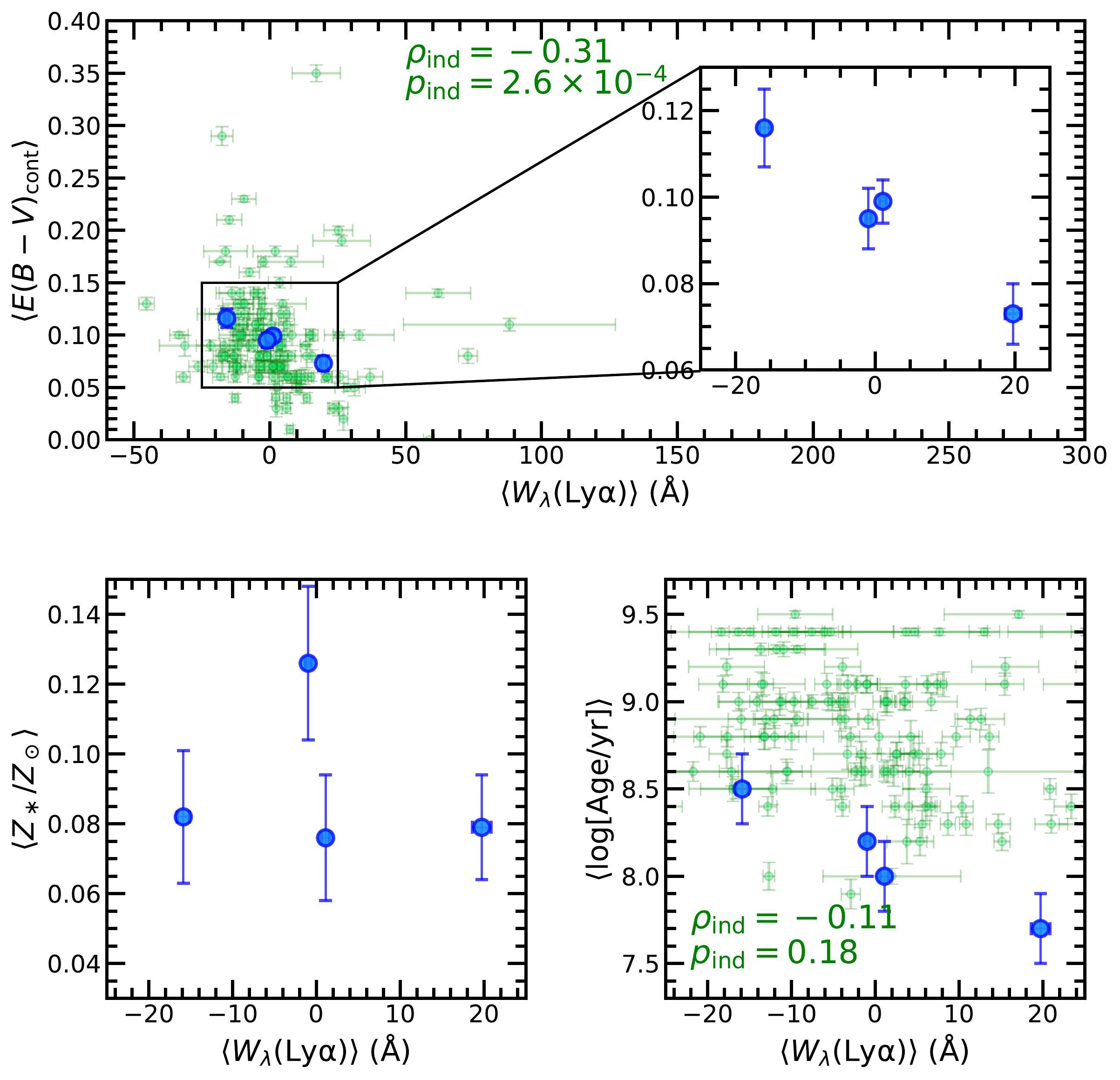}
\caption{Variation of $\langle\ebmvcont\rangle$ (top), $\langle
  Z_\ast/Z_\odot\rangle$ (bottom left), and $\langle \log[{\rm
      Age/yr}]\rangle$ (bottom right) assuming the 100bin models, with
  $\langle \wlya\rangle$ for the A, WT1, WT2, and WT3 subsamples.  The
  $\ebmvcont$ and $\log[{\rm Age/yr}]$ (obtained from SED fitting;
  Section~\ref{sec:sedmodeling}) for individual galaxies are denoted
  by the light green points in the top and bottom right panels,
  respectively.  Also indicated are the Spearman $\rho$ correlation
  coefficient ($\rho_{\rm ind}$) and $p$-value ($p_{\rm ind}$) between
  $\ebmvcont$ and $\wlya$, and $\log[{\rm Age/yr}]$ and $\wlya$, for
  individual galaxies.  The inset panel on the top zooms in on the
  $\langle\ebmvcont\rangle$ and $\langle\wlya\rangle$ measurements
  obtained from the composite FUV spectra.}
\label{fig:wlyavssps}
\end{figure}

The preference for subsolar stellar metallicities persists
irrespective of the age of the stellar population (e.g.,
\citealt{topping20a}).  Additionally, the variations in $Z_\ast$ and
age that are obtained with different BPASS model type are negligible
compared to the random uncertainties in these parameters
\citep{steidel16}.  Furthermore, $Z_\ast$ and age are primarily
determined by the overall level of photospheric line blanketing in the
FUV and, as such, are relatively insensitive to the assumed continuum
dust attenuation curve.  Because the FUV spectrum is dominated by the
light from the youngest stellar populations, the ages derived from the
spectral fitting are a factor of $\simeq 2-3\times$ lower than---but
still correlate with---those derived from the broadband SED fitting
(Section~\ref{sec:sedmodeling}; see also Section~\ref{sec:wlyavssps}).
The former are most relevant for our analysis since the youngest
stellar populations dictate the shape of the ionizing spectrum.

\subsubsection{Correlations between $\wlya$ and SPS Model Parameters}
\label{sec:wlyavssps}

The procedure described in Section~\ref{sec:fuvfittingprocedure} was
used to fit the composite FUV spectra for subsamples A, AL, ALN, and
the nine subsamples constructed in bins of $\wlya$
(Table~\ref{tab:compositestats}).  The best-fit values of $\langle
\ebmvcont\rangle$, $\langle Z_\ast/Z_\odot\rangle$, and $\langle
\log[{\rm Age/yr}]\rangle$ obtained from fitting the 100bin SPSneb
models are listed in Table~\ref{tab:params}.  The variations in
$\langle\wlya\rangle$ with the aforementioned parameters are presented
in Figure~\ref{fig:wlyavssps}.

The reddening of the stellar continuum, parameterized by $\langle
\ebmvcont\rangle$, clearly anti-correlates with $\langle\wlya\rangle$, such
that objects with higher $\langle \wlya\rangle$ exhibit significantly
bluer $\langle\ebmvcont\rangle$.  The composite FUV spectrum with
$\langle\wlya\rangle \approx 20$\,\AA\, (subsample WT3) has
$\langle\ebmvcont\rangle = 0.073\pm 0.007$, roughly $0.043$\,mag bluer
than the composite with $\langle\wlya\rangle \approx -20$\,\AA\,
(subsample WT1).  
Throughout the analysis, we focus on the significance of the
difference in the measurements made on composite spectra that contain
independent sets of galaxies, such as subsamples formed from the lower
and upper third of the $\wlya$ distribution of individual galaxies
(e.g., subsamples WT1 and WT3, subsamples WT1LN and WT3, and so on).

The top panel of Figure~\ref{fig:wlyavssps} also shows the
$\ebmvcont$ and $\wlya$ measured for individual galaxies in the
sample, where the former were inferred from broadband SED fitting
(Section~\ref{sec:sedmodeling}).  The averages of these individual
$\ebmvcont$ measurements in bins of $\wlya$ are similar within the
uncertainties to the $\langle \ebmvcont\rangle$ of the composite
spectra.  The Spearman rank correlation coefficient between
$\ebmvcont$ and $\wlya$ for individual galaxies is $\rho_{\rm ind} =
-0.31$, with a probability of $p_{\rm ind} = 2.6\times 10^{-4}$ that
the two variables are uncorrelated.

As noted in Section~\ref{sec:zstarandageissues}, the modeling of the
composite FUV spectra strongly prefers sub-solar stellar
metallicities: these metallicities vary from $\langle
Z_\ast/Z_\odot\rangle \simeq 0.07$ to $0.12$ for the composites formed
in bins of $\wlya$.  However, there is no
significant correlation between $\langle Z_\ast\rangle$ and $\langle
\wlya\rangle$ for the ensembles analyzed here.  For example, the two
composites with the most extreme values of $\langle\wlya\rangle$,
namely subsamples WT1LN and WT3, have $\langle Z_\ast/Z_\odot \rangle =
0.080\pm 0.021$ and $0.079\pm 0.015$, respectively.  

On the other hand, the composite FUV spectra indicate a marginally
significant anti-correlation between $\langle \log[{\rm
    Age/yr}]\rangle$ and $\langle \wlya\rangle$.  Composites with
$\langle \wlya\rangle > 0$ have $\langle\log[{\rm Age/yr}]\rangle \la
8.0$, while those with net $\lya$ in absorption have $\langle\log[{\rm
    Age/yr}]\rangle \ga 8.3$.  Formally, the number of degrees of
freedom in the fit (determined by the number of wavelength points
considered in the fitting, $\nu = 1060$) implies a difference in
reduced $\chi^2$, $\delta \chi^2_{\rm r} \approx 0.0198$ for
$\Delta\sigma = 1$.  Thus, based on the variation in $\chi^2_{\rm r}$
with $\langle \log[{\rm Age/yr}]\rangle$, the difference in ages of
the two composites with the largest difference in
$\langle\wlya\rangle$ (corresponding to subsamples WT1LN and WT3) is
significant at the $\simeq 2\sigma$ level.

For comparison, the distributions of SED-inferred $\log[{\rm Age/yr}]$
(Section~\ref{sec:sedmodeling}) and $\wlya$ measured for individual
galaxies are shown in the lower right panel of
Figure~\ref{fig:wlyavssps}.  A Spearman rank correlation test
indicates a marginal anti-correlation between $\log[{\rm Age/yr}]$ and
$\wlya$ for individual galaxies, with a probability of $18\%$ that the
two are uncorrelated.  As noted in
Section~\ref{sec:zstarandageissues}, the strengths of the correlations
summarized in Figure~\ref{fig:wlyavssps} are similar to those obtained
with the 300bin, 100sin, and 300sin models.  The ionizing spectrum
assumed for the next step of the fitting process (photoionization
modeling; Section~\ref{sec:optfitting}) was determined by the SPSneb
model that best fits the composite FUV spectrum.

\begin{deluxetable*}{lcccccccc}
\tabletypesize{\footnotesize}
\tablecaption{Optical Nebular Emission Line Measurements}
\tablehead{
\colhead{Subsample} &
\colhead{$\langle\lhb\rangle $\tablenotemark{a}} &
\colhead{$\langle I(\oii)\rangle$\tablenotemark{b}} &
\colhead{$\langle I(\neiii)\rangle$\tablenotemark{c}} &
\colhead{$\langle I(\oiii4960)\rangle$\tablenotemark{d}} &
\colhead{$\langle I(\oiii5008)\rangle$\tablenotemark{e}} &
\colhead{$\langle I(\ha)\rangle$\tablenotemark{f}} &
\colhead{$\langle I(\nii)\rangle$\tablenotemark{g}} &
\colhead{$\langle I(\sii)\rangle$\tablenotemark{h}}}
\startdata
ALN & $277\pm 41$ & $3.18\pm0.37$ & $ 0.29\pm 0.08$ & $1.17\pm0.10$ & $3.56\pm0.25$ & $2.79\pm0.01$ & $0.35\pm0.04$ & $0.55\pm0.06$ \\
\hline
WT1LN & $287\pm119$ & $3.77\pm1.03$ & $ 0.29\pm 0.21$ & $1.02\pm0.25$ & $3.57\pm0.65$ & $2.79\pm0.01$ & $0.33\pm0.05$ & $0.49\pm0.11$ \\
WT2LN & $368\pm128$ & $4.59\pm1.01$ & $ 0.03\pm 0.21$ & $1.07\pm0.21$ & $2.66\pm0.42$ & $2.79\pm0.01$ & $0.48\pm0.07$ & $0.67\pm0.13$ \\
WT3LN & $314\pm 65$ & $3.03\pm0.54$ & $ 0.41\pm 0.16$ & $1.37\pm0.19$ & $4.50\pm0.51$ & $2.79\pm0.01$ & $0.29\pm0.05$ & $0.44\pm0.07$ \\
\hline
ST1 & $158\pm 47$ & $3.43\pm0.83$ & $ 0.31\pm 0.28$ & $0.94\pm0.21$ & $2.64\pm0.49$ & $2.80\pm0.01$ & $0.46\pm0.08$ & $0.56\pm0.14$ \\
ST1L & $160\pm 56$ & $3.41\pm0.91$ & $ 0.25\pm 0.29$ & $1.00\pm0.27$ & $2.64\pm0.59$ & $2.80\pm0.01$ & $0.42\pm0.07$ & $0.54\pm0.13$ \\
ST2 & $400\pm 97$ & $4.44\pm0.93$ & $ 0.28\pm 0.24$ & $0.87\pm0.19$ & $3.04\pm0.47$ & $2.79\pm0.01$ & $0.42\pm0.06$ & $0.58\pm0.08$ \\
ST2L & $426\pm 98$ & $4.74\pm0.88$ & $ 0.30\pm 0.25$ & $0.80\pm0.18$ & $3.00\pm0.51$ & $2.79\pm0.01$ & $0.44\pm0.06$ & $0.60\pm0.09$ \\
ST3 & $491\pm137$ & $2.91\pm0.45$ & $ 0.41\pm 0.10$ & $1.53\pm0.17$ & $4.76\pm0.41$ & $2.79\pm0.01$ & $0.28\pm0.05$ & $0.39\pm0.05$ \\
ST3L & $511\pm173$ & $2.94\pm0.45$ & $ 0.41\pm 0.11$ & $1.48\pm0.17$ & $4.65\pm0.43$ & $2.79\pm0.01$ & $0.28\pm0.05$ & $0.39\pm0.05$ \\
\hline
sST1 & $269\pm 96$ & $3.83\pm0.92$ & $ 0.23\pm 0.19$ & $0.69\pm0.17$ & $2.30\pm0.31$ & $2.80\pm0.01$ & $0.53\pm0.05$ & $0.61\pm0.09$ \\
sST1L & $303\pm104$ & $4.28\pm1.16$ & $ 0.19\pm 0.31$ & $0.72\pm0.20$ & $2.52\pm0.50$ & $2.79\pm0.01$ & $0.51\pm0.05$ & $0.62\pm0.10$ \\
sST2 & $415\pm136$ & $4.12\pm0.78$ & $ 0.49\pm 0.21$ & $1.15\pm0.21$ & $3.60\pm0.60$ & $2.79\pm0.01$ & $0.36\pm0.06$ & $0.50\pm0.10$ \\
sST2L & $400\pm117$ & $3.95\pm0.64$ & $ 0.49\pm 0.20$ & $1.10\pm0.20$ & $3.63\pm0.63$ & $2.79\pm0.01$ & $0.37\pm0.06$ & $0.49\pm0.10$ \\
sST3 & $298\pm 74$ & $2.41\pm0.42$ & $ 0.33\pm 0.11$ & $1.58\pm0.17$ & $4.88\pm0.39$ & $2.79\pm0.01$ & $0.23\pm0.04$ & $0.37\pm0.07$ \\
sST3L & $323\pm 90$ & $2.57\pm0.51$ & $ 0.38\pm 0.11$ & $1.52\pm0.16$ & $4.68\pm0.45$ & $2.79\pm0.01$ & $0.25\pm0.05$ & $0.39\pm0.08$ 
\enddata
\tablenotetext{a}{Dust-corrected $\hb$ luminosity in units of
$10^{40}$\,erg\,s$^{-1}$, assuming the intrinsic $\ha/\hb$
     ratio of the best-fit photoionization model and the
     \citet{cardelli89} extinction curve, for the subsamples with
     complete coverage of the optical nebular emission lines.}
\tablenotetext{b}{Dust-corrected $\oii$\,$\lambda\lambda 3727,3730$ luminosity relative to $\lhb$.}
\tablenotetext{c}{Dust-corrected $\neiii$\,$\lambda 3870$ luminosity relative to $\lhb$.}
\tablenotetext{d}{Dust-corrected $\oiii$\,$\lambda 4960$ luminosity relative to $\lhb$.}
\tablenotetext{e}{Dust-corrected $\oiii$\,$\lambda 5008$ luminosity relative to $\lhb$.}
\tablenotetext{f}{Dust-corrected $\ha$ luminosity relative to $\lhb$.}
\tablenotetext{g}{Dust-corrected $\nii$\,$\lambda 6585$ luminosity relative to $\lhb$.}
\tablenotetext{h}{Dust-corrected $\sii$\,$\lambda\lambda 6718, 6733$ luminosity relative to $\lhb$.}
\label{tab:neblinemeasurements}
\end{deluxetable*}

\subsection{Photoionization Modeling and Results}
\label{sec:optfitting}

\subsubsection{Photoionization Modeling Procedure}
\label{sec:optfittingprocedure}

The Cloudy version 17.02 radiative transfer code \citep{ferland17} was
used to predict the nebular continuum spectra and line intensities for
the SPSneb models that best fit the composite FUV spectra, following
the same procedure described in \citet{topping20a}.  The
photoionization modeling also assumed a plane-parallel geometry with
an electron density $n_{e} = 250$\,cm$^{-3}$, the average value
inferred for MOSDEF galaxies (\citealt{sanders16a}).  \footnote{For
  the subsample containing all galaxies with coverage of the optical
  nebular emission lines (subsample ALN), $\langle n_{e}\rangle =
  288\pm58$\,cm$^{-3}$ based on the ratio of the two lines of the
  $\oii$ doublet, consistent with the $n_{e}=250$\,cm$^{-3}$ assumed
  for the photoionization modeling.}  All elemental abundances were
updated to the values given in \citet{asplund09}.  Additionally, we
assumed $\log({\rm N/O}) = -1.20$ based the N2O2 and N2S2 indices
computed for the composites and using the calibrations of
\citet{steidel16} and \citet{strom17}.  For the range of O abundances
inferred for the composites using lines other than $\nii$, $\log({\rm
  N/O})$ is expected to vary by $\approx 0.14$\,dex according to the
best-fit relation between $\log({\rm N/O})$ and $12+\log({\rm O/H})$
provided in \citet{pilyugin12a}.\footnote{The \citet{pilyugin12a}
  relation between $\log({\rm N/O})$ and electron-temperature-based
  ($T_{\rm e}$-based) $12+\log({\rm O/H})$ was shifted by $\simeq
  0.24$\,dex towards higher $12+\log({\rm O/H})$ to account for the
  tendency of the latter to underestimate O abundances obtained from
  photoionization modeling (e.g., \citealt{esteban14,
    blanc15,steidel16}) and to ensure that the relation simultaneously
  passes through the solar values of $\log({\rm N/O})$ and $\log({\rm
    O/H})$.}  Regardless, the best-fit $\langle Z_{\rm neb}\rangle$
does not change significantly if $\log({\rm N/O})$ is allowed to vary
with O abundance in the photoionization modeling, or if $\nii$ is
excluded altogether from the fitting.

The nebular oxygen abundance was varied from $Z_{\rm neb}/Z_\odot = 0.1$ to 
$1.0$ in steps of $0.1$\,$Z_\odot$.  
Additionally, the intensity of the ionizing
radiation field at the illuminated face of the gas slab, expressed by
the ionization parameter, $U$, where
\begin{eqnarray}
U\equiv \frac{n_\gamma}{n_{\rm H}},
\label{eq:logu}
\end{eqnarray}
was varied from $\log U = -3.5$ to $\log U = -1.6$ in steps of
$0.1$\,dex.  Here, $U$ is the ratio of the number densities of
hydrogen-ionizing photons and hydrogen atoms, where $n_{\rm H}\approx
n_{\rm e}$.

Each photoionization model predicts the intrinsic ratio of
$\ha$-to-$\hb$ which, in concert with the \citet{cardelli89}
extinction curve, was used to calculate the nebular reddening,
$\ebmvneb$ (e.g., \citealt{reddy15}), and dust-correct all the nebular
emission lines.\footnote{\citet{reddy20} find that the Galactic
  extinction curve \citep{cardelli89} provides an adequate description
  of the nebular reddening curve at optical wavelengths for $z\sim 2$
  galaxies in the MOSDEF survey.}  The dust-corrected nebular emission
line intensities relative to dust-corrected $\hb$ for all the
subsamples with complete coverage of the optical nebular emission
lines are listed in Table~\ref{tab:neblinemeasurements}.  The optical
nebular emission line ratios predicted by each photoionization model
were then compared to the dust-corrected emission line ratios measured
from the composite optical spectra to deduce the best-fit combinations
of $Z_{\rm neb}$ and $\log U$.\footnote{The best-fit photoionization
  models predict an intrinsic $\ha$-to-$\hb$ ratio of $(\ha/\hb)_{\rm
    int} = 2.79$ (independent of the BPASS model type used for
  fitting), slightly lower than the canonical value of $(\ha/\hb)_{\rm
    int} = 2.86$ for Case B recombination and $T_{\rm e} = 10,000$\,K
  \citep{osterbrock06}.}  Table~\ref{tab:lineratios} lists a few of
the line diagnostics discussed below and their definitions.  In
identifying the best-fit photoionization model, we considered all of
the following lines: $\oii$\,$\lambda\lambda 3727,3730$,
$\neiii$\,$\lambda 3870$, $\hb$, $\oiii$\,$\lambda\lambda 4960, 5008$,
$\ha$, $\nii$\,$\lambda 6585$, and $\sii$\,$\lambda\lambda 6718,
6733$.  The photoionization modeling constraints on $Z_{\rm neb}$,
$\log U$, and the BPASS model type are discussed below in
Sections~\ref{sec:znebmod}, \ref{sec:logumod}, and
\ref{sec:bpasstypemod}, respectively.

\begin{deluxetable}{lr}
\tablecaption{Line Diagnostics}
\tablehead{
\colhead{Line Diagnostic} &
\colhead{Definition}}
\startdata
O3 & $\log(\oiii 5008/\hb)$ \\
O32 & $\log(\oiii 4960+5008/\oii 3727+3730)$ \\
N2 & $\log(\nii 6585/\ha)$ \\
S2 & $\log(\sii 6718+6733/\ha)$ \\
N2O2 & $\log(\nii 6585/\oii 3727+3730)$ \\
N2S2 & $\log(\nii 6585/\sii 6718+6733)$ \\
Ne3O2 & $\log(\neiii 3870/\oii 3727+3730)$ \\
R23 & $\log((\oiii 4960+5008\, + \oii 3727+3730)/\hb)$ \\
C3O3 & $\log(\ciii 1907+1909/ \interoiii 1661+1666)$ \\
\enddata
\tablenotetext{}{}
\label{tab:lineratios}
\end{deluxetable}

\begin{figure*}
\epsscale{1.2}
\plotone{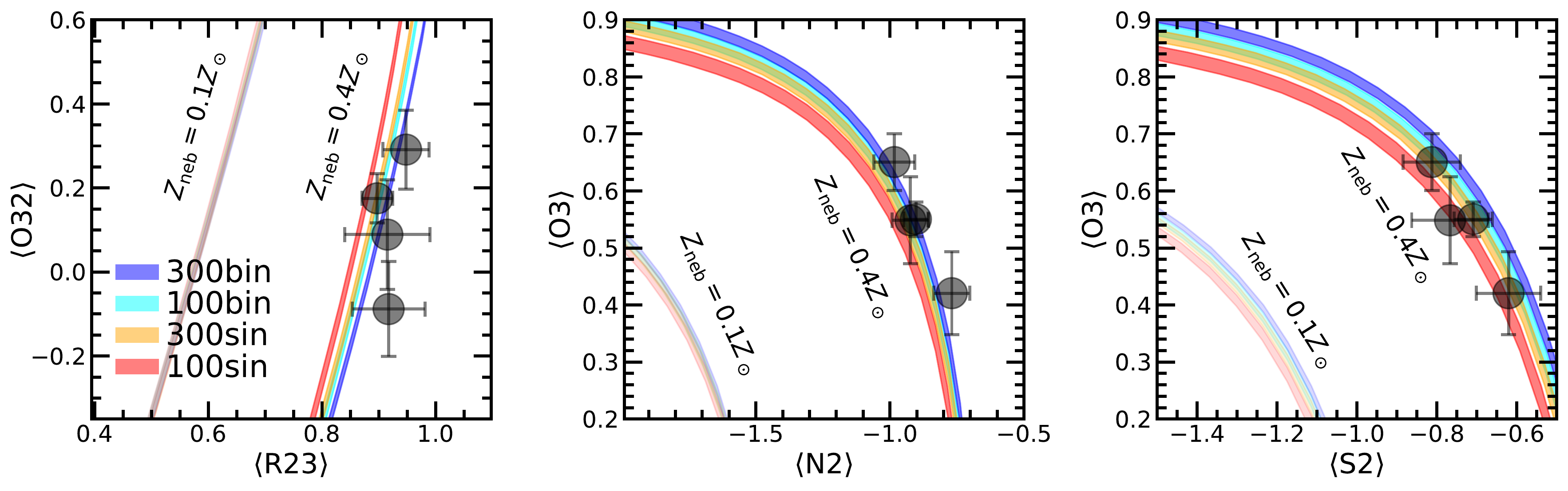}
\caption{O32 versus R23 (left), O3 versus N2 (middle), and O3 versus
  S2 (right) diagnostic diagrams.  The dark blue, cyan, orange, and
  red curves show the predicted line ratios for the 300bin, 100bin,
  300sin, and 100sin BPASS models types, respectively, for $Z_{\rm
    neb} = 0.4$\,$Z_\odot$, while the thickness of the curves
  indicates the range of expected line ratios for a (1) $Z_\ast =
  0.0010$ and $\log[{\rm Age/yr}] = 7.8$, and (2) $Z_\ast = 0.0018$
  and $\log[{\rm Age/yr}] = 8.5$ stellar population.  The set of four
  lighter curves are for the four BPASS model types assuming $Z_{\rm
    neb} = 0.1$\,$Z_\odot$.  The modeling assumes $\log({\rm N/O}) =
  -1.20$ (Section~\ref{sec:optfittingprocedure}).  The points denote
  the line ratios for the ALN, WT1LN, WT2LN, and WT3LN subsamples
  which have complete coverage of the optical nebular emission lines.}
\label{fig:optratios1}
\end{figure*}

\begin{figure*}
\epsscale{1.0}
    \plotone{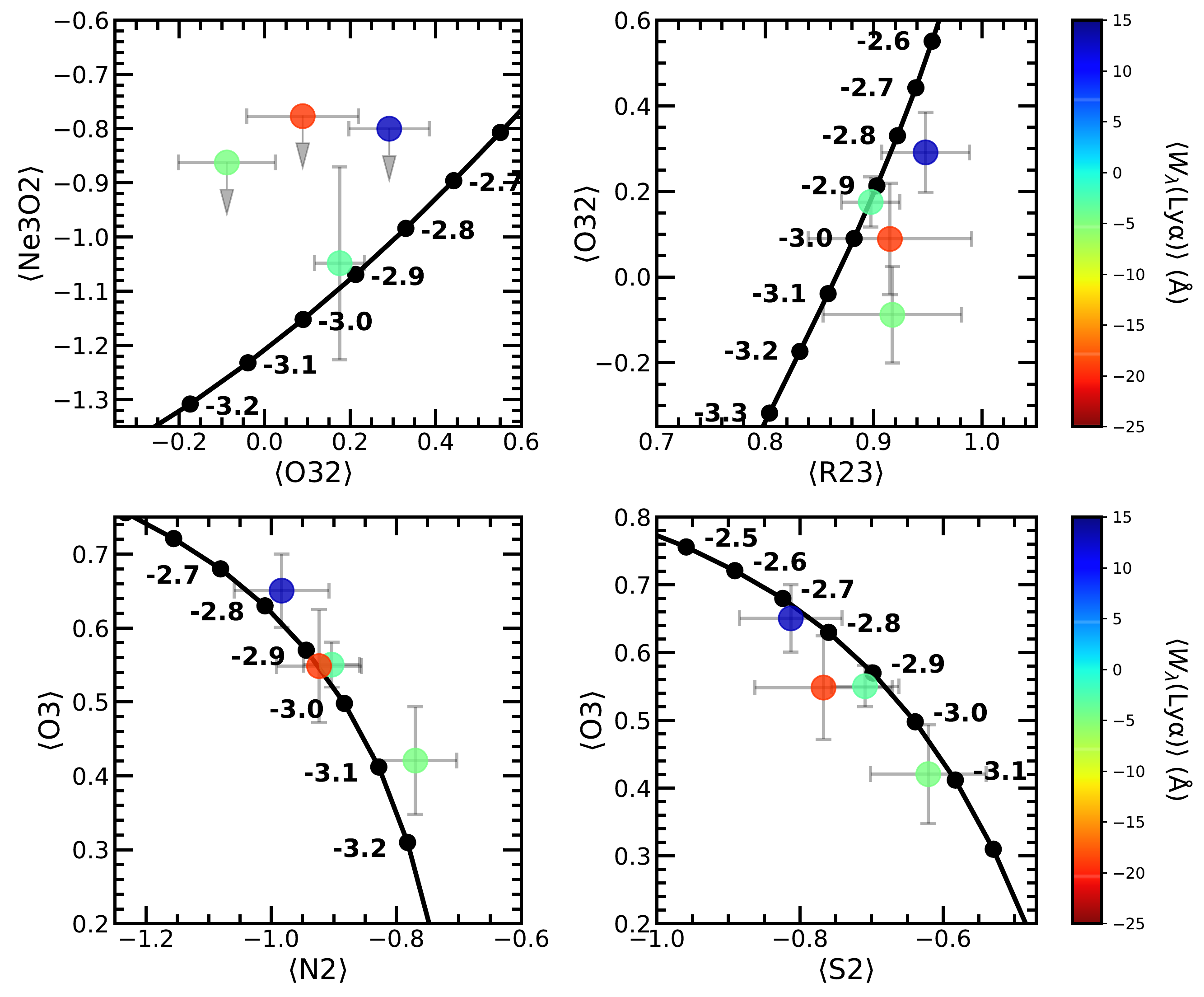}
    \caption{Ne3O2 versus O32 (top left), O32 versus R23 (top right),
      O3 versus N2 (bottom left), and O3 versus S2 (bottom right)
      diagnostic diagrams for the ALN, WT1LN, WT2LN, and WT3LN
      subsamples which have complete coverage of the optical nebular
      emission lines, where the points are color coded according to
      $\langle \wlya\rangle$.  The $3\sigma$ upper limits in Ne3O2 are
      shown for composites where $\neiii$ was not detected with $S/N >
      3$.  The thick black curves show the expected line ratios for a
      100bin stellar population with $Z_\ast = 0.0012$ and $\log[{\rm
          Age/yr}] = 8.0$, and $Z_{\rm neb} = 0.4$\,$Z_\odot$.  The
      points along the curves indicate the labeled values of $\log
      U$.}
    \label{fig:optratios2}
\end{figure*}

\subsubsection{Modeling Constraints on Oxygen Abundance}
\label{sec:znebmod}

A few of the key line ratios used to constrain $Z_{\rm neb}$ are
depicted in Figure~\ref{fig:optratios1}, including the O32 versus R23
diagnostic plane---which provides a clean separation between models of
different $Z_{\rm neb}$ (e.g., \citealt{mcgaugh91, kobulnicky99,
  kewley02})---and the N2 and S2 BPT planes \citep{baldwin81}.  In
these diagrams, each colored curve represents the sequence of $\log U$
at a fixed $Z_{\rm neb}$ for a given BPASS model type.  The widths of
the curves indicate the range of line ratios expected given the range
of $\langle Z_\ast\rangle$ and $\langle\log[{\rm Age/yr}]\rangle$
inferred from modeling the composite FUV spectra
(Section~\ref{sec:fuvfitting}).  Specifically, the upper and lower
edges of each curve indicates the model expectations for the ionizing
spectrum of a stellar population with $Z_\ast = 0.0010$ and $\log[{\rm
    Age/yr}] = 7.8$, and $Z_\ast = 0.0018$ and $\log[{\rm Age/yr}] =
8.5$, covering roughly the range of $Z_\ast$ and ages found in
Section~\ref{sec:fuvfitting}.  The differences in the ionizing spectra
of the best-fit SPSneb models are discussed further in
Section~\ref{sec:primary}, but for the moment we note that these
variations in the ionizing spectrum do little to alter the nebular
line ratios predicted from the photoionization modeling (e.g., see
also \citealt{runco21}).

Figure~\ref{fig:optratios1} clearly demonstrates the preference for
oxygen abundances of $\langle Z_{\rm neb}\rangle \simeq
0.4$\,$Z_\odot$ ($\langle 12+\log({\rm O/H})\rangle \simeq 8.29$)
averaged across the ALN and WT subsamples with complete coverage of
the optical nebular emission lines.  The $\langle Z_{\rm neb}\rangle$
obtained from the photoionization modeling vary in the range $\langle
Z_{\rm neb}\rangle \simeq (0.3-0.5)$\,$Z_\odot$ (see also
\citealt{topping20a, runco21}).  As noted earlier, we adopted
$\log({\rm N/O}) = -1.20$ based on the N2O2 and N2S2 indices.  This
choice yields N2 that are consistent with the predictions of the
photoionization model (with $Z_{\rm neb} \approx 0.4$\,$Z_\odot$) that
reproduces all of the other line ratios under consideration.  The
inferred $\langle Z_{\rm neb}\rangle$ are similar between the
different BPASS model types given the typical uncertainties in the
nebular line ratios.  Photoionization models with $Z_{\rm neb}$
comparable to the {\em stellar} metallicities obtained from fitting
the composite FUV spectra ($\langle Z_\ast\rangle \simeq 0.001$;
Section~\ref{sec:fuvfitting}) are ruled out at the $3-4$\,$\sigma$
level, depending on the subsample (e.g., see also
\citealt{steidel16}).  Appendix~\ref{sec:directmethod} discusses the O
abundances derived from the ``direct'' metallicity method: these
direct-method abundances agree with those obtained from the
photoionization modeling.

\subsubsection{Modeling Constraints on Ionization Parameter}
\label{sec:logumod}

The line diagnostic diagrams in Figure~\ref{fig:optratios1}, in
addition to Ne3O2 versus O32 (e.g., \citealt{nagao06, perez07,
  levesque14, steidel16, strom17, jeong20}), may be used to constrain
the ionization parameter, $U$.  These diagrams are presented in
Figure~\ref{fig:optratios2}, zoomed in on the regions of line
diagnostic space that encompass the measurements for the ALN and WT
subsamples with complete coverage of the optical nebular emission
lines.  For simplicity, we only show the expected line ratios for a
100bin stellar population with $Z_\ast = 0.0012$, $\log[{\rm Age/yr}]
= 8.0$, and with $Z_{\rm neb} = 0.4$\,$Z_\odot$ (thick black curve),
where the points indicate the labeled values of $\log U$.  The
photoionization modeling points to best-fit values of $\langle \log
U\rangle \simeq -3.1$ to $-2.8$.  As per the discussion in
Section~\ref{sec:znebmod} and Figure~\ref{fig:optratios1}, the
best-fit values of $\langle \log U\rangle$ are insensitive to the
BPASS model type and the range of ionizing spectra corresponding to
the best-fit SPSneb models.

\subsubsection{Constraints on the BPASS Model Type (or Binarity)}
\label{sec:bpasstypemod}

\begin{figure}
\epsscale{1.05}
\plotone{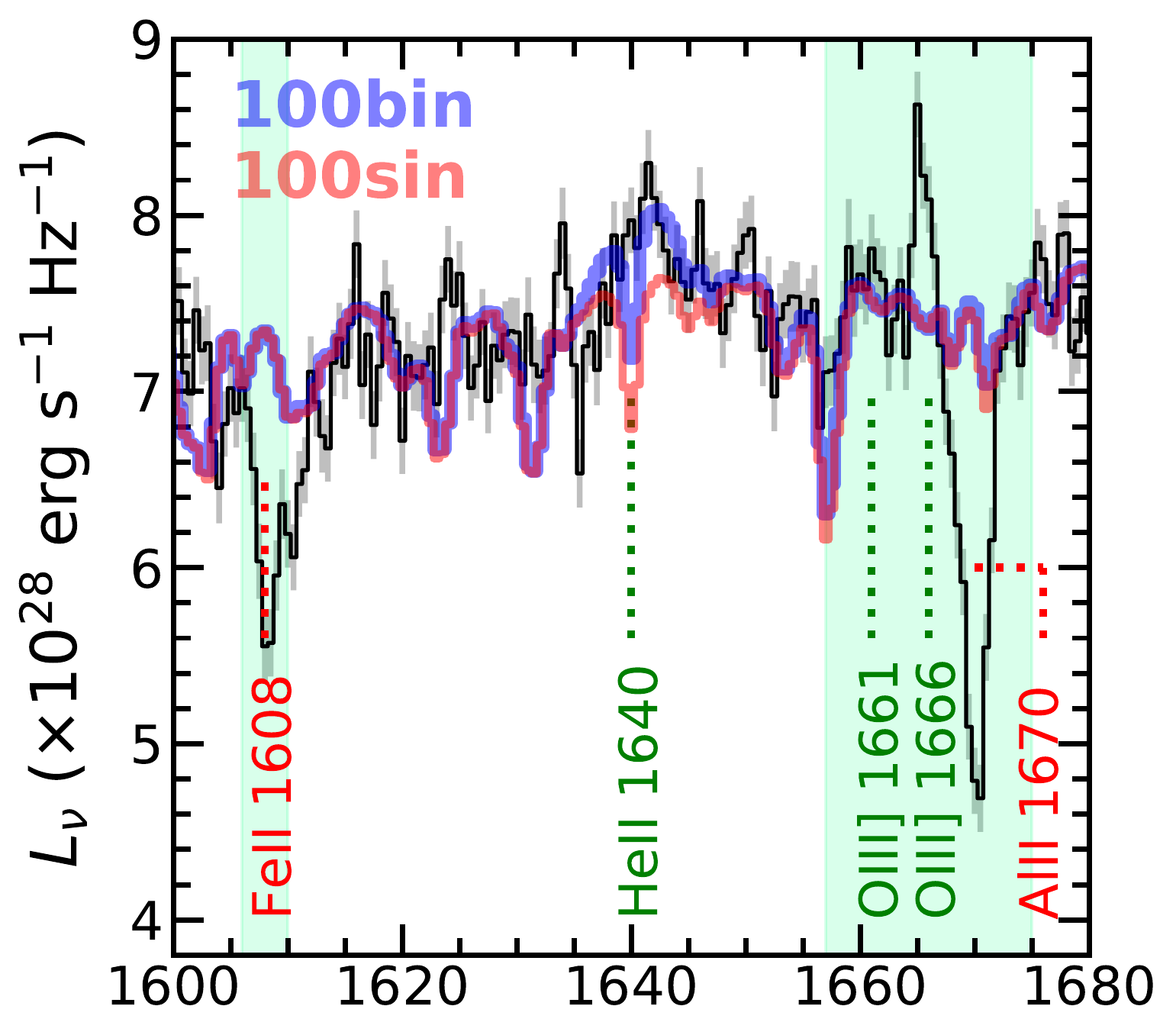}
\caption{Comparison of the best-fit 100bin (i.e., with binaries) and
  100sin (i.e., without binaries) models (blue and red, respectively)
  to the composite FUV spectrum of all galaxies in the sample (black),
  zoomed in on the region around $\heii$\,$\lambda 1640$.  Binary
  models predict significant stellar $\heii$ emission, while this
  emission is absent for comparably-aged models that do not include
  binaries.  Regions not included in the fitting are indicated by the
  light green shaded regions.  }
\label{fig:heiifit}
\end{figure}

The ratio of nebular $\heii$\,$\lambda 1640$ and $\hb$ is one of the
most sensitive discriminators of the BPASS model type
\citep{steidel16}.  In single-star models, $\heii$ emission arises
from the stellar winds of very high mass and short-lived Wolf-Rayet
stars (e.g., \citealt{shirazi12, crowther16}).  On the other hand, the
higher effective temperatures of massive binaries, their harder
ionizing spectra, and the increased main-sequence lifetimes of such
stars, result in significant production of \heii-ionizing photons for
a longer duration compared to single-star models \citep{eldridge17}.
In the context of a continuous star-formation history, binary stellar
evolution results in significant stellar $\heii$ emission, while this
emission is absent in the single-star models for ages comparable to
those that best fit the composite FUV spectra, as shown in
Figure~\ref{fig:heiifit}.  As alluded to in Section~\ref{sec:intro},
the (non-ionizing) FUV spectrum is insufficient to discriminate
between single and binary star population synthesis models.  On the
other hand, binarity results in both a harder and more intense
ionizing (EUV) spectrum whose signature can be probed with FUV and
optical nebular emission line ratios (e.g., \citealt{steidel14,
  steidel16, schaerer19, nanayakkara19, chisholm19}).  Along these
lines, the BPASS model type (i.e., single versus binary star
evolution) can be evaluated by comparing the inferred nebular $\heii$
luminosity with that predicted by the best-fit photoionization model.

\begin{figure}
\epsscale{1.05}
\plotone{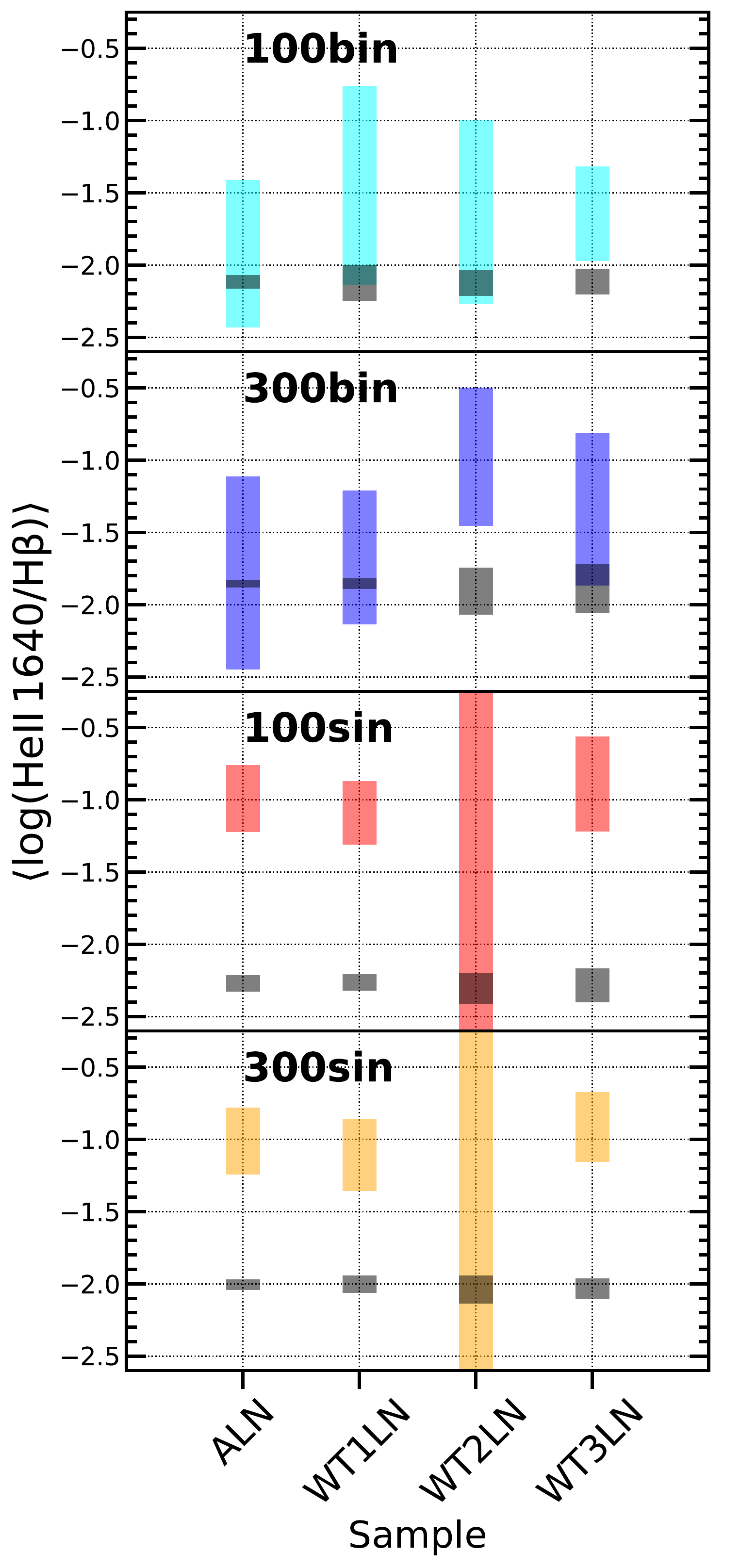}
\caption{Measured intensity of nebular $\heii$\,$\lambda 1640$,
  $\langle \heii/\hb\rangle$, compared to the photoionization model
  predictions for the ALN, WT1LN, WT2LN, and WT3LN subsamples which
  have complete coverage of the optical nebular emission lines.  Each
  panel shows the comparison for a particular BPASS model type
  (100bin, 300bin, 100sin, 300sin).  The colored bars indicate the
  $\pm 1\sigma$ range of the (dust-corrected) residual $\heii$
  intensities obtained by subtracting the best-fit SPSneb models from
  the composite FUV spectra, where the errors include measurement
  uncertainties in the composite FUV spectra, uncertainties in the
  best-fit SPSneb models, and uncertainties in the best-fit $\langle
  \ebmvneb\rangle$ (which is used for the dust correction; see text).
  The gray bars indicate the $\pm 1\sigma$ range of nebular $\heii$
  intensities predicted by the best-fit photoionization models that
  assume the best-fit SPSneb models as the source of the He-ionizing
  photons where, for visibility, the errors have been multiplied by a
  factor of 5.  The errors include the uncertainties in the best-fit
  photoionization models.}
\label{fig:heii}
\end{figure}

Following \citet{steidel16}, we measured the residual $\heii$
luminosity from the composite FUV spectrum after subtracting the
best-fit SPSneb model to that composite spectrum.  Because the SPSneb
model includes stellar $\heii$ emission, the residual $\heii$
luminosity is assumed to be nebular in nature.  The inferred nebular
$\heii$ luminosity was corrected for dust obscuration assuming
$\langle\ebmvneb\rangle$ and the \citet{cardelli89} extinction curve.
These residual values are listed in
Table~\ref{tab:fuvlinemeasurements} for the 100bin model type.
Figure~\ref{fig:heii} shows the relative intensity of nebular $\heii$
to $\hb$, compared to $\langle \heii/\hb\rangle$ predicted by the
best-fit photoionization models for the ALN and WT subsamples with
complete coverage of the optical nebular emission lines.  The errors,
reflected by the length of the bars in this figure, include
measurement uncertainties (and thus uncertainties in the best-fit
SPSneb model and uncertainties in the best-fit photoionization model).
A majority of the subsamples has residual nebular $\heii$ emission
that is formally ``undetected'' for the binary models, for the reasons
explained below.

\begin{deluxetable*}{lcccc}
\tabletypesize{\footnotesize}
\tablecaption{FUV Nebular Emission Line Measurements}
\tablehead{
\colhead{Subsample} &
\colhead{$\langle I(\interoiii)\rangle$\tablenotemark{a}} &
\colhead{$\langle I(\ciii)\rangle$\tablenotemark{b}} &
\colhead{$\langle W_{\lambda}(\ciii)\rangle$\tablenotemark{c}} &
\colhead{$\langle I(\heii)\rangle$\tablenotemark{d}}}
\startdata
ALN & $0.112\pm0.034$ & $0.393\pm0.120$ & $1.330\pm0.244$ & $ 0.012\pm 0.029$ \\
\hline
WT1LN & $0.089\pm0.052$ & $0.280\pm0.245$ & $1.032\pm0.561$ & $ 0.064\pm 0.195$ \\
WT2LN & $0.147\pm0.072$ & $0.847\pm0.531$ & $1.278\pm0.554$ & $ 0.023\pm 0.107$ \\
WT3LN & $0.149\pm0.096$ & $0.478\pm0.201$ & $1.790\pm0.637$ & $ 0.015\pm 0.042$ \\
\hline
ST1 & $0.132\pm0.062$ & $0.286\pm0.183$ & $1.138\pm0.509$ & $ 0.009\pm 0.061$ \\
ST1L & $0.103\pm0.072$ & $0.305\pm0.289$ & $1.053\pm0.474$ & $ 0.010\pm 0.058$ \\
ST2 & $0.122\pm0.085$ & $0.774\pm0.403$ & $1.660\pm0.605$ & $ 0.022\pm 0.090$ \\
ST2L & $0.102\pm0.042$ & $0.861\pm0.372$ & $1.759\pm0.639$ & $ 0.029\pm 0.082$ \\
ST3 & $0.133\pm0.079$ & $0.459\pm0.196$ & $1.546\pm0.503$ & $ 0.038\pm 0.076$ \\
ST3L & $0.133\pm0.099$ & $0.423\pm0.230$ & $1.349\pm0.520$ & $ 0.026\pm 0.061$ \\
\hline
sST1 & $0.268\pm0.101$ & $0.646\pm0.397$ & $1.537\pm0.503$ & $ 0.035\pm 0.115$ \\
sST1L & $0.121\pm0.087$ & $0.831\pm0.587$ & $1.518\pm0.514$ & $ 0.039\pm 0.164$ \\
sST2 & $0.081\pm0.042$ & $0.357\pm0.251$ & $0.999\pm0.548$ & $ 0.029\pm 0.083$ \\
sST2L & $0.054\pm0.049$ & $0.320\pm0.207$ & $0.966\pm0.539$ & $ 0.030\pm 0.075$ \\
sST3 & $0.082\pm0.070$ & $0.410\pm0.144$ & $1.935\pm0.414$ & $ 0.020\pm 0.040$ \\
sST3L & $0.112\pm0.118$ & $0.391\pm0.183$ & $1.691\pm0.451$ & $ 0.019\pm 0.046$
\enddata
\tablenotetext{a}{Dust-corrected $\interoiii$\,$\lambda\lambda 1660,1666$ luminosity relative $\lhb$.}
\tablenotetext{b}{Dust-corrected $\ciii$\,$\lambda\lambda 1907,1909$ luminosity relative to $\lhb$.}
\tablenotetext{c}{Equivalent width of $\ciii$\,$\lambda\lambda 1907,1909$.}
\tablenotetext{d}{Dust-corrected $\heii$\,$\lambda 1640$ luminosity relative to $\lhb$.}
\label{tab:fuvlinemeasurements}
\end{deluxetable*}

Recall that the photoionization modeling assumes the best-fit SPSneb
model as the ionizing source (Section~\ref{sec:optfittingprocedure}).
Hence, the inferred stellar $\heii$ emission {\em and} the nebular
$\heii$ emission predicted by the best-fit photoionization model both
assume the same SPSneb model, providing internal consistency in the
way that $\heii$ is modeled.  The inferred nebular $\heii$ emission is
much larger in the single-star models because of the {\em weaker}
inferred stellar emission, and in fact that nebular emission is
significantly larger than that predicted by the best-fit
photoionization models when considering the same single-star (100sin
and 300sin) models.  In other words, the ionizing spectra associated
with single-star models are unable to account for the level of nebular
$\heii$ emission inferred based on the weak stellar $\heii$ predicted
by the same models.

On the other hand, the binary models yield a residual $\heii$
intensity that is comparable within the uncertainties to that
predicted by the photoionization models.  Based on comparing the
offsets between the predicted and residual nebular $\heii$ emission
from subsample to subsample, it may be tempting to conclude that the
binary models also tend to underestimate the nebular $\heii$ emission.
However, it is important to keep in mind that many of the subsamples
have galaxies in common, and therefore the offsets observed for a
given subsample are not necessarily independent of the offsets for a
different subsample.  When considering all galaxies in our subsample
with complete coverage of the optical nebular emission lines (i.e.,
subsample ALN), the inferred offset between the predicted and residual
nebular $\heii$ emission for the binary models is consistent with
zero, while it at least $1$\,dex for the single-star models.

In conclusion, we find that the binary models produce
internally-consistent inferences of the nebular $\heii$ emission for
the composites, while the single-star models do not.  Similar
conclusions were reached by \citet{steidel16} in their analysis of the
KBSS composite spectrum.  Here, we have shown that the binary models
appear consistent with the composite spectra irrespective of
$\langle\wlya\rangle$ (see Section~\ref{sec:wlyavsphot} for further
discussion).  In the present context, the most pertinent impact of the
BPASS model type is on the intrinsic production of H-ionizing photons,
which is addressed in Section~\ref{sec:primary}).  As per the
discussion in Sections~\ref{sec:zstarandageissues}, \ref{sec:znebmod},
and \ref{sec:logumod}, the BPASS model type has little effect on the
derived $Z_\ast$, $\log[{\rm Age/yr}]$, $Z_{\rm neb}$, and $\log U$.

\subsubsection{Correlations between $\wlya$ and Photoionization Model Parameters}
\label{sec:wlyavsphot}

\begin{figure*}
\epsscale{1.15}
\plotone{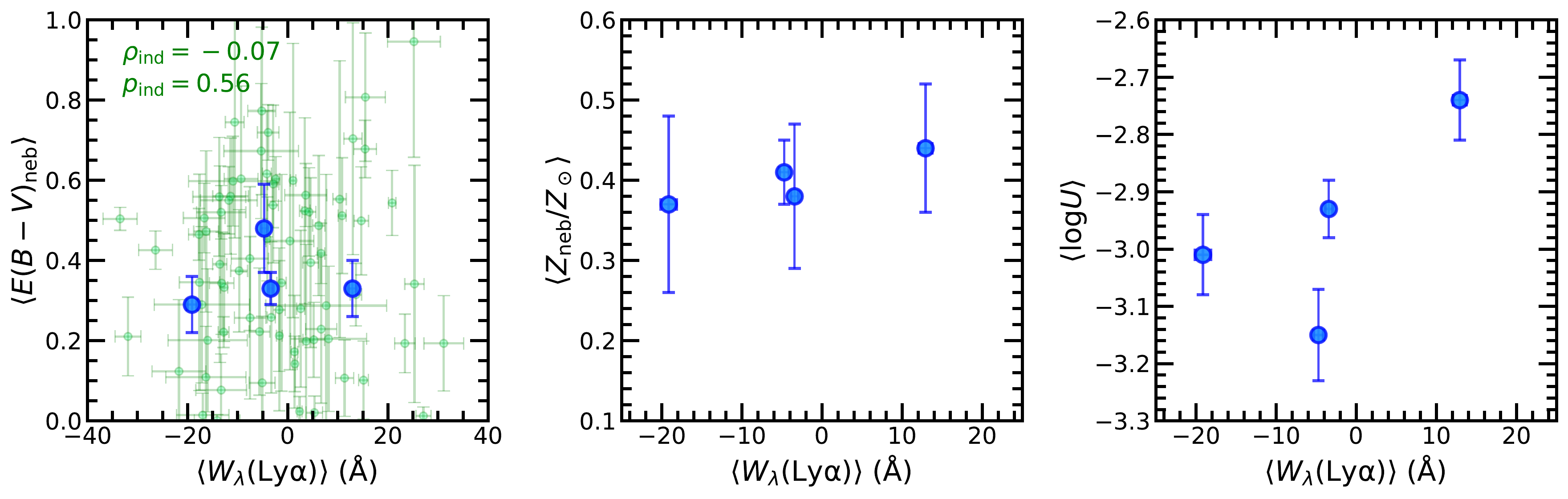}
\caption{Variation of $\langle\ebmvneb\rangle$ (left), $\langle Z_{\rm
    neb}/Z_\odot\rangle$ (middle), and $\langle \log U \rangle$
  (right) assuming the 100bin models, with $\langle \wlya\rangle$ for
  the ALN, WT1LN, WT2LN, and WT3LN subsamples which have complete
  coverage of the optical nebular emission lines.  The $\ebmvneb$
  calculated for individual galaxies are denoted by the light green
  points in the left panel.  Also indicated are the Spearman $\rho$
  correlation coefficient ($\rho_{\rm ind}$) and $p$-value ($p_{\rm
    ind}$) between $\ebmvneb$ and $\wlya$ for individual galaxies.}
\label{fig:wlyavsphot}
\end{figure*}

Figure~\ref{fig:wlyavsphot} summarizes the correlations between the
best-fit photoionization model parameters (Table~\ref{tab:params}) and
$\langle \wlya \rangle$.  For comparison, the $\ebmvneb$ calculated
for individual galaxies with significant ($S/N\ge 3$) detections of
$\ha$ and $\hb$ are also shown in the left panel of the figure.  These
individual $\ebmvneb$ were calculated using the same method to
calculate $\langle \ebmvneb\rangle$ for the ensembles.  In calculating
$\ebmvneb$ for individual galaxies, we assumed $(\ha/\hb)_{\rm int} =
2.79$ based on the results of the photoionization modeling
(Section~\ref{sec:optfittingprocedure}).  
Over the dynamic range probed by our sample, we do not find a
significant trend between $\langle\ebmvneb\rangle$ and
$\langle\wlya\rangle$.  Furthermore, while the averages of individual
$\ebmvneb$ measurements in bins of $\wlya$ are consistent with the
$\langle \ebmvneb\rangle$ measured from the composite spectra, a
Spearman correlation test indicates a high probability ($p_{\rm ind} =
0.56$) of a null correlation between $\ebmvneb$ and $\wlya$ for
individual galaxies (c.f., \citealt{scarlata09}), perhaps as a result
of the larger uncertainties on the individual $\ebmvneb$ measurements.

The best-fit $\langle Z_{\rm neb}\rangle$ also do not appear to
correlate strongly with $\langle \wlya\rangle$, in the sense that
galaxies with high $\langle \wlya\rangle \ga 10$\,\AA\, have $\langle
Z_{\rm neb}\rangle$ consistent within the uncertainties with that of
galaxies with lower $\langle \wlya\rangle$.  The ostensible lack of
correlation between $\langle Z_{\rm neb}\rangle$ and $\langle
\wlya\rangle$ can be appreciated from Figures~\ref{fig:optratios1} and
\ref{fig:optratios2}, where all composites irrespective of $\langle
\wlya\rangle$ have optical nebular line ratios that are generally
consistent with photoionization models where $Z_{\rm neb}\approx
0.4$\,$Z_\odot$.

The only signficant difference found among the photoionization
modeling parameters for galaxies with low and high $\wlya$ is $\log
U$.  Composites formed from the lower and upper third of the $\wlya$
distribution (i.e., WT1LN and WT3LN) indicate $\langle \log
U\rangle\simeq -3.01\pm0.07$ and $-2.74\pm0.07$, respectively, a
difference significant at the $\simeq 3\sigma$ level (see also
Figure~\ref{fig:optratios2}).

\subsection{Neutral ISM Modeling and Results}
\label{sec:ismfitting}

\subsubsection{Neutral ISM Modeling Procedure}
\label{sec:ismfittingprocedure}

The third and last step of the spectral modeling involved modifying
the SPSneb models to include interstellar $\hi$ absorption with
varying column density and line-of-sight reddening.  The method
employed here is identical to that of \citet{reddy16a,reddy16b} and
\citet{steidel18}, where the observed spectrum consists of two
components: (a) the intrinsic spectrum and (b) the spectrum that
emerges after passing through foreground $\hi$ and dust.  These two
components are then weighted according to the $\hi$ gas covering
fraction, $\fcovhi$, to model a non-unity covering fraction of
optically-thick $\hi$.  This model of the neutral ISM, referred to as
the ``clumpy'' model (also called the ``holes'' model in
\citealt{steidel18}) can be expressed as:
\begin{eqnarray}
m_{\rm final} & = & \fcovhi \times m_{\hi} \times 10^{-0.4\ebmvlos k(\lambda)} + \nonumber \\
& & [1-\fcovhi]\times m_0.
\label{eq:fcov}
\end{eqnarray}
Here, $m_0$ is the intrinsic (unreddened) stellar spectrum determined
from fitting the composite FUV spectrum with the SPSneb models
(Section~\ref{sec:fuvfittingprocedure}).  $\ebmvlos$ is the
line-of-sight reddening (i.e., the reddening of the covered portion of
the continuum), where a range $\ebmvlos = 0.000 - 0.300$ and the SMC
extinction curve, $k(\lambda)$, was assumed.  Lyman series absorption
lines were added to the intrinsic (unreddened) stellar spectrum with a
Doppler parameter of $b=125$\,km\,s$^{-1}$ and column densities in the
range $\lognhi = 18.0 - 23.0$.  $\nhi$ is primarily constrained by the
$\hi$ damping wings and is therefore insensitive to the particular
choice of $b$.  Furthermore, the assumed range of $\lognhi$ is
motivated by the results of \citet{reddy16b} (e.g., see their
Figure~2) that show that it is not possible to fit the red damping
wings of the $\hi$ lines, particularly that of $\lya$, unless $\lognhi
\ga 19$.  The resulting $\hi$-absorbed spectrum is denoted by $m_{\rm
  \hi}$ in Equation~\ref{eq:fcov}.  Accordingly, we obtained a grid of
neutral ISM models with varying $\ebmvlos$, $\lognhi$, and $\fcovhi$.

The clumpy model makes the physically-motivated assumption that the
mechanisms that establish low-column-density channels in the ISM also
result in negligible dust attenuation along those channels, such that
only the light passing through the high-column-density $\hi$ is
reddened by dust (c.f., \citealt{borthakur14}).  However, because
  we cannot rule out the possibility of non-negligible columns of dust
  in these channels, we also consider the more extreme possibility of
  a ``screen'' model where all light is attenuated by a foreground
  screen of dust (e.g., \citealt{steidel18, gazagnes18}).  In this
  case, the ISM model can be expressed as:
\begin{eqnarray}
m_{\rm final} & = & [\fcovhi \times m_{\hi} + (1-\fcovhi)\times m_0] \times \nonumber \\ 
& & 10^{-0.4\ebmvcont k(\lambda)}.
\label{eq:screen}
\end{eqnarray}
Covering fractions derived from the screen model are systematically
lower than those derived from the clumpy model (see
Section~\ref{sec:wlyavsfcov} for further discussion) .  Throughout the
subsequent discussion, we assume by default the $\fcovhi$ derived from
the clumpy model, and discuss the results from the screen model where
relevant.

The best-fit values of $\ebmvlos$, $\lognhi$, and $\fcovhi$ are
determined by the neutral ISM (clumpy) model with the minimum $\chi^2$ relative
to the composite FUV spectrum in the wavelength windows specified in
\citet{steidel18}---slightly modified from the windows used in
\citet{reddy16b}---where only those windows with $\lambda_0 \ga
1000$\,\AA\, were considered, for the following reason.  While the
best constraints on $\fcovhi$ come from fitting multiple Lyman series
absorption lines, doing so would require us to only consider galaxies
for which those lines are redshifted above the atmospheric cutoff at
$\lambda \simeq 3100$\,\AA.  All of these lines, and the LyC continuum
region at $\lambda \simeq 900$\,\AA, can be observed from the ground
at $z>2.7$, a limit that excludes the bulk of our sample which lies at
lower redshifts (Figure~\ref{fig:zhistwlyahist}).  Alternatively,
adopting a redshift cutoff of $z>2.12$ ensures coverage of at least
$\lyb$ while still retaining sufficient numbers of galaxies in the
smallest subsamples to reduce uncertainties in the foreground IGM+CGM
opacity to $\la 10\%$ \citep{steidel18}.  For this reason, the neutral
ISM models are compared to the composite spectra in wavelength windows
lying above $\lambda_0 \ga 1000$\,\AA.  Consequently, $\fcovhi$ is
primarily determined by the depth of $\lyb$\footnote{$\lya$ absorption
  provides little constraint on $\fcovhi$ owing to emission-filling.},
while $\lognhi$ is constrained by the damping wings of $\lya$ and
$\lyb$.  Note that the depths of the Lyman series lines are sensitive
to the $\hi$ gas covering fraction only if the lines are saturated.
Below, we present evidence that this is the case for galaxies in our
sample.

\begin{figure*}
\epsscale{1.15}
    \plotone{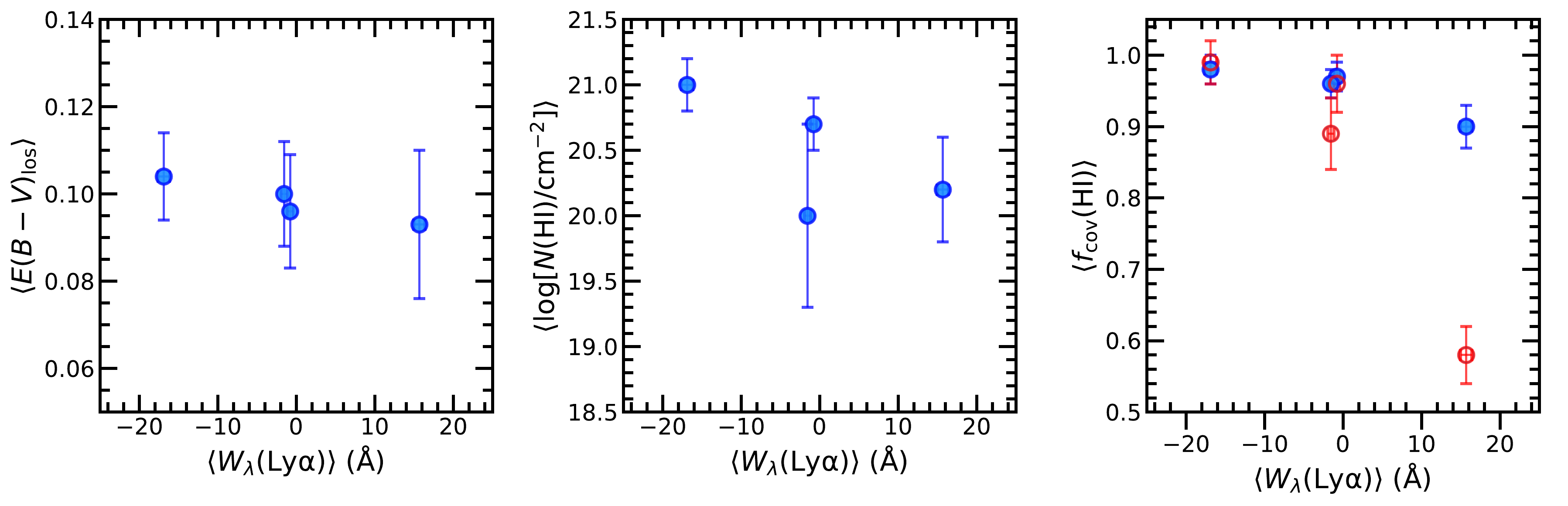}
    \caption{Variation of $\langle\ebmvlos\rangle$ (left), $\langle
      \lognhi \rangle$ (middle), and $\langle \fcovhi \rangle$ (right)
      assuming the 100bin models, with $\langle \wlya\rangle$ for the
      AL, WT1L, WT2L, and WT3L subsamples with coverage of $\lyb$.
      The blue and red points in the right panel denote covering
      fractions derived in the clumpy and screen models,
      respectively.}
    \label{fig:wlyavsfcov}
\end{figure*}

\begin{figure*}
  \epsscale{1.15}
    \plotone{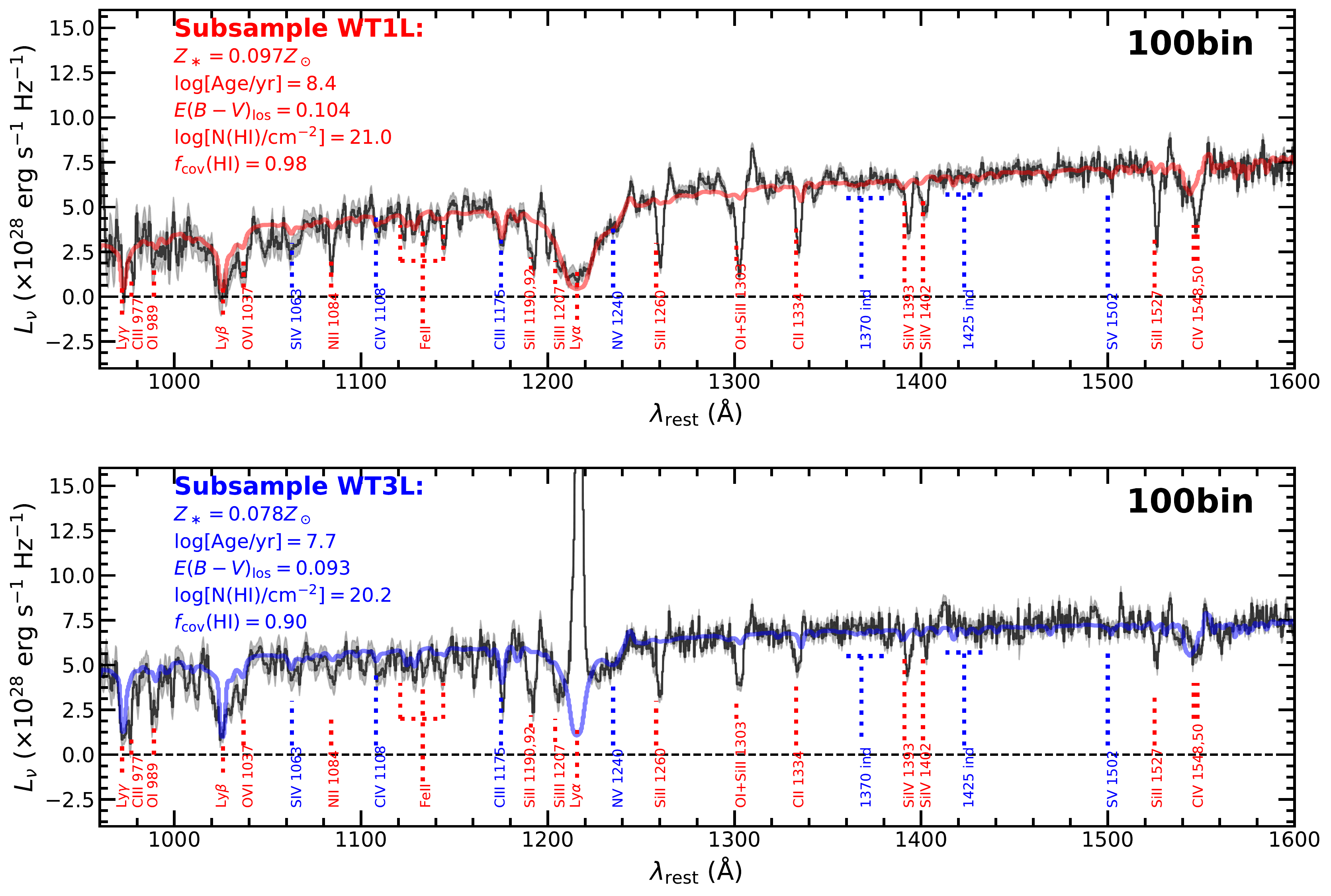}
    \caption{Composite FUV spectra for subsamples WT1L (top) and WT3L
      (bottom) shown in black ($1\sigma$ error in gray), along with
      the best-fit neutral ISM model fits in red (top) and blue
      (bottom).  These neutral ISM models assume the best-fit $\langle
      Z_\ast\rangle$ and $\langle \log[{\rm Age/yr}]\rangle$ obtained
      from the 100bin SPSneb model fitting
      (Section~\ref{sec:fuvfittingprocedure}).  Indicated in each
      panel are the $\ebmvlos$, $\lognhi$, and $\fcovhi$ for the
      models shown.  Line labeling is similar to
      Figure~\ref{fig:compspec}.}
    \label{fig:fcovdemo}
\end{figure*}

\subsubsection{Correlations between $\wlya$ and Neutral ISM Model Parameters}
\label{sec:wlyavsfcov}

The procedure described in Section~\ref{sec:ismfittingprocedure} was
used to determine the best-fit $\langle\ebmvlos\rangle$,
$\langle\lognhi\rangle$, and $\langle \fcovhi\rangle$ for the AL, ALN,
and WT subsamples with coverage of $\lyb$
(Table~\ref{tab:compositestats}).  These best-fit values are listed in
Table~\ref{tab:params}, and their correlations with $\langle
\wlya\rangle$ are shown in Figure~\ref{fig:wlyavsfcov}.
Figure~\ref{fig:fcovdemo} demonstrates the neutral ISM model fits to
the composite FUV spectra for subsamples WT1L and WT3L.  Although the
gas covering fraction is primarily constrained by the depth of $\lyb$,
the inferred gas covering fractions provide a good match to the depth
of $\lyg$, for which only the subset of galaxies with $z\ga 2.3$ has
coverage.

The neutral ISM model fitting indicates no significant correlation
between $\langle\ebmvlos\rangle$ and $\langle\wlya\rangle$.  There are
marginal differences in the $\hi$ column densities for galaxies with
low and high $\wlya$ (middle panel of Figure~\ref{fig:wlyavsfcov}),
such that galaxies with high $\langle\wlya\rangle$ appear to have
somewhat lower $\langle\lognhi\rangle$.  However, the uncertainties in
the latter are sufficently large to prevent us from coming to any
strong conclusions regarding any correlation between
$\langle\wlya\rangle$ and $\langle\lognhi\rangle$.  Regardless, the
column densities even for composites with the highest $\langle
\wlya\rangle$ (i.e., those containing the strongest $\lya$ emitters)
indicate gas that is optically-thick in all the Lyman series lines and
the Lyman continuum.  Thus, the presence of significant $\lya$
emission spatially coincident with the stellar continuum (i.e., $\lya$
emission that is detected within the spectroscopic aperture) suggests
that the $\lya$ photons must be escaping the galaxies through
optically-thin (i.e., ionized or low-column-density) channels in the
ISM, or are scattered off of gas with sufficient velocity to be
redshifted out of resonance and pass unimpeded through foreground
$\hi$.

Evidence for a non-unity covering fraction of optically-thick gas is
illustrated in the comparison of the neutral ISM model fits to
subsamples WT1L and WT3L (Figure~\ref{fig:fcovdemo}), consisting of
galaxies in the lower and upper third of the $\wlya$ distribution,
respectively (Table~\ref{tab:compositestats}).  In the former case,
there is little residual flux under the core of the $\lyb$ line---and
$\lyg$ for the subset of higher redshift galaxies where this line is
covered in the spectra, while the damping wings of $\lya$
imply high-column-density gas.  Thus, the model fit to subsample WT1L
suggests a very high covering fraction of high-column-density $\hi$.
The neutral ISM model fit for subsample WT3L suggests somewhat lower
(but still high) column densities of $\lognhi = 20.2\pm0.4$, with
significant residual flux detected under the $\lyb$ core, and $\lyg$
for the subset of galaxies where this line is covered.  This residual
flux is a principal signature of the non-unity covering fraction of
optically-thick $\hi$ gas (e.g., \citealt{reddy16b}).

More generally, galaxies with $\langle\wlya\rangle< 0$\,\AA\, have
$\langle\fcovhi\rangle \ga 0.96$, while those with
$\langle\wlya\rangle> 0$\,\AA\, have $\langle\fcovhi\rangle \la 0.93$.
This difference in $\langle\fcovhi\rangle$ is significant at the
$\simeq 2.5\sigma$ level.  However, the constraints on
$\langle\fcovhi\rangle$ are sufficient to rule out (to $\ga 3\sigma$)
unity covering fraction for galaxies with high $\langle\wlya\rangle$.
A more significant anti-correlation between $\langle \fcovhi\rangle$
and $\langle \wlya\rangle$ was found for UV-selected galaxies at
$z\sim 3$ \citep{reddy16b} where multiple Lyman series lines were used
to constrain $\langle\fcovhi\rangle$.  The right panel of
Figure~\ref{fig:wlyavsfcov} also shows $\langle \fcovhi\rangle$
inferred from the screen model (red circles), which are for the most
part systematically lower than $\fcovhi$ derived from the clumpy
model, but still correlate with $\langle \wlya\rangle$.  At any rate,
the inference of high-column-density $\hi$, $\lyb$ that is not black
at line center, and a similar level of residual flux under the $\lyb$
and $\lyg$ lines suggest that the Lyman series lines are saturated and
that their depths are sensitive to the covering fraction of
high-column-density $\hi$.

It is worth noting that for some of the composites considered in this
study, a screen model implies $\langle \fcovhi\rangle$ that are up to
a factor of $\simeq 2\times$ lower than the values implied by the
clumpy model.  Such low covering fractions imply high escape fractions
of ionizing photons that lead to suppressed Balmer line luminosities.
Accounting for the fraction of ionizing photons that escapes results
in $\ha$-inferred SFRs ($\sfrha$) that exceed the UV-inferred SFRs
($\sfruv$; e.g., see discussion in Section~\ref{sec:sigsfrcalc} and
Appendix~\ref{sec:sfrcomparison}) by up to a factor of $\simeq 2$,
even when using age-dependent conversions between $\ha$/UV luminosity
and SFR (i.e., to account for the age of the stellar population when
converting luminosity to SFR).  On the other hand,
$\langle\fcovhi\rangle$ inferred from the clumpy model lead to
consistent $\ha$- and UV-inferred SFRs.  These results suggest that
the $\langle\fcovhi\rangle$ inferred using the clumpy model are more
realistic than those inferred with the screen model.  Thus, the dust
column densities in optically-thin channels in the ISM are likely
lower than what is assumed for the screen model, and closer to what is
assumed in the clumpy model (i.e., negligible dust column densities).

There are a few salient points to keep in mind regarding the
inferences of gas covering fractions.  First, foreground contamination
\citep{vanzella10b, vanzella12, nestor11, nestor13, mostardi15,
  siana15} can in principle lead to residual flux under the Lyman
series lines in the spectra of the targets of interest.  However, as
discussed in \citet{reddy16b}, FUV spectra of the typical depth
attained in this study have proven to be quite effective in
identifying and removing spectroscopic blends of target and foreground
galaxies, particularly when coupled to deep optical (i.e., MOSFIRE)
spectra than enable the identification of low-redshift contaminants
over a broader range of redshifts via a second set of (typically
strong) optical nebular lines.  The final sample for this work
excludes galaxies for which either the MOSFIRE or LRIS spectrum
indicated there may be foreground contamination in the spectroscopic
aperture (Section~\ref{sec:finalsample}).  The purity of these samples
has been confirmed with spatially-resolved deep {\em HST} observed
optical and near-infrared imaging where foreground contaminants can be
identified through photometric redshifts of galaxy subcomponents
(\citealt{mostardi15, pahl21}).

Second, redshift uncertainties can artificially broaden and weaken the
Lyman series absorption lines, resulting in a larger residual flux
under the cores of the lines.  \citet{reddy16b} used a sample of
$z\sim 3$ UV-selected galaxies to show that uncertainties in the
systemic redshifts---which were derived from combining absorption line
and $\lya$ emission redshifts in a manner similar to that discussed in
Section~\ref{sec:redshifts}---cannot explain the level of residual
flux observed in composite FUV spectra, particularly for inferred
$\nhi \ga 10^{19.5}$\,cm$^{-2}$.  Relative to the previous work of
\citet{reddy16b}, our analysis has the advantage of direct
measurements of $z_{\rm sys}$ from the strong optical nebular emission
lines, resulting in a factor of $\simeq 2$ lower uncertainties in
$z_{\rm sys}$.  Thus, based on the precise $z_{\rm sys}$ and the large
column densities inferred from the damping wings of $\lya$, the
residual flux observed in the FUV composite spectra cannot be due to
redshift uncertainties.

Third, there are a few reasons why the covering fraction inferred from
the $\hi$ lines may be a lower limit on the true covering fraction of
high-column-density $\hi$.  First, limited spectral resolution will
mask the presence of narrow ($\Delta v \la 300$\,km\,s$^{-1}$)
absorption components that may be opaque to LyC (and $\lya$) radiation
but not detectable because of the lack of prominent damping wings,
corresponding to column densities of $18\la \lognhi\la 20$.
Section~\ref{sec:primary} argues that the presence of this moderate
column density and unresolved gas is likely correlated with the
covering fraction of optically-thick $\hi$.  Second, if different
velocity components of the $\hi$ gas are not spatially-coincident,
then the derived covering fraction will underestimate the true
covering fraction (e.g., \citealt{vasei16, reddy16b}).  Finally,
independent of the $\hi$ column density, $\lya$ and LyC photons may be
further attenuated if there is significant dust in the
low-column-density channels of the ISM (e.g., \citealt{borthakur14}),
as is assumed in the screen model of the ISM.  Consequently, the
line-of-sight escape fraction of $\lya$ photons may be viewed as a
more direct indicator of the effective gas and dust covering fractions
(Section~\ref{sec:primary}).

\subsubsection{Gas Covering Fractions from Saturated
Low-Ionization Interstellar Absorption Lines}
\label{sec:fcovmetal}

\begin{deluxetable}{lccc}
\tabletypesize{\footnotesize}
\tablecaption{Interstellar Absorption Line Measurements.}
\tablehead{
\colhead{Subsample} &
\colhead{$\langle\rsiione\rangle$\tablenotemark{a}} &
\colhead{$\langle\rsiitwo\rangle$\tablenotemark{b}} &
\colhead{$\langle\rcii\rangle$\tablenotemark{c}}}
\startdata
A & $0.46\pm0.04$ & $0.53\pm0.04$ & $0.52\pm0.04$ \comment{& $0.62\pm0.05$} \\
AL & $0.43\pm0.04$ & $0.53\pm0.04$ & $0.51\pm0.05$ \comment{& $0.62\pm0.05$} \\
ALN & $0.40\pm0.05$ & $0.48\pm0.04$ & $0.48\pm0.04$ \comment{& $0.61\pm0.04$} \\
\hline
WT1 & $0.30\pm0.07$ & $0.40\pm0.05$ & $0.40\pm0.06$ \comment{& $0.54\pm0.08$} \\
WT1L & $0.32\pm0.09$ & $0.38\pm0.07$ & $0.38\pm0.07$ \comment{& $0.56\pm0.08$} \\
WT1LN & $0.24\pm0.10$ & $0.47\pm0.07$ & $0.31\pm0.11$ \comment{& $0.59\pm0.09$} \\
\\
WT2 & $0.43\pm0.06$ & $0.47\pm0.06$ & $0.46\pm0.05$ \comment{& $0.60\pm0.07$} \\
WT2L & $0.40\pm0.07$ & $0.44\pm0.05$ & $0.45\pm0.07$ \comment{& $0.63\pm0.07$} \\
WT2LN & $0.30\pm0.07$ & $0.42\pm0.07$ & $0.43\pm0.09$ \comment{& $0.55\pm0.07$} \\
\\
WT3 & $0.51\pm0.07$ & $0.64\pm0.08$ & $0.71\pm0.05$ \comment{& $0.64\pm0.07$} \\
WT3L & $0.47\pm0.07$ & $0.64\pm0.09$ & $0.65\pm0.07$ \comment{& $0.59\pm0.09$} \\
WT3LN & $0.52\pm0.06$ & $0.53\pm0.06$ & $0.55\pm0.07$ \comment{& $0.56\pm0.07$} \\
\hline
ST1 & $0.36\pm0.09$ & $0.35\pm0.09$ & $0.45\pm0.09$ \comment{& $0.57\pm0.09$} \\
ST1L & $0.35\pm0.10$ & $0.37\pm0.09$ & $0.47\pm0.09$ \comment{& $0.55\pm0.09$} \\
\\
ST2 & $0.33\pm0.08$ & $0.40\pm0.07$ & $0.49\pm0.09$ \comment{& $0.55\pm0.08$} \\
ST2L & $0.29\pm0.07$ & $0.39\pm0.06$ & $0.39\pm0.08$ \comment{& $0.56\pm0.07$} \\
\\
ST3 & $0.46\pm0.06$ & $0.49\pm0.07$ & $0.48\pm0.07$ \comment{& $0.63\pm0.06$} \\
ST3L & $0.43\pm0.07$ & $0.48\pm0.07$ & $0.47\pm0.09$ \comment{& $0.59\pm0.06$} \\
\hline
sST1 & $0.31\pm0.11$ & $0.41\pm0.06$ & $0.47\pm0.08$ \comment{& $0.56\pm0.08$} \\
sST1L & $0.33\pm0.10$ & $0.42\pm0.06$ & $0.46\pm0.07$\comment{ & $0.56\pm0.08$} \\
\\
sST2 & $0.40\pm0.10$ & $0.37\pm0.10$ & $0.39\pm0.13$ \comment{& $0.57\pm0.07$} \\
sST2L & $0.35\pm0.10$ & $0.36\pm0.09$ & $0.35\pm0.13$ \comment{& $0.54\pm0.08$} \\
\\
sST3 & $0.45\pm0.05$ & $0.50\pm0.07$ & $0.53\pm0.05$ \comment{& $0.64\pm0.06$} \\
sST3L & $0.41\pm0.06$ & $0.48\pm0.07$ & $0.52\pm0.06$ \comment{& $0.64\pm0.06$}
\enddata
\tablenotetext{a}{Residual flux at line center of $\siii$\,$\lambda 1260$ relative to the continuum.}
\tablenotetext{b}{Residual flux at line center of $\siii$\,$\lambda 1527$ relative to the continuum.}
\tablenotetext{c}{Residual flux at line center of $\cii$\,$\lambda 1334$ relative to the continuum.}
\label{tab:isabs}
\end{deluxetable}

As discussed in Section~\ref{sec:intro}, saturated line transitions of
metal ions---in particular, low-ionization interstellar absorption
lines---have been widely-used to infer $\hi$ covering fractions given
their relative ease of detection.  As noted elsewhere \citep{henry15,
  rivera15, reddy16b, vasei16, steidel18, gazagnes18, chisholm18,
  du21}, covering fractions derived from the low-ionization metal
lines may significantly underpredict the $\hi$ covering fraction if,
for example, the metal-bearing gas traces only the regions with
highest gas densities, the $\hi$ gas is metal poor, or
partially-ionized $\hi$ provides significant LyC opacity and where
metal ions are primarily in the doubly-ionized state.  Additionally,
there may be some degree of emission filling of the absorption from
line photons scattered along the line of sight (e.g.,
\citealt{prochaska11, scarlata15}), though adjacent fine-structure
transitions can mitigate this effect (e.g., \citealt{steidel16}).
Nevertheless, several studies have found that covering fractions
deduced from the low-ionization metal lines correlate significantly
with those derived directly from the $\hi$ lines \citep{reddy16b,
  gazagnes18}, and thus the former could potentially be used as a
proxy for the $\hi$ gas covering fraction (e.g., \citealt{gazagnes18,
  chisholm18}).

With that in mind, we explored how the depths of three saturated
low-ionization metal lines in the composite FUV
spectra---$\siii$\,$\lambda 1260$, $\siii$\,$\lambda 1527$, and
$\cii$\,$\lambda 1334$---correlate with $\wlya$ and
$\fcovhi$.\footnote{\citet{du21} present a detailed analysis of the
  relationships between $\wlya$, equivalent widths of the
  low-ionization metal lines, and $\ebmvcont$ for MOSDEF-LRIS
  galaxies.  Here, we just focus on using the low-ionization metal
  lines as proxies for the gas covering fraction.}  In the
optically-thin limit, the ratio of the equivalent widths of the two
$\siii$ transitions is $W_{1260}/W_{1527} \ga 6$.  However, the
observed ratio is $W_{1260}/W_{1527} \simeq 1$, indicating that the
lines are saturated and their depths are sensitive to the covering
fraction of $\siii$-enriched gas.  The similarity in velocity width
and depth of the low-ionization transition $\cii$\,$\lambda 1334$ to
those of the saturated $\siii$ transitions implies that the former
arises from the same gas, and that its depth is also sensitive to the
covering fraction.  To measure these lines, we divided the composite
FUV spectra by their corresponding best-fit SPSneb models, thus
allowing us to both normalize the line absorption relative to the
continuum and remove any stellar absorption components (e.g.,
$\cii$\,$\lambda 1334$).  The residual fluxes, $\langle R\rangle$, at
the absorption line centers were then measured from the
continuum-normalized composite FUV spectra.  These values are listed
in Table~\ref{tab:isabs}.

\begin{figure}
\epsscale{1.15}
\plotone{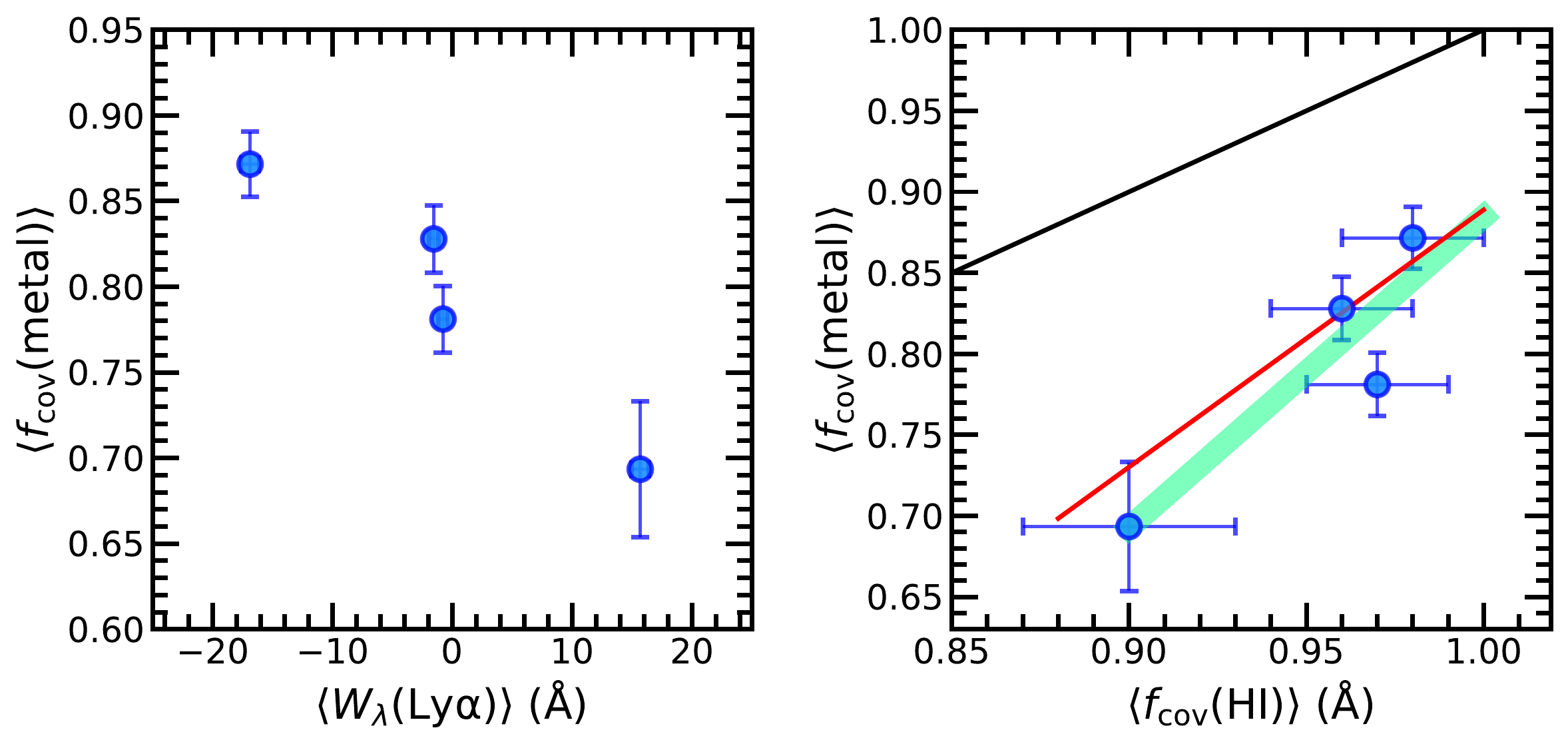}
\caption{ Inferred covering fraction of the metal-bearing gas versus
  $\langle\wlya\rangle$ ({\em left}) and $\langle \fcovhi\rangle$
  ({\em right}) for the clumpy geometry.  Measurements are shown for
  the AL, WT1L, WT2L, and WT3L subsamples which have coverage of
  $\lyb$.  In addition, the thick green line indicates a linear fit
  between $\langle\fcovmetal\rangle$ and $\langle\fcovhi\rangle$
  (Equation~\ref{eq:fcovmetalvsfcovhi}).  The red line indicates the
  best-fit linear relation between the covering fraction of
  $\siii$-enriched gas and $\hi$ for a sample of 18, mostly local,
  star-forming galaxies with Lyman series observations
  \citep{gazagnes18}.  An equivalent gas covering fraction of the
  metal-bearing gas and $\hi$ is indicated by the black line.}
\label{fig:fcovmetals}
\end{figure}

These residual fluxes were converted to covering fractions based on
assuming the two-component model discussed in
Section~\ref{sec:ismfittingprocedure} (see also \citealt{gazagnes18}).
The variations in the metal-line covering fractions, $\langle
\fcovmetal\rangle$ with $\langle\wlya\rangle$ and
$\langle\fcovhi\rangle$ are shown in Figure~\ref{fig:fcovmetals}.
These results show that galaxies with higher $\langle \wlya\rangle$
have lower covering fractions of metal-bearing gas.  The direct
comparison between the metal and $\hi$ covering fractions suggests the
former is systematically smaller than the latter (see discussion
above), indicating that some portion of the optically-thick $\hi$ may
be metal-poor.  Because of its potential utility, we provide the
following empirical calibration between $\langle\fcovmetal\rangle$ and
$\langle\fcovhi\rangle$ for the independent WTL subsamples analyzed
here:
\begin{equation}
\langle\fcovmetal\rangle = 1.50\langle\fcovhi\rangle - 0.62,
\label{eq:fcovmetalvsfcovhi}
\end{equation}
for $0.90\la\langle\fcovhi\rangle \la 1.00$ (shown as a thick green
line in the bottom right panel of Figure~\ref{fig:fcovmetals}).  We
have not formally calculated errors on the slope and intercept of the
relation as the subsamples are not independent of each other.  In any
case, while direct inferences of $\fcovmetal$ are useful, they do
require {\em a priori} knowledge of $\fcovhi$ and $\ebmvlos$ for the
clumpy model of the ISM discussed above.  For the screen model of the
ISM, both $\langle \fcovhi\rangle$ and $\langle\fcovmetal\rangle$ are
systematically lower than the corresponding values for the clumpy
model, while $\langle\fcovhi\rangle$ is still systematically larger
than $\langle\fcovmetal\rangle$.  At any rate, for the same reasons
given in Section~\ref{sec:wlyavsfcov}, the limited resolution of the
spectra implies that the derived $\fcovmetal$ may be a lower limit on
the true covering fraction of the metal-bearing gas.

\subsection{Summary of Composite-Fitting Results and 
Comparisons to Previous Works}
\label{sec:fittingsummary}

We performed comprehensive modeling of composite FUV and optical
spectra of galaxies in the MOSDEF-LRIS survey to deduce key properties
of the massive stars (Section~\ref{sec:fuvfitting}), ionized ISM
(Section~\ref{sec:optfitting}), and neutral ISM
(Section~\ref{sec:ismfitting}) of star-forming galaxies at redshifts
$1.85\le z\le 3.49$; and determined how these properties vary with
$\wlya$.  The results of the spectral modeling are summarized below.

\subsubsection{Reddening Measures}
\label{sec:physicalreddening}

We find a significant anti-correlation between $\langle \wlya\rangle$
and $\langle \ebmvcont\rangle$, similar to results from other studies
(e.g., \citealt{shapley03, gawiser06, atek08, atek09, pentericci09,
  pentericci10, finkelstein11, guaita11, jones12, atek14, hathi16,
  du18, du21}).  Generally, this trend has been interpreted in a
framework where $\lya$ photons are preferentially attenuated (owing to
their resonant scattering) compared to continuum photons, resulting in
lower $\wlya$ with increasing $\ebmvcont$.  Alternatively,
\citet{reddy16b} find that $\ebmvcont$ correlates with $\fcovhi$.
This correlation arises because a galaxy with higher $\fcovhi$ has a
larger fraction of sightlines with non-negligible dust content,
translating to a higher $\ebmvcont$; i.e., the reddening measured in
the case of a foreground screen of dust, namely $\ebmvcont$, is
effectively the line-of-sight reddening, $\ebmvlos$, weighted by the
gas covering fraction.  In this framework, the decrease in $\wlya$
with increasing $\ebmvcont$ may be more directly tied to the larger
fraction of $\lya$ photons scattering out of the line of sight with
increasing $\fcovhi$ (see Section~\ref{sec:primary}).

We do not find a significant correlation between $\langle
\wlya\rangle$ and $\langle \ebmvlos\rangle$.  This lack of correlation
is perhaps not unexpected if the majority of $\lya$ photons are
escaping through low-column-density (and low-reddening) channels in
the ISM, in which case $\wlya$ would be insensitive to the reddening
of the (high-column-density) {\em covered} portion of the continuum,
$\ebmvlos$.

Finally, $\langle\ebmvneb\rangle$ does not appear to correlate
significantly with $\langle\wlya\rangle$, at least on an individual
galaxy basis, with only a marginal correlation present for the
composite measurements.  At face value, these results run counter to
those reported by \citet{scarlata09}, who found {\em lower} $\lya/\ha$
ratios (i.e., roughly translating to lower $\wlya$) with increasing
$\ha/\hb$, or nebular reddening (see also \citealt{trainor16}).
However, this study was based exclusively on strong $\lya$ emitters at
$z=0.3$, and it is possible that the inclusion of similarly strong
LAEs in our $z\sim 2$ sample would reveal a trend that is otherwise
difficult to discern based on the small fraction of LAEs in our
sample.

For all the relevant subsamples presented here, $\langle
\ebmvneb\rangle > \langle\ebmvcont\rangle$.  Other studies report a
similar offset in the reddening of the nebular emission lines relative
to the stellar continuum (e.g., \citealt{fanelli88, calzetti94,
  calzetti00, forster09, kashino13, reddy10, kreckel13, price14,
  reddy15, battisti16, theios19, shivaei20, reddy20}), though the
extent of the difference depends on the attenuation curves assumed for
the two (e.g., \citealt{shivaei20, reddy20}).  As noted in many of
these same studies, this difference in nebular and stellar reddening
may arise from the higher column densities of dust along the
sightlines to the youngest stellar populations that dominate the
emission line luminosities.

\subsubsection{Ages}

Subsamples with $\langle\wlya\rangle > 10$\,\AA\, have best-fit ages a
factor of $\simeq 3-4\times$ younger (less than $\approx 100$\,Myr)
than those of subsamples with $\langle\wlya\rangle < 0$\,\AA\,
($\approx 300$\,Myr).  There is a marginal correlation between age and
$\wlya$ when examined on an individual galaxy basis, such that
galaxies with stronger $\lya$ emission are younger.  The relative
youth of galaxies with strong $\lya$ emission has been suggested in
many previous studies (e.g., \citealt{gawiser06, pentericci07,
  finkelstein07, guaita11, hagen14}), though there appears to be a
fairly large scatter in the relationship between $\lya$ strength and
age (e.g., \citealt{pentericci09, finkelstein09, kornei10}).  The
implication for these age differences on the shape of the ionizing
spectrum is discussed further in Section~\ref{sec:primary}.

\subsubsection{Stellar and Gas-Phase Metallicities}

There is no apparent significant variation of $\langle Z_\ast\rangle$
and $\langle Z_{\rm neb}\rangle$ with $\langle\wlya\rangle$.  The
typical stellar metallicity across the subsamples is $\langle
Z_\ast\rangle \approx 0.08$\,$Z_\odot$, and the typical nebular
abundance inferred from photoionization modeling is $\langle Z_{\rm
  neb}\rangle \approx 0.4$\,$Z_\odot$.  The direct-method $T_{\rm
  e}$-based estimates from $\interoiii$\,$\lambda\lambda 1660, 1666$
  point to $\langle Z_{\rm neb}\rangle $ consistent with those
  obtained from the photoionization modeling of all the strong optical
  nebular emission lines (Appendix~\ref{sec:directmethod}).  The lack
  of any significant correlation between $\wlya$ and either the
  stellar or gas-phase metallicity may be related to the limited
  dynamic range of metallicity probed by the MOSDEF-LRIS sample.  In
  particular, \citet{topping20b} find a range of $Z_{\rm neb}\simeq
  0.3-1.0Z_\odot$ for individual objects in the MOSDEF-LRIS sample,
  while strong LAEs at similar redshifts ($z\sim 2-3$) are found to
  have lower $Z_{\rm neb}\approx 0.2Z_\odot$ \citep{trainor16}.
  Similarly, the broader stellar mass and $\wlya$ ranges probed by
  \citet{cullen20} for a large spectroscopic sample of $3\le z\le 5$
  galaxies drawn from the VANDELS survey \citep{mclure18b,
    pentericci18a} allowed these authors to uncover a trend between
  $\wlya$ and $Z_\ast$.

Regardless, the apparent insensitivity of $\wlya$ to $Z_\ast$ for
MOSDEF-LRIS galaxies strongly suggests that the observed variation in
$\wlya$ within the sample may be tied to factors unrelated to the
intrinsic stellar population.  We revisit this issue in
Section~\ref{sec:primary}.  For the time being, we note that the
offset in the sample-averaged values of $\langle Z_\ast/Z_\odot\rangle
\simeq 0.08$ and $\langle Z_{\rm neb}/Z_\odot\rangle \simeq 0.40$
implies a ratio of (O/Fe) that is roughly $5\times$ the solar ratio,
placing these galaxies at the theoretical upper limit for (O/Fe) from
core-collapse (Type II) supernovae (SNe) enrichment \citep{nomoto06}.
The apparent $\alpha$-enhancement inferred for these galaxies is
consistent with their spectrally-derived stellar population ages of
less than a few hundred Myr (Section~\ref{sec:wlyavssps}), placing
them at a stage prior to the onset of significant Fe enrichment from
Type Ia SNe (e.g., see also \citealt{steidel16, cullen19, topping20a,
  matthee18}).

\subsubsection{Ionization Parameter, $\log U$}

The ionization parameter, $U$, exhibits significant variation with
$\wlya$: subsamples of galaxies with $\langle\wlya\rangle < 0$\,\AA\,
have low average ionization parameters of $\langle \log U \rangle \la
-2.9$ (and as low as $\langle \log U\rangle \simeq -3.2$), while those
with $\langle\wlya\rangle > 0$\,\AA\, have high average ionization
parameters of $\langle\log U \rangle \ga -2.8$.  Additional evidence
for this increase in ionization parameter with $\wlya$ comes from the
correlation between the latter and $\ciii$\,$\lambda\lambda
1907,1909$, one which cannot be explained by abundance variations
(Appendix~\ref{sec:c3}).  These results concur with previous studies
that have suggested high ionization parameters for LAEs relative to
galaxies with weaker $\lya$ emission at similar redshifts (e.g.,
\citealt{trainor16}).  A possible physical interpretation of the trend
between $\langle\wlya\rangle$ and $\langle\log U\rangle$ is discussed
in Section~\ref{sec:sigmavsU}.

\subsubsection{Evidence of Binary Stellar Populations}

SPS models that include the effects of binary stellar evolution (and
with a high-mass cutoff of the IMF of $100$ and $300$\,$M_\odot$)
predict significant stellar $\heii$\,$\lambda 1640$ emission which,
when subtracted from the observed $\heii$ emission in the composite
FUV spectra, yields residual (nebular) $\heii$ emission consistent
with the predictions from photoionization modeling.  On the other
hand, SPS models that do not include the effects of binary stellar
evolution, regardless of the high-mass cutoff of the IMF, lead to
inferences of nebular $\heii$ emission that are significantly higher
than the predictions of the photoionization models (see also
\citealt{steidel16}).  Only the SPS models that include stellar
binarity can self-consistently explain the observed $\heii$\,$\lambda
1640$ emission, a conclusion that applies to all of the subsamples
considered in this work, irrespective of $\wlya$.

\subsubsection{Column Densities and Gas Covering Fractions}

There is marginal evidence that the mean gas column density decreases
with increasing $\langle\wlya\rangle$, where the former varies from
$\langle\lognhi\rangle \simeq 21.1$ for subsamples with the lowest
$\langle\wlya\rangle$ to $\langle\lognhi\rangle \simeq 19.5$ for
subsamples with the highest $\langle\wlya\rangle$.  In all cases,
however, the best-fit $\lognhi$ imply gas that is optically thick in
all the Lyman series lines and the Lyman continuum.  As a result,
Ly$\alpha$ photons spatially coincident with the stellar continuum are
likely escaping through optically-thin (i.e., ionized or
low-column-density) channels in the ISM, or shifted out of resonance
by gas with non-zero velocity.

For both the clumpy and screen geometries of the ISM considered above,
we find that subsamples with $\langle\wlya\rangle < 0$\,\AA\, have
mean $\hi$ covering fractions, $\langle\fcovhi\rangle$, that are close
to unity, while the modeling of subsamples with $\langle\wlya\rangle >
0$\,\AA\, rules out unity covering fractions at the $\ga 3\sigma$
level.  Additional constraints on the covering fraction of
metal-bearing neutral $\hi$, $\fcovmetal$, comes from an examination
of saturated low-ionization interstellar absorption lines.  Covering
fractions inferred from these lines are strongly anti-correlated with
$\langle \wlya\rangle$, where subsamples with the lowest $\langle
\wlya\rangle$ have $\langle\fcovmetal\rangle \simeq 0.91$ while those
with the highest $\langle\wlya\rangle$ have $\langle\fcovmetal\rangle
\simeq 0.78$.  This strong anti-correlation between $\wlya$ and the
depths of the low-ionization interstellar absorption lines (or their
equivalent widths) has been noted in a large number of previous
studies (e.g., \citealt{shapley03, pentericci07, pentericci09, erb10,
  berry12, jones12, du18, marchi19, pahl20, du21}), and implies a
lower covering fraction of metal-bearing gas with increasing $\wlya$.

\section{\bf DISCUSSION}
\label{sec:discussion}

With the modeling results of Section~\ref{sec:modelfitting} in hand,
we are in a position to evaluate the primary mechanisms responsible
for the escape of $\lya$ photons for galaxies in our sample.  The
escape fraction of $\lya$ photons is discussed in
Section~\ref{sec:fesclya}.  Section~\ref{sec:primary} focuses on the
role of massive stars and the gas covering fraction on the escape of
$\lya$ photons.  The relations between gas covering fraction and the
distribution of star formation and galaxy potential is discussed in
Section~\ref{sec:sfrdistribution}, along with the connection between
star-formation-rate surface density and ionization parameter.
Section~\ref{sec:scatter} briefly addresses the scatter between
$\wlya$ and several other parameters examined in this work.  We then
conclude with a discussion of the implications of our analysis for the
escape of ionizing radiation at high redshift
(Section~\ref{sec:lycimplications}).

\subsection{$\lya$ Emission Equivalent Width and Escape Fraction}
\label{sec:fesclya}

The correlations presented up to this point have been cast in terms of
$\wlya$, as this quantity can be easily measured for individual
galaxies and ensembles of galaxies.  However, as noted in
Section~\ref{sec:compspec}, the connection between the
production/escape of $\lya$ photons and $\wlya$ is complicated by
virtue of the method used to compute $\wlya$.  $\wlya$ computed using
the procedures of \citet{kornei10} will not only depend on the level
of $\lya$ emission relative to the continuum, but also on the
underlying absorption.  This absorption is detected in some individual
galaxies (typically the brighter ones), ubiquitous in composite FUV
spectra (e.g., Figures~\ref{fig:compspec} and \ref{fig:fcovdemo}) and,
as per the discussion of Section~\ref{sec:ismfitting}, due primarily
to absorption from interstellar $\hi$.  Though the {\em net}
$\langle\wlya\rangle$ for a given subsample (or individual galaxy) may
be negative, there may be some residual leakage of $\lya$ photons.
Hence, to more directly connect the production and escape of $\lya$
photons to many of the galaxy properties discussed up to this point,
we calculated two additional quantities: the emission-line $\wlya$ as
\begin{equation}
\wlyaem = \frac{\llyaobs}{L_{\lambda}^{\rm red}},
\label{eq:wlyaem}
\end{equation}
and the ``escape'' fraction of $\lya$
photons as
\begin{equation}
\fesclya = \frac{\llyaobs}{\llyaint},
\label{eq:fesclya}
\end{equation}
where $\llyaobs$ and $\llyaint$ are the observed and intrinsic $\lya$
emission luminosities, respectively, and $L_{\lambda}^{\rm red}$ is
the mean luminosity density of the continuum just redward of $\lya$
(see \citealt{kornei10}).  Note that $\fesclya$ {\em only} refers to
the fraction of $\lya$ photons exiting along the sightline, and does
not account for photons resonantly scattered out of the spectroscopic
aperture.  As discussed in \citet{steidel11}, the total escape
fraction of $\lya$ when summing over all sightlines will typically be
$\simeq 3\times$ larger than $\fesclya$.

The observed $\lya$ emission luminosity was computed as follows.
First, the best-fit neutral ISM model
(Section~\ref{sec:ismfittingprocedure}) was subtracted from the
composite FUV spectrum, effectively removing both stellar and
interstellar $\hi$ absorption.  The resulting model-subtracted
spectrum was integrated between the wavelength points where the $\lya$
emission line intersects the absorption trough in the original
composite FUV spectrum, yielding $\langle \llyaobs\rangle$.  The
intrinsic $\lya$ luminosity was computed by multiplying the
dust-corrected $\ha$ luminosity, $\lha$ (see
Section~\ref{sec:optfittingprocedure}), by the intrinsic $\lya/\ha$
ratio predicted by the best-fit photoionization model.  This intrinsic
ratio varies in the range $(\lya/\ha)_{\rm int} = 8.99 - 9.58$
depending on the specific SPSneb model, photoionization model, and
BPASS model type, and is slightly larger than the canonically-assumed
value of $(\lya/\ha)_{\rm int} = 8.7$ for an $\hii$ region temperature
of $T_{\rm e} = 10,000$\,K.  While $\llyaint$ can also be computed
from the best-fit SPSneb model (i.e., by intergrating the model to
obtain the ionizing photon production rate, $N({\rm H^0})$;
Section~\ref{sec:primary}), the adopted method of computing
$\llyaint$---i.e., based on $\lha$---obviates the need to account for
the fraction of LyC photons that is either absorbed by dust (e.g.,
\citealt{inoue01a}) or escapes the ISM of the galaxy, and thereby
remains unavailable to photoionize hydrogen.

\begin{figure}
  \epsscale{1.15}
    \plotone{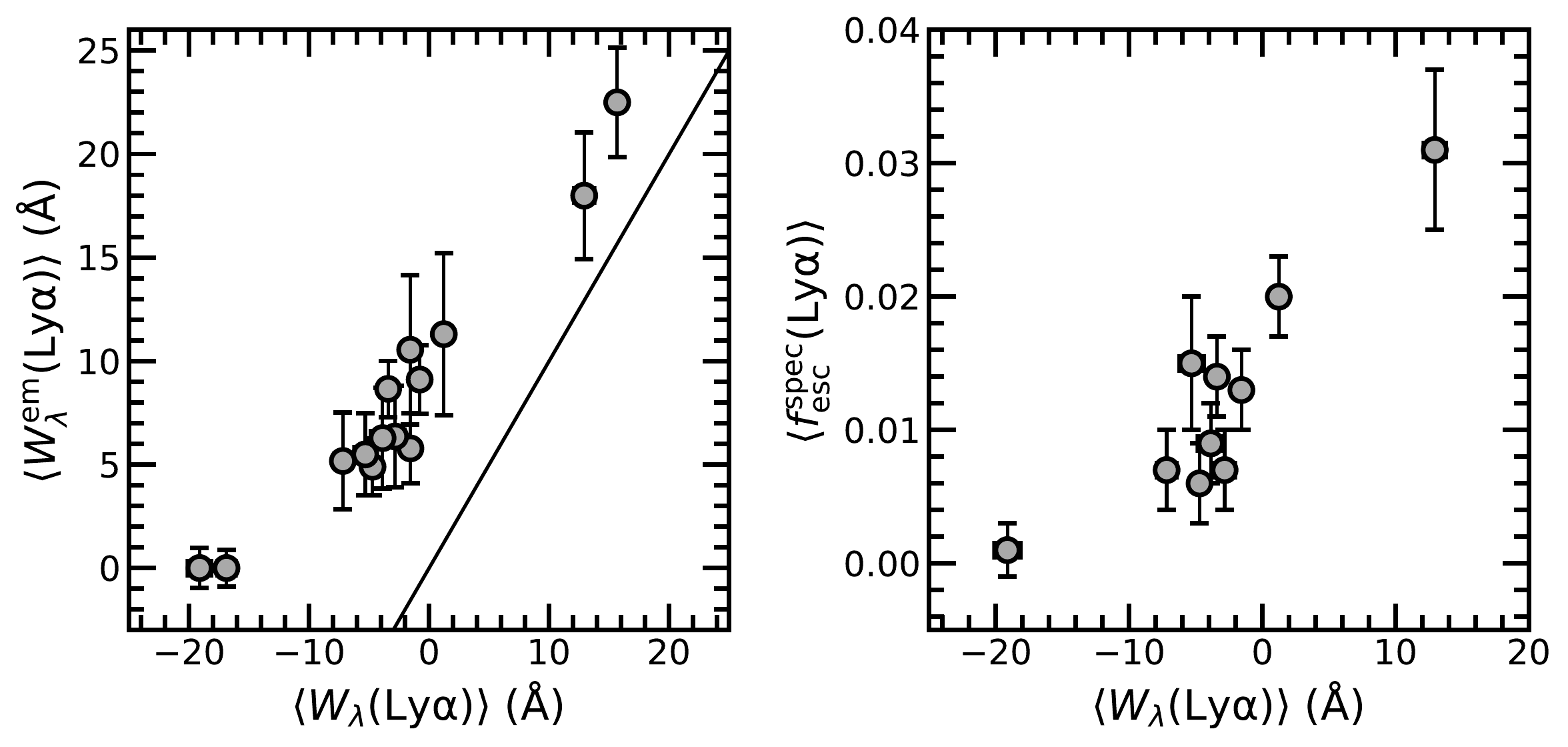}
    \caption{ Left: Emission-line $\wlya$, denoted by $\langle
      \wlyaem\rangle$, versus $\langle \wlya\rangle$ for the 14
      subsamples with complete coverage of $\lyb$
      (Table~\ref{tab:compositestats}).  The solid line indicates the
      one-to-one relation.  Right: Escape fraction of $\lya$,
      denoted by $\langle\fesclya\rangle$, versus
      $\langle\wlya\rangle$ for the 10 subsamples with complete
      coverage of the optical nebular emission lines and $\lyb$
      (Table~\ref{tab:compositestats}).  The values shown assume the
      100bin SPSneb models.}
    \label{fig:fesclyavswlya}
\end{figure}

\begin{figure*}
  \epsscale{1.15}
    \plotone{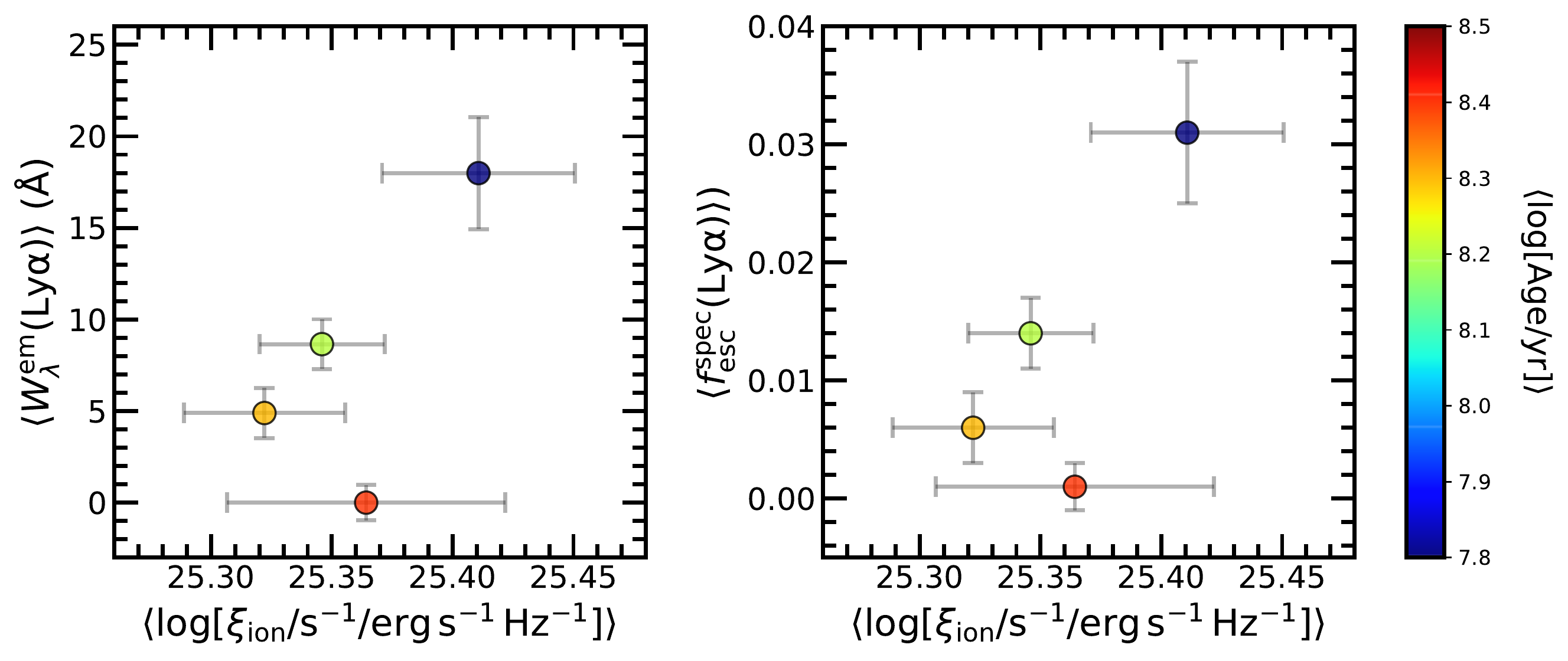}
    \caption{$\langle\wlyaem\rangle$ (left) and
      $\langle\fesclya\rangle$ (right) versus $\langle\logxi\rangle$
      for subsamples ALN, WT1LN, WT2LN, and WT3LN subsamples which
      have complete coverage of $\lyb$ and the optical nebular
      emission lines.  The values assume the 100bin SPSneb models.
      Points are color coded according to the best-fit $\log[{\rm
          Age/yr}]$ (Section~\ref{sec:wlyavssps}).}
    \label{fig:xiion}
\end{figure*}

For reference, $\langle \wlyaem\rangle$ and $\langle \fesclya\rangle$
are listed in Table~\ref{tab:params} and shown in
Figure~\ref{fig:fesclyavswlya}, relative to the corresponding values
of $\langle\wlya\rangle$, for the 100bin models.  The calculation of
$\langle\wlyaem\rangle$ and $\langle\fesclya\rangle$ depend on an
accounting of underlying $\lya$ absorption---constrained through the
neutral ISM modeling discussed in Section~\ref{sec:ismfitting}---and
thus require coverage of $\lyb$.  In addition, $\langle
\fesclya\rangle$ requires coverage of the optical nebular emission
lines to ensure an accurate estimate of the dust-corrected $\langle
\lha\rangle$.  Thus, Figure~\ref{fig:fesclyavswlya} shows the
relations between $\langle\wlyaem\rangle$ and $\langle\wlya\rangle$,
and between $\langle\fesclya\rangle$ and $\langle\wlya\rangle$, for
the 14 subsamples with complete coverage of $\lyb$ and the 10
subsamples with complete coverage of $\lyb$ and the optical nebular
emission lines, respectively.  Not surprisingly, many of the
subsamples with $\langle\wlya\rangle<0$\,\AA\, have a non-negligible
leakage of $\lya$ photons, where $\langle \wlyaem\rangle >0$\,\AA\,
and $\langle \fesclya\rangle > 0$.  As $\llyaobs$ increases and
dominates over the underlying absorption, $\langle\wlya\rangle$
approaches $\langle\wlyaem\rangle$.  Additionally, $\langle
\fesclya\rangle$ ranges from $\langle \fesclya\rangle \simeq 0$ to
$0.03$ for ensembles with the lowest and highest measured
$\langle\wlya\rangle$, respectively.  Evidently, the vast majority
($\ga 96\%$) of $\lya$ photons are either resonantly scattered out of
the spectroscopic aperture or attenuated by dust.  The computed values
of $\langle\wlyaem\rangle$ and $\langle\fesclya\rangle$ are examined
in the context of the shape of the ionizing spectrum and gas covering
fraction in the next section.

\subsection{Primary Modulator of Escaping $\lya$ Photons}
\label{sec:primary}

As noted at the outset, a central focus of this work is to quantify
the role of massive stars (i.e., the shape of the ionizing spectrum)
and gas covering fraction on the emergent $\lya$ emission of
high-redshift galaxies.  There are a few salient points worth
mentioning in the current context.  First, $\wlyaem$ is sensitive to
the shape of the ionizing spectrum, which determines the production of
$\lya$ relative to the non-ionizing FUV continuum, provided there is
sufficient $\hi$ gas to reprocess LyC photons into $\lya$ photons.
Second, $\wlyaem$ is also sensitive to the dust column density and
covering fraction of optically-thick $\hi$, as they determine the
fraction of $\lya$ photons that emerge along the observer's sightline.
On the other hand, $\fesclya$ is insensitive to the ionizing spectrum
for a fixed $\hi$ gas covering fraction and dust column density.  In
reality, of course, the ionizing spectrum may affect the $\hi$ gas
covering fraction: e.g., a harder ionizing spectrum may lead to a
larger fraction of ionized sightlines through which $\lya$ and LyC
leakage can occur (i.e., a lower $\hi$ gas covering fraction; e.g.,
\citealt{erb16, trainor16}).  Additionally, there is evidence that the
dust column density, or reddening, correlates with $\hi$ covering
fraction (e.g., \citealt{reddy16b}).  These points are addressed
below.

\subsubsection{Quantifying the Shape of the Ionizing Spectrum}

The shape of the ionizing spectrum of massive stars is determined by
the BPASS model type (binary versus single star evolution and IMF
slope), $Z_\ast$, $\log[{\rm Age/yr}]$, and the star-formation history
(e.g., \citealt{stanway16, bouwens16a, steidel16, shivaei18,
  chevallard18, chisholm19, topping20b}).  In
Section~\ref{sec:fuvfittingprocedure}, we motivated the choice of a
constant star formation model in fitting the average FUV spectrum of
an ensemble of $z\sim 2$ galaxies.  Further,
Section~\ref{sec:bpasstypemod} presents evidence that SPS models
including the effects of stellar binarity (100bin and 300bin models)
are able to self-consistently explain the observed level of
$\heii$\,$\lambda 1640$ emission in the composite FUV spectra, while
the single star models cannot.  Section~\ref{sec:wlyavssps} noted a
marginal correlation between $\langle \log[{\rm Age/yr}]\rangle$ and
$\langle\wlya\rangle$, while $\langle Z_\ast/Z_\odot\rangle$ appears
to be uncorrelated with $\langle\wlya\rangle$ (e.g.,
Figure~\ref{fig:wlyavssps}).  Based on these constraints, we can more
directly examine the connection between the escape of $\lya$ photons
and the shape (or hardness) of the ionizing spectrum by computing a
commonly-used proxy for the latter, namely the ionizing photon
production efficiency, $\xi_{\rm ion}$ (e.g., \citealt{robertson13,
  bouwens16a, shivaei18, theios19}):
\begin{equation}
\xi_{\rm ion} = \frac{N({\rm H^0})}{L_{\rm FUV}}\,{\rm [s^{-1}/erg\,s^{-1}\,Hz^{-1}]},
\label{eq:xiion}
\end{equation}
where $N({\rm H^0})$ is the ionizing photon rate in s$^{-1}$ and
$L_{\rm FUV}$ is the luminosity density at $1500$\,\AA\, in
erg\,s$^{-1}$\,Hz$^{-1}$.  $\xi_{\rm ion}$ is typically constrained by
combining dust-corrected H$\alpha$ and FUV luminosities (e.g.,
\citealt{bouwens16a, matthee17b, shivaei18, emami20}).  In this work,
$\xi_{\rm ion}$ was computed directly from the best-fit intrinsic
SPSneb model, a route that obviates the need to apply potentially
uncertain dust corrections to $\ha$ and FUV luminosities
\citep{shivaei18}, account for LyC photons that may escape the ISM or
be absorbed by dust \citep{inoue01a}, or implement conversions between
$\ha$ luminosity and $N({\rm H^0})$ that may not apply to the galaxies
in question (e.g., \citealt{reddy16b, theios19}).  The ionizing photon
rate was obtained by integrating the best-fit intrinsic SPSneb model:
\begin{equation}
N({\rm H^0}) = \int_{0}^{912} \frac{\lambda l_\lambda}{hc} d\lambda,
\end{equation}
where $\lambda$ is in \AA, and $l_\lambda$ is the luminosity density
of the best-fit intrinsic SPSneb model in erg\,s$^{-1}$\,\AA$^{-1}$.
Notwithstanding the aforementioned uncertainties in computing
$\xi_{\rm ion}$ from dust-corrected $\ha$ and FUV luminosities, such
estimates \citep{shivaei18} are consistent with those derived directly
from the SPSneb models for fixed assumptions of the nebular and
stellar dust attenuation curves.

\subsubsection{Relationships between $\lya$ Escape, $\xi_{\rm ion}$, and
Gas Covering Fraction}

The relationships between $\langle\wlyaem\rangle$ and
$\langle\logxi\rangle$, and between $\langle\fesclya\rangle$ and
$\langle\logxi\rangle$, are shown in Figure~\ref{fig:xiion}.  We do
not find significant trends in either case, and the scatter in
$\langle\logxi\rangle$ at a fixed $\langle\wlyaem\rangle$ (or
$\langle\fesclya\rangle$) appears to be driven by differences in age
of the stellar population, as demonstrated by the age color coding of
the points shown in Figure~\ref{fig:xiion}.  Subsamples with the
youngest ages, $\langle\log[{\rm Age/yr}]\rangle \la 8$, tend to have
higher $\langle\wlyaem\rangle$ and $\langle\fesclya\rangle$ on
average, and also have $\langle\logxi\rangle$ that are marginally
($\simeq 0.1$\,dex) larger than those of subsamples with the oldest
ages, $\langle\log[{\rm Age/yr}]\rangle \simeq 8.5$.  A similar
increase in $\xi_{\rm ion}$ for strong LAEs relative to
continuum-selected galaxies has also been reported at $z\sim 3$
\citep{nakajima18b}.  At any rate, the lack of a significant
correlation between $\langle\wlyaem\rangle$ and
$\langle\logxi\rangle$, or between $\langle\fesclya\rangle$ and
$\langle\logxi\rangle$, for galaxies in our sample strongly suggests
that the observed variation in $\wlya$ is driven by factors other
than---but which could still be influenced by---the shape of the
ionizing spectrum.  We return to a discussion of the scatter in the
relationship between $\langle f_{\rm cov}\rangle$ and $\langle
\logxi\rangle$ below.

\begin{figure}
  \epsscale{1.15}
    \plotone{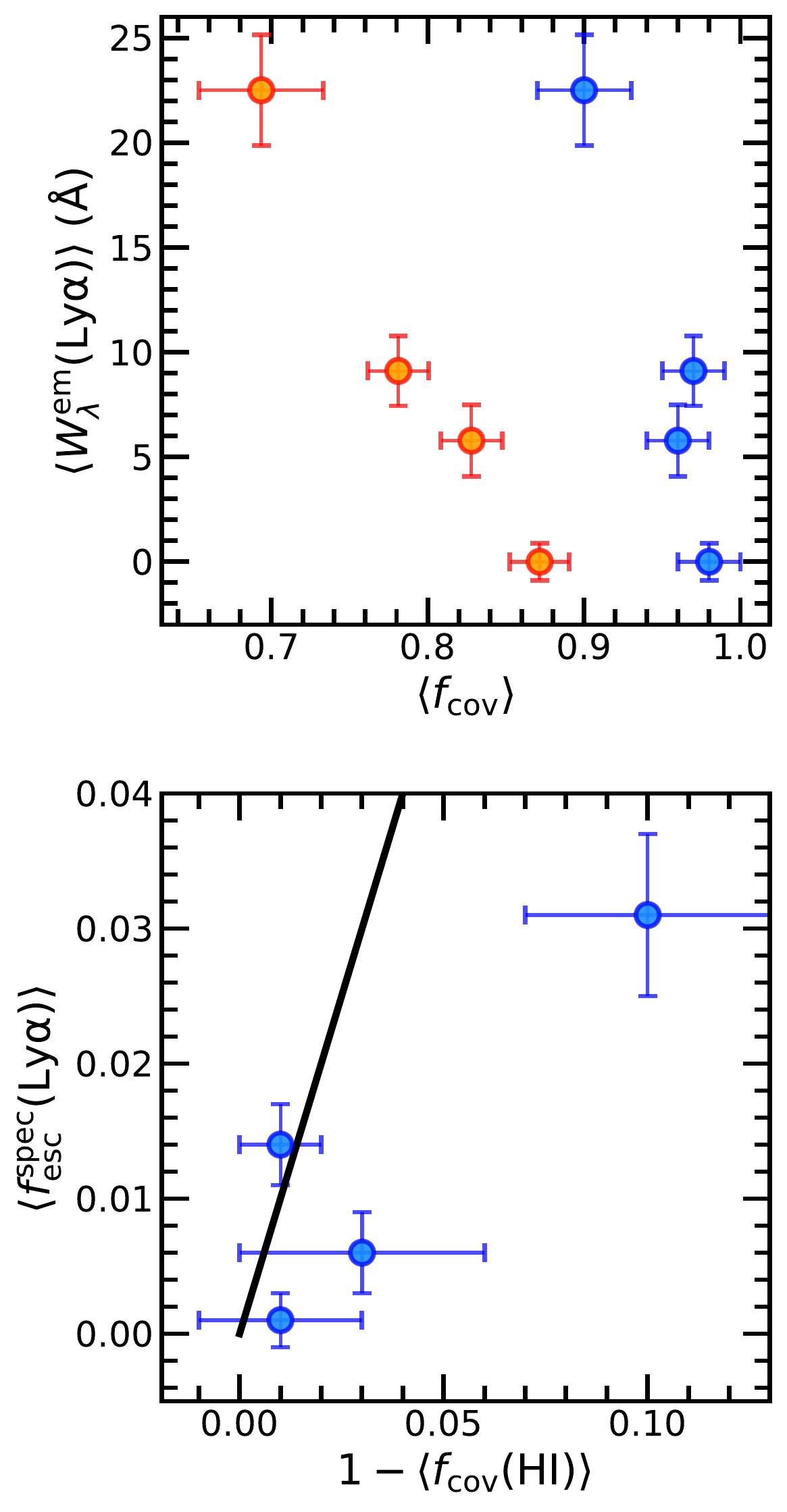}
    \caption{Top: $\langle\wlyaem\rangle$ versus
      $\langle\fcovhi\rangle$ (blue) and $\langle\fcovmetal\rangle$
      (red) for the AL, WT1L, WT2L, and WT3L subsamples which have
      coverage of $\lyb$, assuming the clumpy ISM model.  Bottom:
      $\langle\fesclya\rangle$ versus the uncovered fraction,
      $1-\langle \fcovhi\rangle$ for the ALN, WT1LN, WT2LN, and WT3LN
      subsamples which have coverage of $\lyb$ and the optical nebular
      emission lines.  The solid line indicates $\fesclya = 1 -
      \fcovhi$. }
    \label{fig:lyafracvsfcov}
\end{figure}

The variations of $\langle\wlyaem\rangle$ and $\langle\fesclya\rangle$
with gas covering fraction are directly shown in
Figure~\ref{fig:lyafracvsfcov} for the clumpy model of the ISM.
Subsamples with $\langle\wlyaem\rangle > 10$\,\AA\, exclusively have
$\langle\fcovhi\rangle \la 0.94$, while those with
$\langle\wlyaem\rangle < 10$\,\AA\, have $\langle\fcovhi\rangle \ga
0.94$.  Similar results are noted when considering
$\langle\fcovmetal\rangle$: subsamples with $\langle\wlyaem\rangle >
10$\,\AA\, exclusively have $\langle\fcovmetal\rangle \la 0.84$, while
those with $\langle\wlyaem\rangle < 10$\,\AA\, have
$\langle\fcovmetal\rangle \ga 0.84$.  The significance of the
difference in $\langle\fcovmetal\rangle$ for independent subsamples
(e.g., subsamples WT1 and WT3) is $\ga 6\sigma$.

Unlike the case with $\langle\logxi\rangle$, changes in $\langle
f_{\rm cov}\rangle$ are more than sufficient to account for the full
range of $\langle\wlyaem\rangle$.  If the latter is solely dependent
on the uncovered portion of the continuum, then
$\langle\wlyaem\rangle$ should be directly proportional to
$1-\langle\fcovhi\rangle$.  Furthermore, if $\lya$ photons are
preferentially leaking from ionized or optically-thin channels in the
ISM, then $\langle \fesclya\rangle \approx 1 - \langle\fcovhi\rangle$.
The bottom panel of Figure~\ref{fig:lyafracvsfcov} shows that the $\ga
10\times$ variation in the uncovered portion of the continuum for the
subsamples is more than sufficient to account for the $\ga 3\times$
variation in $\langle\fesclya\rangle$.  In the context of the screen
model, there is significant dust attenuation of $\lya$ photons (in the
optically-thin channels) that further modulates
$\langle\fesclya\rangle$.

\subsubsection{Role of Gas Covering Fraction in $\lya$ Escape}
\label{sec:physicalrole}

The results presented in the previous section imply that the covering
fraction of high-column-density $\hi$ is likely a principal factor
modulating $\langle\fesclya\rangle$.  This is perhaps not surprising
given that even very small changes in $\fcovhi$ can lead to large
variations in $\wlyaem$: e.g., for a fixed ionizing spectral shape, an
$\approx 3\%$ decrease in $\fcovhi$ from $0.97$ to $0.94$ results in a
factor of $2$ increase in $\langle\wlyaem\rangle$ and
$\langle\fesclya\rangle$, assuming $\langle\fesclya\rangle =
1-\langle\fcovhi\rangle$.  In other words, it is precisely because of
the very large gas covering fraction that a small change in this
fraction can lead to a dramatic variation in the line-of-sight (or
down-the-barrel) $\lya$ luminosity.  The bottom panel of
Figure~\ref{fig:lyafracvsfcov} demonstrates that our finding that a
vast majority ($\ga 95\%$) of $\lya$ photons are scattered out of (or
removed from) the line of sight is consistent with the high covering
fraction of optically-thick gas inferred from the depths of the Lyman
series absorption lines (Section~\ref{sec:wlyavsfcov}).

Moreover, there are a number of subsamples where
$1-\langle\fcovhi\rangle$ actually overpredicts $\langle
\fesclya\rangle$, i.e., $\langle\fesclya\rangle \la 1-
\langle\fcovhi\rangle$, implying that $\langle\fcovhi\rangle$ is a
lower limit to the true covering fraction for the reasons discussed in
Section~\ref{sec:wlyavsfcov}.  For example, the presence of unresolved
moderate column density gas may further scatter $\lya$ photons out of
the line of sight.  Figure~\ref{fig:lyafracvsfcov} shows that
$\langle\fesclya\rangle$ still correlates with
$1-\langle\fcovhi\rangle$, and that the (additional) opacity provided
by moderate column density gas (or dust) increases with decreasing
$\langle\fcovhi\rangle$ (e.g., see also \citealt{gazagnes20,
  kakiichi21}).\footnote{\citet{reddy16b} examined the relationship
  between $\langle\fesclya\rangle$ and $\langle\fcovhi\rangle$ for
  composites constructed in bins of $\ebmvcont$ for $\simeq L^\ast$
  UV-selected galaxies at $z\sim 3$.  They found a range of
  $\langle\fesclya\rangle \simeq 0.01 - 0.18$ and
  $\langle\fesclya\rangle \approx 1-\langle\fcovhi\rangle$ for all
  $\ebmvcont$ bins except the bluest one for which $\langle
  \fesclya\rangle > 1-\langle\fcovhi\rangle$.  In general, the
  $\langle\fesclya\rangle$ derived here are lower than those obtained
  by \citet{reddy16b} at a fixed $\langle\fcovhi\rangle$ as the
  intrinsic $\lha$ is larger at a fixed SFR when assuming the binary
  BPASS models.}

Several previous investigations have also emphasized the role of gas
covering fraction in the escape of $\lya$ and LyC photons (e.g.,
\citealt{rivera15, trainor15, dijkstra16, reddy16b, gazagnes18,
  chisholm18, steidel18, jaskot19, gazagnes20, matthee21}).  We also
highlight the recent study of \citet{cullen20}, which finds that the
variation in $Z_\ast$ among the galaxies in their sample is
insufficient to account for the range of observed $\wlya$, and that
some other factor (i.e., gas covering fraction) likely dominates the
escape of $\lya$.  Our analysis points to a similar conclusion for
typical star-forming galaxies at $z\sim 2$, based on the direct
modeling of the interstellar $\hi$ absorption lines and the depths of
saturated interstellar metal absorption lines in the composite FUV
spectra.

\begin{figure}
 \epsscale{1.15}
 \plotone{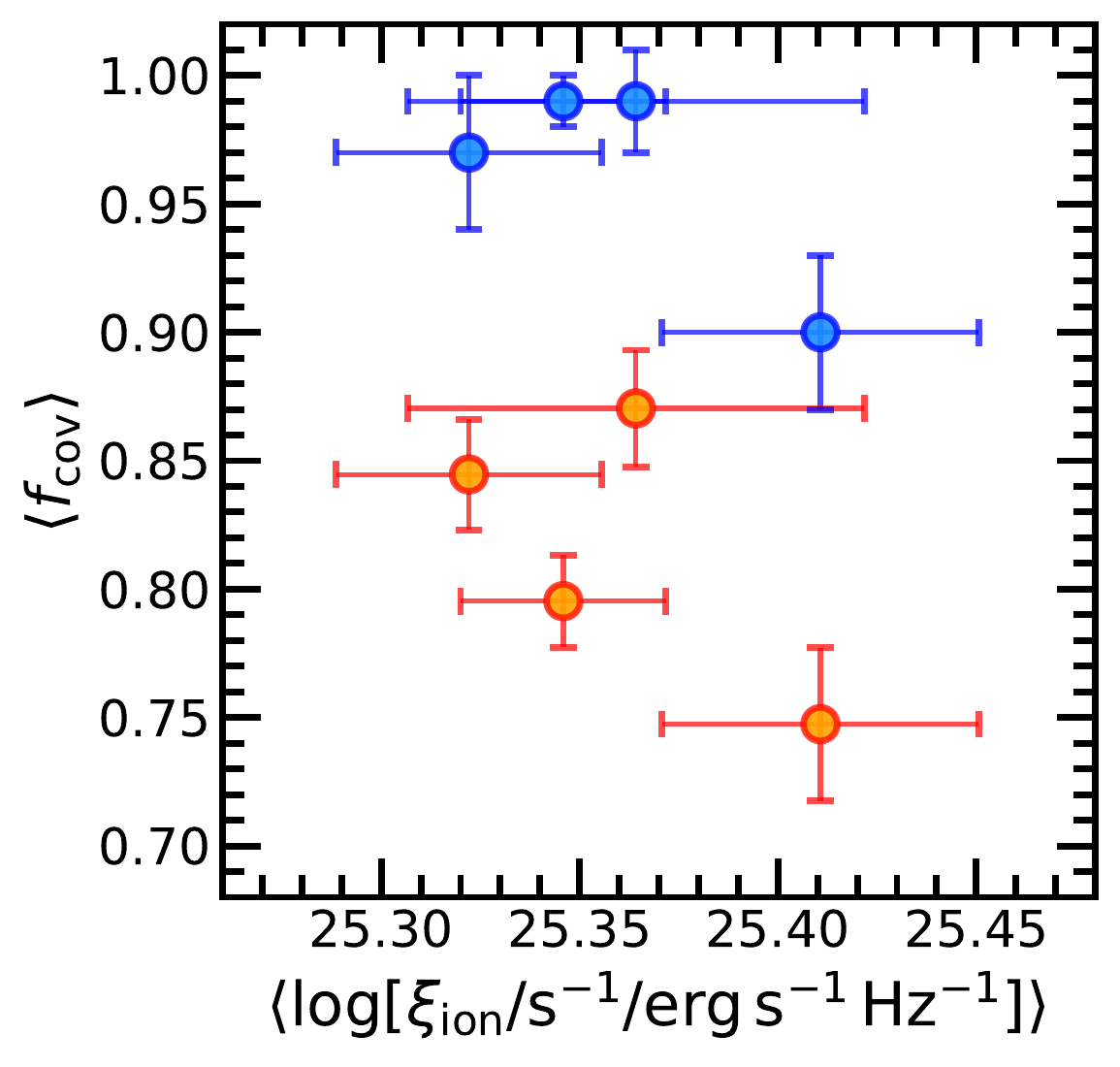}
    \caption{$\langle\fcovhi\rangle$ versus $\langle\logxi\rangle$
      (blue), and $\langle\fcovmetal\rangle$ versus
      $\langle\logxi\rangle$ (red), for the ALN, WT1LN, WT2LN, and
      WT3LN subsamples which have coverage of $\lyb$ and the optical
      nebular emission lines.}
    \label{fig:fcovvsxiion}
\end{figure}

Finally, we briefly comment here on the large scatter in
$\langle\fcovhi\rangle$ and $\langle\fcovmetal\rangle$ at a given
$\langle\logxi\rangle$ (e.g., Figure~\ref{fig:fcovvsxiion}).  In
particular, subsamples with the highest $\hi$ (near unity) or metal
covering fractions have $\langle\logxi\rangle$ that are consistent
within the sampling errors with the values obtained for galaxies with
lower $\langle \fcovhi\rangle$.  These results suggest that the shape
of the ionizing spectrum is not the sole determinative factor in the
covering fraction.
Section~\ref{sec:sfrdistribution} presents evidence that another
factor, namely the compactness of star formation, likely plays an
important role in determining the covering fraction.

In summary, the preference for binary stellar evolution models
irrespective of $\langle \wlya\rangle$
(Section~\ref{sec:bpasstypemod}), the absence of any significant
variation of $\langle Z_\ast\rangle$ with $\langle \wlya\rangle$
(Section~\ref{sec:wlyavssps}), and the modest anti-correlation between
$\langle\log[{\rm Age/yr}]\rangle$ and $\langle \wlya\rangle$
(Section~\ref{sec:wlyavssps}) together imply a narrow range of
$\langle\logxi\rangle \simeq 25.30 - 25.40$.  This range is
insufficient to account for the large variation in
$\langle\wlyaem\rangle$ observed for galaxies in our sample.  On the
other hand, the inferred range of $\langle\fcovhi\rangle$
(Section~\ref{sec:wlyavsfcov}) is more than sufficient to account for
the variation in $\langle\wlyaem\rangle$ observed for galaxies in our
sample.  Furthermore, $\langle\fesclya\rangle$ correlates with
$1-\langle\fcovhi\rangle$ in the manner one would expect if most of
the $\lya$ opacity is due to foreground optically-thick $\hi$ (or
dust, in the case of the screen model).  Accordingly, the covering
fraction of optically-thick $\hi$ (or dust) is the dominant factor in
modulating $\lya$ emission within the spectroscopic aperture, while
the shape of the ionizing spectrum---parameterized by $\xi_{\rm
  ion}$---plays a minor role in modulating the emergent $\lya$
emission of the galaxies analyzed here.

\subsection{Impact of the SFR Surface Density and Gravitational
Potential on the Escape of $\lya$ and the Ionization
Parameter}
\label{sec:sfrdistribution}

Section~\ref{sec:primary} presents evidence that the covering fraction
of optically-thick $\hi$ is the dominant factor in modulating
$\langle\wlyaem\rangle$ and $\langle\fesclya\rangle$ among the
galaxies in our sample.  The question remains as to the properties of
galaxies that regulate the $\hi$ covering fraction.  The radiative,
thermal, and mechanical feedback from star formation, stellar winds,
and/or supernovae can promote the formation of ionized and/or
low-column-density channels in the ISM (and CGM), providing pathways
through which $\lya$ and LyC photons can escape galaxies (e.g.,
\citealt{gnedin08, ma16, kimm19, ma20, cen20, kakiichi21}).  Both
theoretical and observational work suggest the impact of these
``feedback'' effects on the surrounding ISM, and the subsequent escape
of $\lya$ and LyC photons, can be enhanced in regions of compact star
formation (e.g., \citealt{ma16, sharma16, verhamme17, marchi19, cen20,
  naidu20}), typically expressed by the SFR surface density,
$\Sigma_{\rm SFR}$, in units of $M_\odot$\,yr$^{-1}$\,kpc$^{-2}$.  The
deep CANDELS imaging in the MOSDEF survey fields allows us to examine
the connection between $\lya$ escape and $\Sigma_{\rm SFR}$ through
the measurement of galaxy sizes.  In addition, while previous analyses
have focused almost exclusively on the impact of $\Sigma_{\rm SFR}$ on
the escape of $\lya$ and LyC photons, the galaxy potential may also
play an important role (e.g., \citealt{kim20}).  For example, the
feedback associated with a fixed $\Sigma_{\rm SFR}$ may be more
efficient in clearing sightlines in the ISM of a low-mass
(low-escape-velocity) galaxy relative to a high-mass (high-escape
velocity) galaxy.  This possibility is explored below by also
considering $\Sigma_{\rm SFR}$ normalized by stellar mass, $M_\ast$.
Section~\ref{sec:sigsfrcalc} describes the calculation of $\Sigma_{\rm
  SFR}$ and its normalization by $M_\ast$, while the correlation
between $\lya$ escape and these quantities is discussed in
Section~\ref{sec:sigmavslya}.  Section~\ref{sec:sigmavsU} concludes
with a discussion of the correlation between the compactness of star
formation and ionization parameter.

\subsubsection{Parameterization of the Distribution of SFR and Galaxy Potential}
\label{sec:sigsfrcalc}

The distribution, or compactness, of star formation is typically parameterized
by the SFR surface density, $\Sigma_{\rm SFR}$:
\begin{equation}
\Sigma_{\rm SFR} = \frac{\sfrha}{2\pi R_{\rm e}^2},
\end{equation}
where $R_{\rm e}$ is the effective radius within which half the total
light of the galaxy is contained.  These half-light radii were
obtained from the single- component S\'{e}rsic fits to the {\em
  HST}/F160W images of the MOSDEF-LRIS galaxies, as compiled in
\citet{vanderwel14}.  Note that $\Sigma_{\rm SFR}$ was computed assuming
$\sfrha$.  Adopting the $\sfrsed$ inferred from broadband SED modeling
(Section~\ref{sec:sedmodeling}) does not significantly affect any of
our conclusions.

For individual galaxies for which $\ha$ and $\hb$ are both detected
with $S/N\ge 3$, $\sfrha$ was computed by multiplying the
dust-corrected $\lha$ for each galaxy
(Section~\ref{sec:optfittingprocedure}) by the factor $2.12\times
10^{-42}$\,$M_\odot$\,yr$^{-1}$\,erg$^{-1}$\,s, appropriate for the
$Z_\ast = 0.001$ 100bin SPSneb models---i.e., the same models assumed
in fitting the broadband photometry (Section~\ref{sec:sedmodeling}).
For ensembles containing galaxies with complete coverage of the
optical nebular emission lines, $\langle\sfrha\rangle$ was computed by
multiplying the dust-corrected $\langle\lha\rangle$ by the factor
derived from the best-fitting SPSneb model.  This factor is similar to
the one used for individual galaxies, and depends on the exact age and
metallicity of the SPSneb model that best fits the composite FUV
spectrum.  Furthermore, these values of $\langle\sfrha\rangle$ were
divided by $\langle\fcovhi\rangle$ inferred from the clumpy ISM model
(Section~\ref{sec:ismfittingprocedure}) to account for the fraction of
ionizing photons that escapes, and is therefore unavailable for
generating Balmer recombination photons.  $\Sigma_{\sfrha}$ for
individual galaxies and ensembles were then computed from the
$\sfrha$, assuming the individual and mean $R_{\rm e}$, respectively.

To examine the dependence of $\lya$ escape on gravitational potential,
we computed $\Sigma_{\sfrha}$ normalized by stellar mass:
\begin{equation}
\Sigma_{\rm sSFR} = \frac{\Sigma_{\rm SFR}}{M_\ast} = 
\frac{\sfrha}{2\pi R_{\rm e}^2 M_\ast},
\end{equation}
where the subscript ``${\rm sSFR}$'' refers to the specific SFR based
on $\ha$, and $M_\ast$ is used as a proxy for the galaxy
potential.\footnote{Galaxy potentials based on dynamical and baryonic
  masses will be considered elsewhere.  For the time being, we note
  that \citet{price20} find a highly significant correlation between
  dynamical and stellar mass for $z\sim 2$ galaxies in the MOSDEF
  sample: a Spearman test of the correlation between these two
  variables yields $\rho = 0.67$ with a probability of $p=5.19\times
  10^{-44}$ of a null correlation.  Thus, we use $M_\ast$ as a proxy
  for galaxy potential, given that this parameter is available for
  every galaxy in the sample.}  For simplicity, the factor of 2 in the
denominator is retained---a choice that does not affect any of the
relative trends examined in this work---as the impact of feedback on
the ISM is likely to be sensitive to the entire galaxy mass, not just
that contained within the half-light radius.

Galaxies were grouped together based on their individually-measured
$\Sigma_{\rm SFR}$ and $\Sigma_{\rm sSFR}$ to form subsamples ST1
through ST3 and sST1 through sST3, respectively
(Table~\ref{tab:compositestats}).\footnote{The grouping of galaxies
  based on their $\Sigma_{\rm SFR}$ and $\Sigma_{\rm sSFR}$ requires
  that they have detected $\ha$ and $\hb$ emission lines.  However, we
  note that grouping galaxies based on $\Sigma_{\sfrsed}$ and
  $\Sigma_{\rm s\sfrsed}$ (which do not require individual $\ha$ and
  $\hb$ detections) yields results similar to those presented here.}
The best-fit parameters obtained from fitting the composite spectra of
these subsamples are listed in Table~\ref{tab:params}; and the average
sizes, stellar masses, $\sfrha$, $\Sigma_{\rm SFR}$, and $\Sigma_{\rm
  sSFR}$ for all the relevant subsamples (i.e., those with coverage of
$\lyb$ and the optical nebular emission lines) are given in
Table~\ref{tab:sigmasfr}.

\begin{deluxetable*}{lccccc}
\tabletypesize{\footnotesize}
\tablecaption{Galaxy Sizes, $\sfrha$, $\Sigma_{\rm SFR}$, and $\Sigma_{\rm sSFR}$}
\tablehead{
\colhead{Subsample} &
\colhead{$\langle R_{\rm e}\rangle$\tablenotemark{a}} &
\colhead{$\langle M_\ast\rangle$\tablenotemark{b}} &
\colhead{$\langle \sfrha\rangle$\tablenotemark{c}} &
\colhead{$\langle \Sigma_{\rm SFR}\rangle$\tablenotemark{d}} &
\colhead{$\langle \Sigma_{\rm sSFR}\rangle$\tablenotemark{e}} \\
\colhead{} &
\colhead{(kpc)} &
\colhead{($10^{10}$\,$M_\odot$)} &
\colhead{($M_\odot$\,yr$^{-1}$)} &
\colhead{($M_\odot$\,yr$^{-1}$\,kpc$^{-2}$)} &
\colhead{(Gyr$^{-1}$\,kpc$^{-2}$)}}
\startdata
ALN & $2.32\pm0.02$ & $1.11\pm0.03$ & $ 16.14\pm  2.49$ & $0.479\pm0.083$ & $ 0.044\pm 0.008$ \\
\hline
WT1LN & $2.31\pm0.02$ & $1.24\pm0.03$ & $ 17.08\pm  7.55$ & $0.521\pm0.258$ & $ 0.043\pm 0.020$ \\
WT2LN & $2.67\pm0.02$ & $1.17\pm0.03$ & $ 23.15\pm  8.34$ & $0.507\pm0.174$ & $ 0.046\pm 0.018$ \\
WT3LN & $2.05\pm0.01$ & $0.99\pm0.02$ & $ 19.27\pm  4.18$ & $0.751\pm0.191$ & $ 0.080\pm 0.031$ \\
\hline
ST1L & $3.12\pm0.04$ & $0.97\pm0.02$ & $ 10.08\pm  3.75$ & $0.163\pm0.051$ & $ 0.017\pm 0.006$ \\
ST2L & $2.57\pm0.01$ & $1.63\pm0.04$ & $ 26.96\pm  6.31$ & $0.660\pm0.155$ & $ 0.041\pm 0.012$ \\
ST3L & $1.30\pm0.00$ & $0.94\pm0.02$ & $ 32.51\pm 11.60$ & $3.012\pm1.032$ & $ 0.318\pm 0.097$ \\
\hline
sST1L & $3.39\pm0.02$ & $1.82\pm0.05$ & $ 17.64\pm  6.08$ & $0.241\pm0.076$ & $ 0.013\pm 0.004$ \\
sST2L & $2.37\pm0.02$ & $1.23\pm0.03$ & $ 26.23\pm  7.87$ & $0.755\pm0.250$ & $ 0.061\pm 0.018$ \\
sST3L & $1.36\pm0.00$ & $0.51\pm0.01$ & $ 20.37\pm  5.96$ & $1.752\pm0.538$ & $ 0.355\pm 0.114$
\enddata
\tablenotetext{a}{Mean effective radius.}
\tablenotetext{b}{Mean stellar mass.}
\tablenotetext{c}{Mean $\sfrha$, corrected for the fraction of ionizing photons that escape the galaxies based on $\fcovhi$ inferred from the clumpy ISM model (see text).}
\tablenotetext{d}{Mean $\Sigma_{\rm SFR}$, assuming $\sfrha$.}
\tablenotetext{e}{Mean $\Sigma_{\rm sSFR}$, assuming $\sfrha$.}
\label{tab:sigmasfr}
\end{deluxetable*}

\subsubsection{Connection between $\lya$ Escape and the SFR Distribution and Galaxy Potential}
\label{sec:sigmavslya}

\begin{figure}
    \plotone{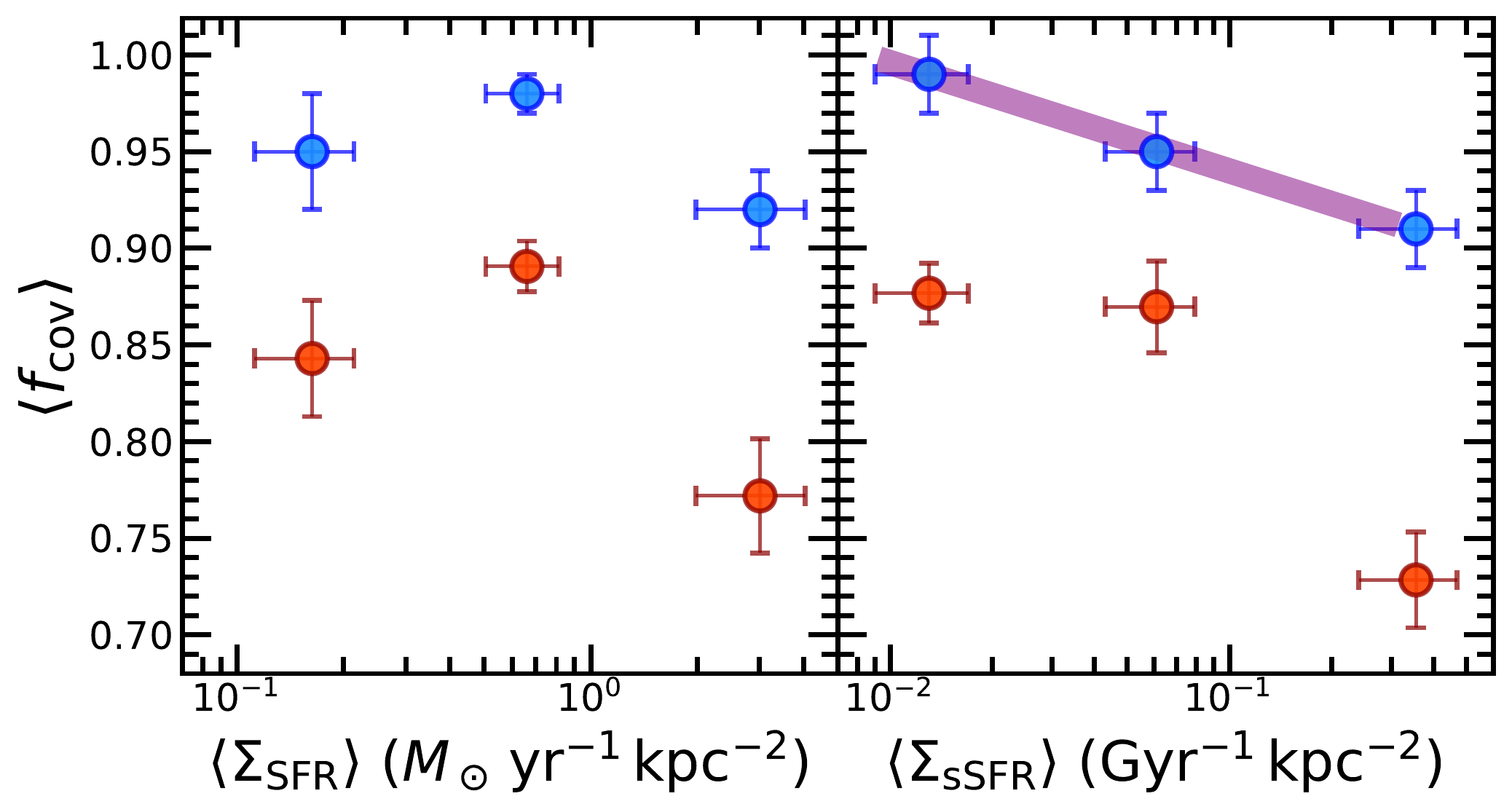}
    \caption{Dependence of $\hi$ (blue) and metal covering fraction
      (red) on $\Sigma_{\rm SFR}$ (left) and $\Sigma_{\rm sSFR}$
      (right) for, respectively, the ST and sST subsamples with
      coverage of $\lyb$, assuming the clumpy ISM model.  The thick purple
      line in the top right panel denotes a linear fit to the
      independent subsamples constructed in bins of $\Sigma_{\rm
        sSFR}$, indicating $\langle\fcovhi\rangle \propto
      -0.056\log[\langle\Sigma_{\rm sSFR}\rangle/{\rm
          Gyr^{-1}\,kpc^{-2}}]$ (see
      Equation~\ref{eq:fcovvssigmassfr}).}
    \label{fig:fcovvssigma}
\end{figure}

The correlations between $\wlya$ and $R_{\rm e}$, SFR, and $M_\ast$
for individual galaxies are presented in
Appendix~\ref{sec:othercorrelations}.  Here, we focus on the
individual and composite measurements of $\Sigma_{\rm SFR}$ and
$\Sigma_{\rm sSFR}$ and their impact on $\lya$ escape.  If
concentrated star formation and a shallower galaxy potential lead to
conditions favorable for the formation of ionized or
low-column-density channels in the ISM, then we expect the
gas covering fraction to correlate with $\Sigma_{\rm SFR}$
and $\Sigma_{\rm sSFR}$.  Figure~\ref{fig:fcovvssigma} shows the
variation of $\langle\fcovhi\rangle$ and $\langle\fcovmetal\rangle$
with $\langle \Sigma_{\rm SFR}\rangle$ and $\langle \Sigma_{\rm
  sSFR}\rangle$.

The left panel of Figure~\ref{fig:fcovvssigma} shows that there is no
significant difference in $\langle\fcovhi\rangle$ for galaxies in the
lowest and highest $\Sigma_{\rm SFR}$ bins.  There is a marginally
significant difference ($\simeq 2\sigma$) in
$\langle\fcovmetal\rangle$ for the same two bins: galaxies in the
upper third of the $\Sigma_{\rm SFR}$ distribution have lower
$\langle\fcovmetal\rangle$ than galaxies in the lower third of the
$\Sigma_{\rm SFR}$ distribution.  There does not appear to be a clear
monotonic relation between either $\langle\fcovhi\rangle$ or
$\langle\fcovmetal\rangle$ and $\langle\Sigma_{\rm SFR}\rangle$ over
the dynamic range probed by the MOSDEF-LRIS sample.  

In contrast to their lack of obvious correlation with $\langle
\Sigma_{\rm SFR}\rangle$, $\fcovhi$ and $\fcovmetal$ appear to be
significantly anti-correlated with $\langle \Sigma_{\rm sSFR}\rangle$.
For example, galaxies in the upper third of the $\Sigma_{\rm sSFR}$
distribution have $\langle\fcovhi\rangle = 0.91\pm 0.02$, while those
in the lower third have $\langle\fcovhi\rangle = 0.99\pm 0.02$, a
difference that is significant at the $\approx 3\sigma$ level.
Similarly, there is a large and significant difference in
$\langle\fcovmetal\rangle$ for galaxies in the upper and lower third
of $\Sigma_{\rm sSFR}$.  Similar conclusions are reached if we
consider $\langle\fcovhi\rangle$ and $\langle\fcovmetal\rangle$ from
the screen model.

\begin{figure}
  \epsscale{1.15}
    \plotone{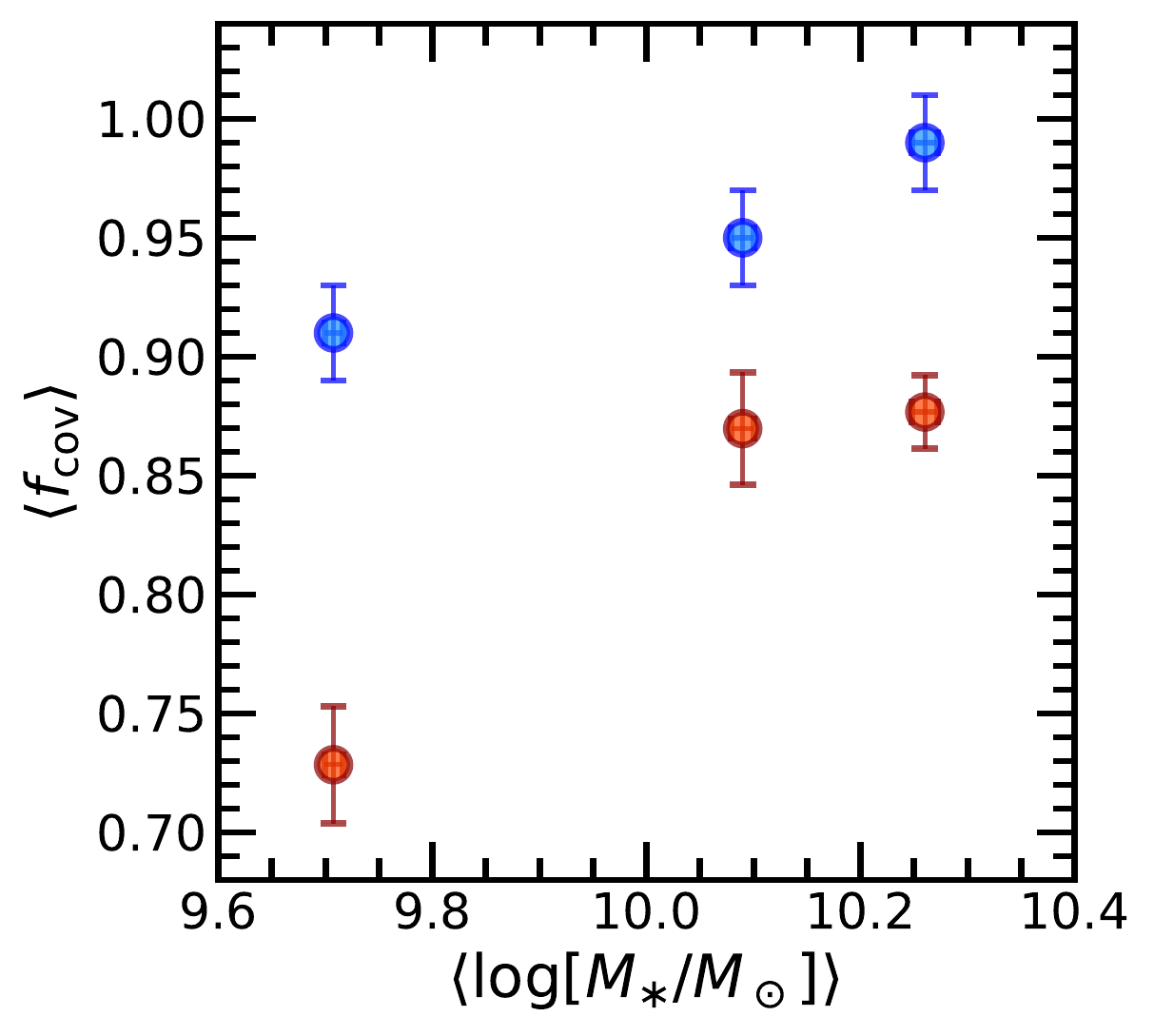}
    \caption{Correlation of $\langle \fcovhi\rangle$ (blue points) and
      $\langle\fcovmetal\rangle$ (red points) with average stellar
      mass, $\langle M_\ast\rangle$, for subsamples sST1L, sST2L, and
      sST3L.}
    \label{fig:fcovvsmass}
\end{figure}

The difference in $\langle\fcovhi\rangle$ (and
$\langle\fcovmetal\rangle$) between the upper and lower third of the
$\Sigma_{\rm sSFR}$ distribution is both larger and more significant
that the difference in the covering fraction in the upper and lower
third of the $\Sigma_{\rm SFR}$ distribution.  Additionally, there is
a clear monotonic trend between covering fraction and
$\langle\Sigma_{\rm sSFR}\rangle$ while no such trend is apparent
between covering fraction and $\langle\Sigma_{\rm SFR}\rangle$.  These
results point to a tighter anti-correlation between
$\langle\fcovhi\rangle$ and $\langle\Sigma_{\rm sSFR}\rangle$ than
between $\langle\fcovhi\rangle$ and $\langle\Sigma_{\rm SFR}\rangle$,
in accordance with the expectation that the gravitational potential
may play a role in the impact of feedback on gas covering fraction.
Fitting the independent subsamples in bins of $\Sigma_{\rm sSFR}$
yields the following relation:
\begin{equation}
\langle\fcovhi\rangle = -0.056\log\left[\frac{\langle\Sigma_{\rm
      sSFR}\rangle}{{\rm Gyr}^{-1}\,{\rm kpc}^{-2}}\right]  + 0.884,
\label{eq:fcovvssigmassfr}
\end{equation}
for $0.01\la \Sigma_{\rm sSFR}\la 0.3$\,Gyr$^{-1}$\,kpc$^{-2}$ (thick
purple line in the right panel of Figure~\ref{fig:fcovvssigma}).  

The possibility that covering fraction anti-correlates more tightly
with $\Sigma_{\rm sSFR}$ than with $\Sigma_{\rm SFR}$ suggests that
the covering fraction should correlate with $M_\ast$.  As shown in
Figure~\ref{fig:fcovvsmass}, galaxies with lower stellar masses have
lower $\hi$ and metal covering fractions.  Ensembles with $\langle
M_\ast\rangle \ga 10^{10}$\,$M_\odot$ have $\langle \fcovhi\rangle \ga
0.95$, while those with $\langle M_\ast\rangle \la 10^{10}$\,$M_\odot$
have $\langle \fcovhi\rangle \la 0.95$.  For subsamples with
$\langle\log[M_\ast/M_\odot]\rangle \la 9.8$, unity covering fractions
of $\hi$ can be ruled out with $\ga 3\sigma$ confidence.  The
difference in the covering fraction of metal-bearing gas between low-
and high-mass galaxies is even more pronounced: subsamples with
$\langle\log[M_\ast/M_\odot]\rangle \la 10.0$ have
$\langle\fcovmetal\rangle \la 0.80$, signifying a $\ga 10\sigma$
difference from the $\langle\fcovmetal\rangle\simeq 0.90$ for
subsamples with $\langle\log[M_\ast/M_\odot]\rangle \ga 10.2$.  In the
present framework, the correlation between covering fraction and
$M_\ast$ is driven by the higher $\Sigma_{\rm SFR}$ and lower
gravitational potential associated with low-mass galaxies, conditions
favorable for an increased ISM porosity.

A more detailed investigation of the effects of compact star formation
and the galaxy potential on gas kinematics will be presented
elsewhere.  For the time being, note that the anti-correlation between
covering fraction and $\Sigma_{\rm sSFR}$ comports with the
expectation that compact star formation in a low gravitational
potential has the effect of decreasing the covering fraction of
optically-thick $\hi$.  As a consequence, $\wlya$, $\wlyaem$, and
$\lyafrac$ are all expected to increase in such cases.

\begin{figure}
  \epsscale{1.15}
    \plotone{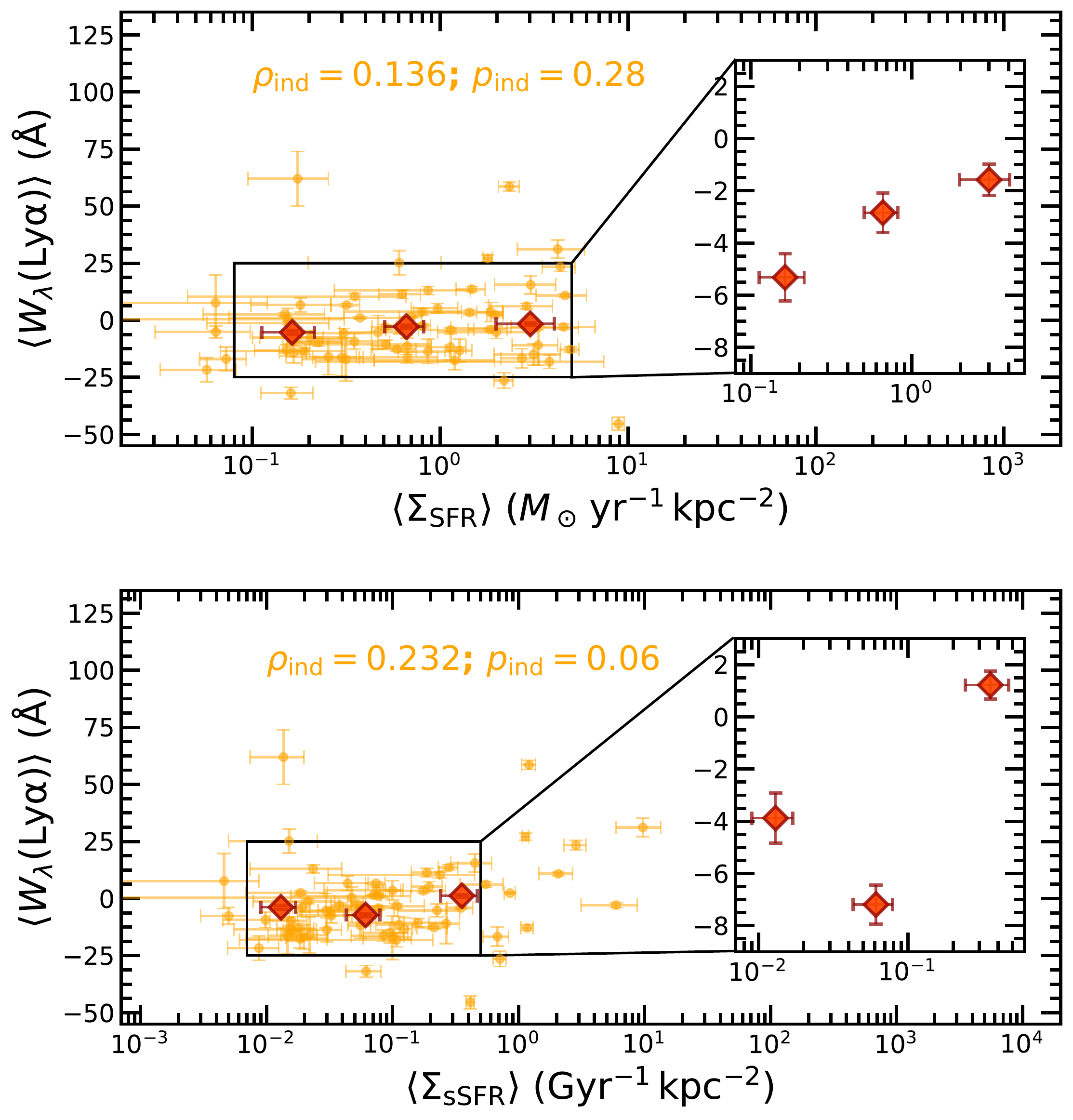}
    \caption{$\wlya$ versus $\Sigma_{\rm SFR}$ (top) and $\Sigma_{\rm
        sSFR}$ (bottom) for individual galaxies and subsamples.  Small
      orange points indicate individual galaxies.  The Spearman
      correlation coefficient ($\rho_{\rm ind}$) and $p$-value
      ($p_{\rm ind}$) between $\wlya$ and $\Sigma_{\rm SFR}$ and
      between $\wlya$ and $\Sigma_{\rm sSFR}$ are indicated in each
      panel.  Large red diamonds indicate values for subsamples
      constructed in bins of $\Sigma_{\rm SFR}$ and $\Sigma_{\rm
        sSFR}$.}
    \label{fig:wlyavssigma}
\end{figure}

The correlations between $\wlya$ and both $\Sigma_{\rm SFR}$ and
$\Sigma_{\rm sSFR}$ are shown for individual objects and subsamples in
Figure~\ref{fig:wlyavssigma}.  There is a high probability ($p_{\rm
  ind}=0.28$) of a null correlation between $\wlya$ and $\Sigma_{\rm
  SFR}$ for individual galaxies, though we do find more significant
differences in the average values computed for the subsamples.  For
example, the subsample with the highest $\langle \Sigma_{\rm
  SFR}\rangle$ has $\langle\wlya\rangle\ga -1.5$\,\AA, while the one
with the lowest $\langle \Sigma_{\rm SFR}\rangle$ has
$\langle\wlya\rangle\simeq -5$\,\AA, a difference in
$\langle\wlya\rangle$ that is significant at the $\simeq 3.5\sigma$ level.
Individual galaxies exhibit a correlation between $\wlya$ and
$\Sigma_{\rm sSFR}$ that is more significant than the one between
$\wlya$ and $\Sigma_{\rm SFR}$.  Furthermore, the subsample with the
highest $\langle\Sigma_{\rm sSFR}\rangle$ has $\langle\wlya\rangle
\simeq 1$\,\AA\, compared to $\langle\wlya\rangle \simeq -4$ to
$-7$\,\AA\, for subsamples with lower $\langle\Sigma_{\rm
  sSFR}\rangle$, a difference that is significant at the $\ga 4\sigma$
level.  Interestingly, we do not find a monotonic trend between
$\langle\wlya\rangle$ and $\langle\Sigma_{\rm sSFR}\rangle$ for the
composite measurements, suggesting a large scatter in the relationship
between these two quantities (see discussion in
Section~\ref{sec:scatter}).

\begin{figure}
  \epsscale{1.15}
    \plotone{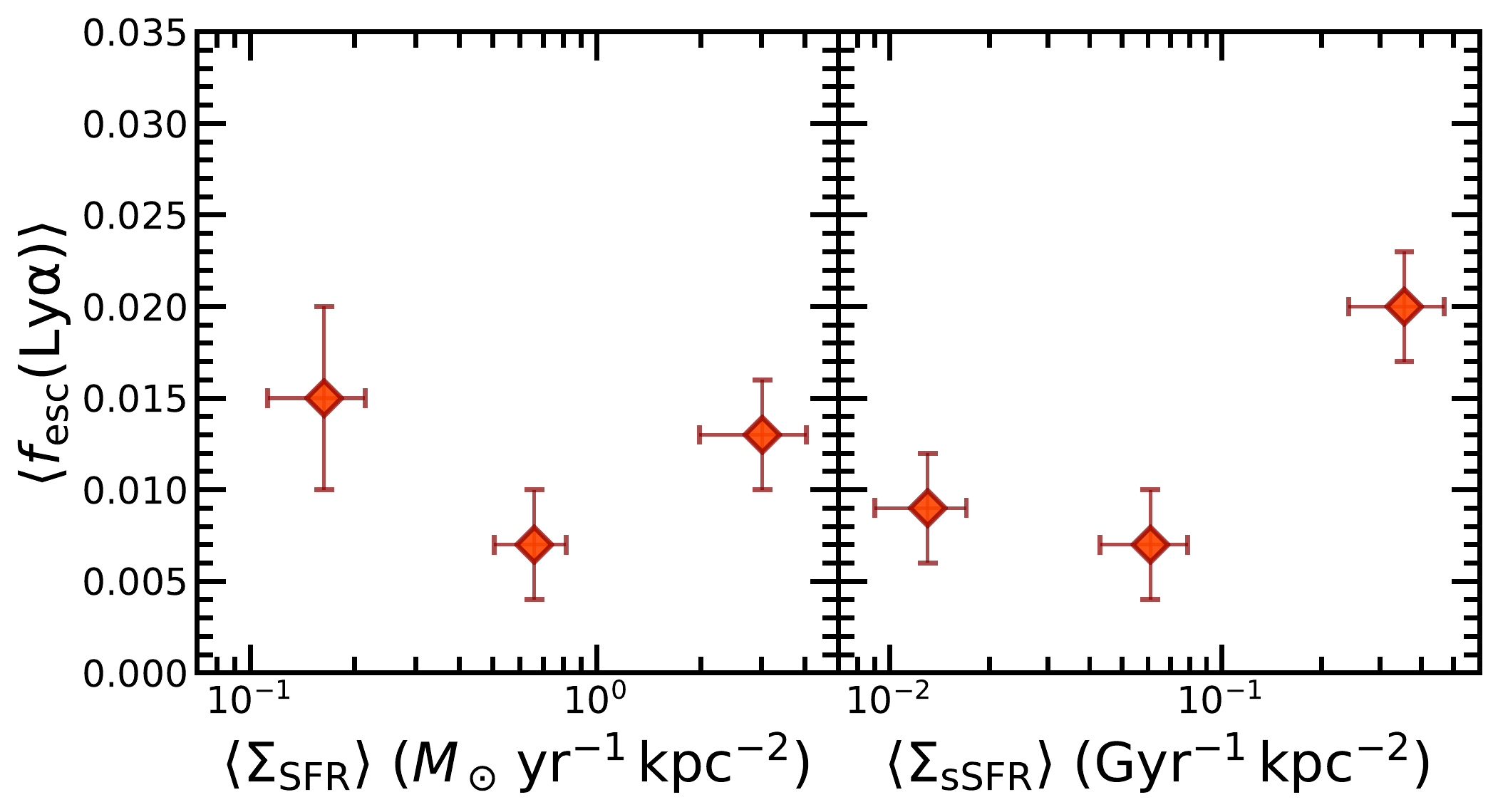}
    \caption{ $\langle\fesclya\rangle$ as a function of
      $\langle\Sigma_{\rm SFR}\rangle$ (left) and $\langle\Sigma_{\rm
        sSFR}\rangle$ (right) for subsamples constructed in bins of
      $\langle\Sigma_{\rm SFR}\rangle$ and $\langle\Sigma_{\rm
        sSFR}\rangle$, respectively, and with coverage of $\lyb$.}
    \label{fig:fesclyavssigma}
\end{figure}

Figure~\ref{fig:fesclyavssigma} shows the variation of
$\langle\fesclya\rangle$ with $\langle\Sigma_{\rm SFR}\rangle$ and
$\langle\Sigma_{\rm sSFR}\rangle$.  The difference in
$\langle\fesclya\rangle$ between the upper and lower third of the
$\Sigma_{\rm SFR}$ distribution is not as pronounced as it is between
the upper and lower third of the $\Sigma_{\rm sSFR}$ distribution.
The significance of the difference in $\langle \wlya\rangle$ and
$\langle\fesclya\rangle$ for galaxies in the upper and lower thirds of
the $\Sigma_{\rm sSFR}$ distribution relative to that obtained for the
upper and lower thirds of the $\Sigma_{\rm SFR}$ distribution suggest
that the gravitational potential may be a relevant factor in the
escape of $\lya$ photons (see also \citealt{kim20}), consistent with a
framework in which compact star formation in a shallow potential
results in lower gas covering fractions (Figure~\ref{fig:fcovvssigma}
and associated discussion).  Datasets spanning a larger dynamic range
in $\wlya$, $\fcovhi$, $\fcovmetal$, $\Sigma_{\rm SFR}$, and
$\Sigma_{\rm sSFR}$ will be needed to confirm these preliminary
results regarding the role of the gravitational potential in
modulating $\fcovhi$ and, as a consequence, $\fesclya$.

\subsubsection{Impact of the SFR Distribution on Ionization Parameter}
\label{sec:sigmavsU}

One significant difference between subsamples with high and low
$\langle\wlya\rangle$ is in their ionization parameters
(Section~\ref{sec:wlyavsphot}), where subsamples with higher
$\langle\wlya\rangle$ also have higher $\langle\log U\rangle$ (e.g.,
Figure~\ref{fig:wlyavsphot}).  As per the discussion in
Sections~\ref{sec:znebmod} and \ref{sec:logumod}, and
Figures~\ref{fig:optratios1} and \ref{fig:optratios2}, $\langle \log
U\rangle$ is insensitive to the BPASS model type and the range of
ionizing spectra indicated by the best-fit SPSneb models.  The
question then remains as to the physical cause of the change in
$\langle \log U\rangle$ with $\langle \wlya\rangle$.
Section~\ref{sec:sigmavslya} presents evidence that compact star
formation in a shallow potential yields conditions (i.e., lower gas
covering fractions) favorable for the escape of $\lya$ photons.  Here,
we suggest that changes in the compactness of star formation may also
be responsible for the observed variation in $\langle\log U\rangle$.

In particular, \citet{shimakawa15} find a significant correlation
between $\Sigma_{\rm SFR}$ and $n_e$ for $z=2.5$ galaxies, such
that
\begin{equation}
n_e\propto \Sigma_{\rm SFR}^{1/S_{n_e}},
\label{eq:shimakawa}
\end{equation}
where $S_{n_e} = 1.7\pm 0.3$.  This correlation may be a consquence of
the relationship between star formation activity and interstellar pressure
in $\hii$ regions \citep{shimakawa15}, and may further imply
a connection between the electron and cold gas densities, even though
the two are sensitive to gas on different physical scales (i.e.,
$50-100$\,pc scales for the densities of $\hii$ regions versus $\sim
1$\,kpc scales for the density of cold gas; e.g., \citealt{shirazi14,
  shimakawa15, jiang19, davies21}).\footnote{The $n_{\rm e}$ derived in bins
    of $\Sigma_{\rm SFR}$ for the present sample are too uncertain to
    independently confirm a trend between $\Sigma_{\rm SFR}$ and
    $n_e$.}

The SFR density within a sphere of radius $R$ can be written as
\begin{equation}
n_{\rm SFR} \propto \frac{\Sigma_{\rm SFR}}{R} \propto \Sigma_{\rm SFR}^{[1-1/S_{R}]},
\label{eq:nsfr}
\end{equation}
where $S_R = -2.82\pm 0.16$ based on fitting the relationship between
$\log[\Sigma_{\rm SFR}/M_\odot\,{\rm yr}^{-1}\,{\rm kpc}^{-2}]$ and
$\log[R/{\rm kpc}]$ for the independent bins of $\Sigma_{\rm SFR}$.
The number density of ionizing photons is proportional
to the product of $n_{\rm SFR}$ and $\xi_{\rm ion}$:
\begin{equation}
n_\gamma \propto n_{\rm SFR}\xi_{\rm ion};
\label{eq:ngamma}
\end{equation}
i.e., such that an increase in $n_{\rm SFR}$ by some factor results in
a similar factor increase in $n_\gamma$ at a fixed $\xi_{\rm ion}$
and, likewise, an increase in $\xi_{\rm ion}$ by some factor results
in a similar factor increase in $n_\gamma$ at a fixed $n_{\rm SFR}$.
We do not find a significant correlation between $\xi_{\rm ion}$ and
$\Sigma_{\rm SFR}$, and therefore assume
\begin{equation}
n_\gamma \propto \Sigma_{\rm SFR}^{[1-1/S_{R}]}
\label{eq:intermediate}
\end{equation}
based on Equation~\ref{eq:nsfr}.  

\begin{figure}
  \epsscale{1.15}
    \plotone{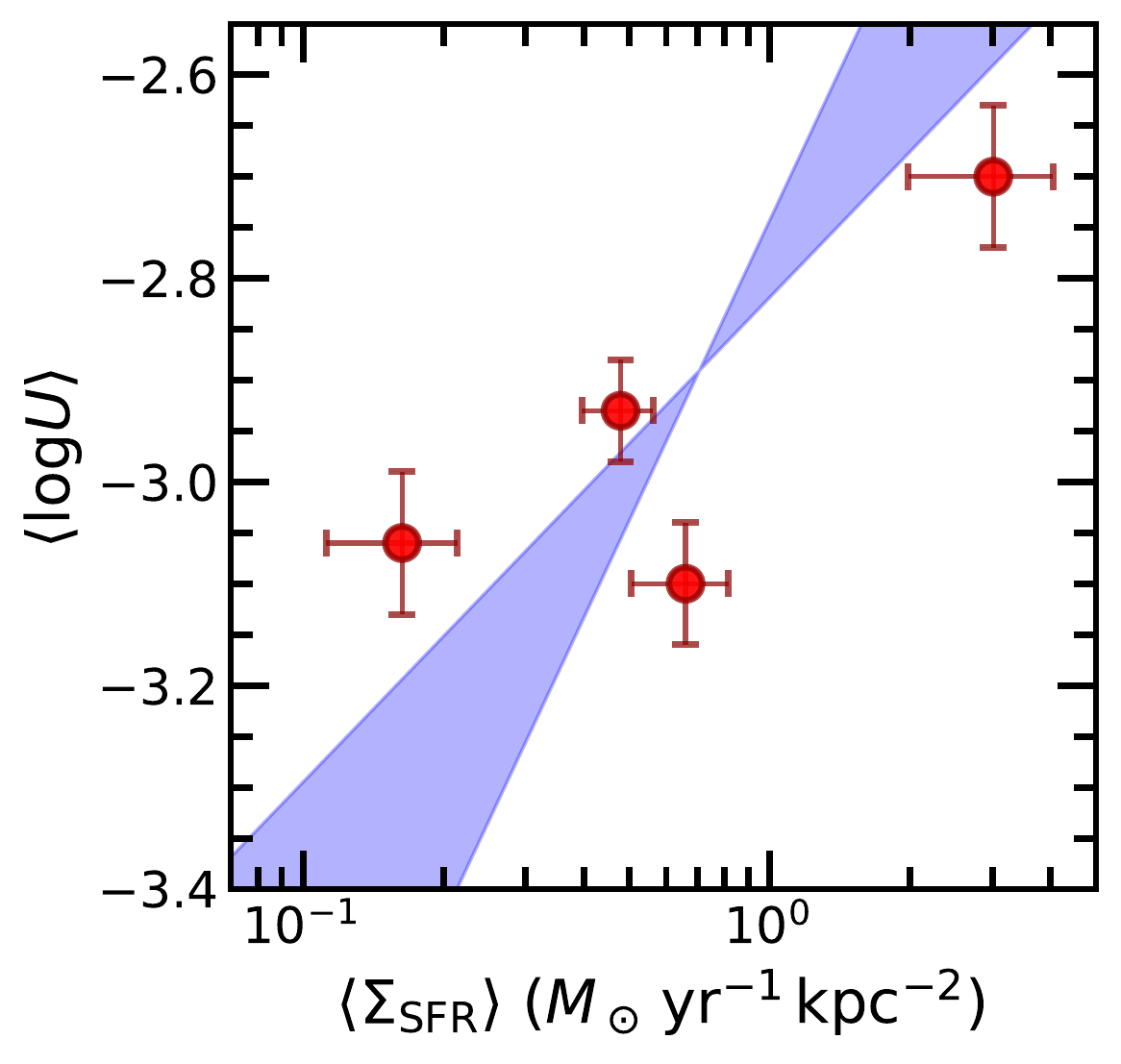}
    \caption{$\langle\log U\rangle$ versus $\langle\Sigma_{\rm
        SFR}\rangle$ for the ALN, ST1L, ST2L, and ST3L subsamples
      which have coverage of $\lyb$ and the optical nebular emission
      lines.  The shaded region indicates the $95\%$ confidence
      interval on the predicted relationship between $\log U$ and
      $\log[\Sigma_{\rm SFR}/M_\odot\,{\rm yr}^{-1}\,{\rm kpc}^{-2}]$
      given in Equation~\ref{eq:final}, normalized to pass through the
      point ($\langle\Sigma_{\rm SFR}\rangle_0$,$\langle\log
      U\rangle_0$) = ($0.5$\,$M_\odot$\,yr$^{-1}$\,kpc$^{-2}$,
      $-3.0$).}
    \label{fig:logUvssigma}
\end{figure}

Combining Equations~\ref{eq:shimakawa} and \ref{eq:intermediate} with
the definition of the ionization parameter as $U \equiv n_\gamma/n_e$
yields the following
relationship:
\begin{equation}
\log U \propto \left[1 - \frac{1}{S_R} - \frac{1}{S_{n_e}} \right]
\log[\Sigma_{\rm SFR}/M_\odot\,{\rm yr}^{-1}\,{\rm kpc}^{-2}].
\label{eq:final}
\end{equation}
The shaded region in Figure~\ref{fig:logUvssigma} depicts the $95\%$
confidence interval on the slope of Equation~\ref{eq:final}, where the
relationship has been normalized to pass through the point
($\langle\Sigma_{\rm SFR}\rangle_0$,$\langle\log U\rangle_0$) =
($0.5$\,$M_\odot$\,yr$^{-1}$\,kpc$^{-2}$, $-3.0$).

The relationship between $\log U$ and $\Sigma_{\rm SFR}$ is
undoubtedly more complicated than what we have approximated here given
the large differences in physical scales associated with the ionized
and cold gas, and ambiguity in the physical interpretation of $\log U$
for geometries that depart from the simple ones (plane-parallel or
spherical) assumed in the photoionization modeling
(Section~\ref{sec:optfittingprocedure}).  In spite of these
complications, the simplified assumptions adopted here predict a
relationship between $\log U$ and $\Sigma_{\rm SFR}$ that lies within
a factor of $\la 2$ of the measurements for most of the subsamples
considered in this work
(Figure~\ref{fig:logUvssigma}).\footnote{\citet{runco21} find a
  similar trend between $\log U$ and $\Sigma_{\rm SFR}$ for MOSDEF
  galaxies which form a superset of the galaxies considered here.}
Thus, the apparent super-linear dependence of $\Sigma_{\rm SFR}$ on
$n_e$ (Equation~\ref{eq:shimakawa}) could plausibly explain at least
part of the dependence of $\log U$ on $\Sigma_{\rm SFR}$.

\subsection{Scatter between Proxies for $\lya$ Escape and Other
Galaxy Properties}
\label{sec:scatter}

\begin{figure}
  \epsscale{1.15}
    \plotone{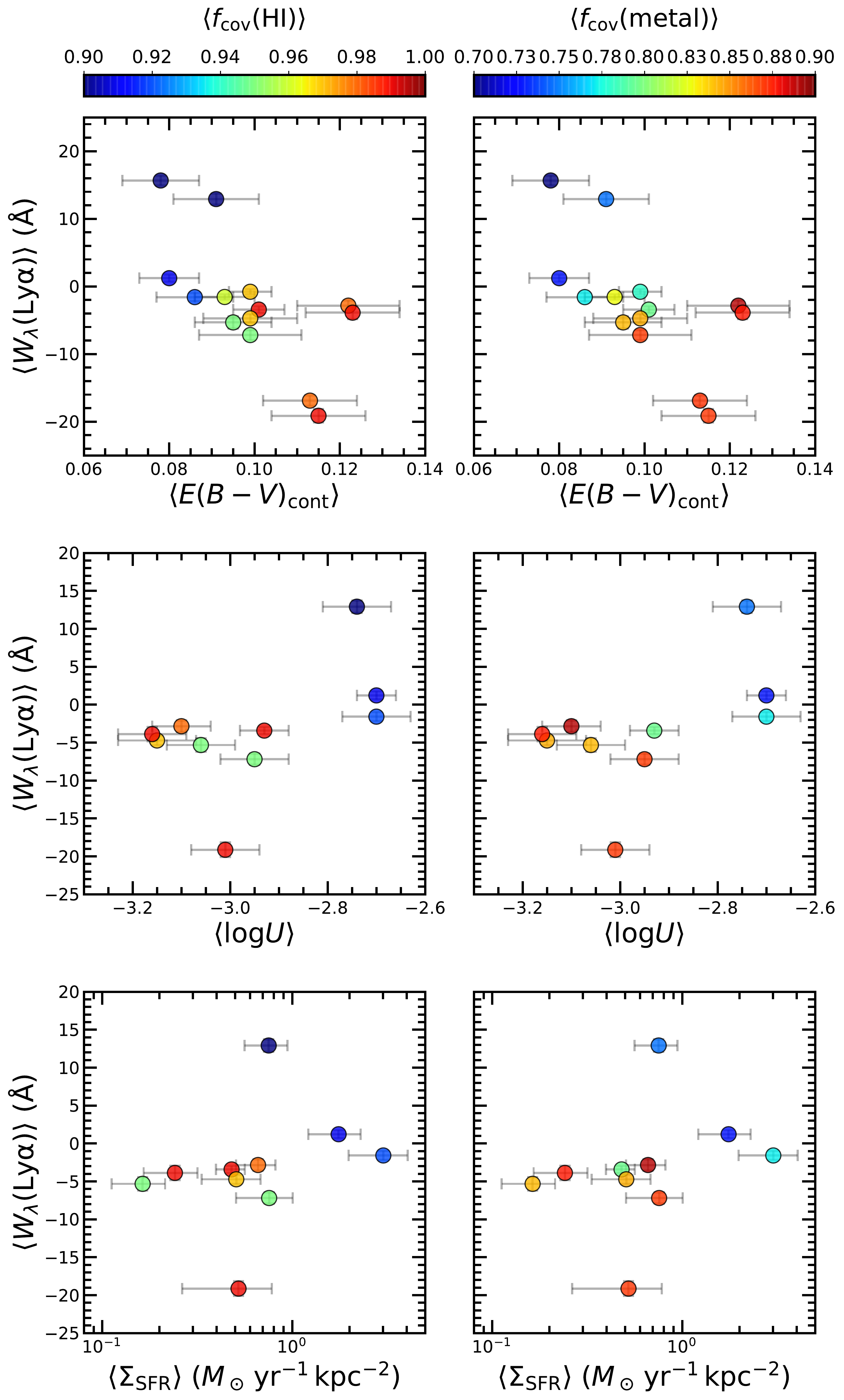}
    \caption{$\langle\wlya\rangle$ versus $\langle \ebmvcont\rangle$
      (top), $\langle \log U\rangle$ (middle), and $\langle\Sigma_{\rm
        SFR}\rangle$ (bottom), where the points have been color coded
      by $\langle\fcovhi\rangle$ (left column) and
      $\langle\fcovmetal\rangle$ (right column).  The top panels only
      include subsamples with coverage of $\lyb$ (required for a
      robust inference of $\fcovhi$) and the middle and bottom panels
      only include subsamples with complete coverage of the
      optical nebular emission lines and $\lyb$.}
    \label{fig:scatterdemo}
\end{figure}

A principal consequence of the dependence of $\wlya$ on covering
fraction (Section~\ref{sec:primary}) is that $\lya$ visibility is a
highly stochastic function of viewing angle, and that a high $\wlya$
may signal the fortuitous alignment of a low-column-density and/or
ionized channel in the ISM with the observer's line of sight as has
been suggested by radiative transfer simulations (e.g., see also
\citealt{verhamme12, behrens14, behrens19, smith19, mauerhofer21}).
This stochasticity may be responsible for much of the scatter between
$\wlya$ and other galaxy properties noted in previous studies.
Examples of this scatter are shown in Figure~\ref{fig:scatterdemo},
which in some cases only becomes apparent when binning galaxies by
properties independent of those on the abscissa and ordinate.  For
instance, the top two panels show the relationship between
$\langle\wlya\rangle$ and $\langle \ebmvcont\rangle$, including
subsamples that were not constructed in bins of either $\wlya$ or
$\ebmvcont$ (i.e., the ST and sST subsamples).  Similarly, the bottom
two panels show the relationship between $\langle\wlya\rangle$ and
$\langle\Sigma_{\rm SFR}\rangle$, including those subsamples not
constructed in bins of either $\wlya$ or $\Sigma_{\rm SFR}$ (i.e., the
sST subsamples).  These results suggest that the spread in $\wlya$ at
a fixed $\ebmvcont$, $\log U$, or $\Sigma_{\rm SFR}$ may in part be
driven by changes in $\hi$/metal covering fraction, as demonstrated by
the color coding of the points in Figure~\ref{fig:scatterdemo}.  For
instance, subsamples WT1LN and WT3LN have a relatively small
separation in both $\langle\log U\rangle$ and $\langle\Sigma_{\rm
  SFR}\rangle$, yet they contain galaxies at the extreme ends of the
$\wlya$ distribution.  In this case, the difference in
$\langle\wlya\rangle$ between the two samples is likely due to changes
in gas covering fraction.

It is worth noting that several of the correlations investigated here
are not expected to be monotonic, such as that between $\wlya$ and
$\fcovhi$, and between $\wlya$ and other parameters that influence
$\fcovhi$ (e.g., $\Sigma_{\rm SFR}$, $\Sigma_{\rm sSFR}$), for
galaxies with lower covering fractions.  A low covering fraction
implies a significant escape of LyC photons that will result in a
corresponding reduction in $\hi$ recombination line (e.g., $\lya$,
$\ha$, $\hb$, etc.) strengths.  If $\fesclya = f_{\rm esc}({\rm
  LyC})$, then one may expect $\wlya \propto \fcovhi (1-\fcovhi)$,
where the first term reflects the fraction of LyC photons that
photoionize hydrogen.  While this function reaches a maximum value at
$\fcovhi = 0.50$, additional scattering/attenuation of $\lya$ photons
from moderate-column-density gas/dust---such that $\fesclya < f_{\rm
  esc}({\rm LyC})$---may result in a turnover of the function at
higher $\fcovhi$.  On the other hand, if $\fesclya > f_{\rm esc}({\rm
  LyC})$ (e.g., \citealt{dijkstra16, izotov21b})---as might be
expected if a significant fraction of the $\lya$ flux observed within
the spectroscopic aperture has resonantly scattered through many mean
free paths (e.g., Figure~\ref{fig:lyafracvsfcov})---then the turnover
in the relation between $\wlya$ and $\fcovhi$ may occur at lower
$\fcovhi$.  The non-monotonic dependence of $\wlya$ on $\fcovhi$ and
variations in the turnover of this function from galaxy to galaxy
would naturally contribute additional scatter in the relations between
$\wlya$ and other properties that correlate with $\fcovhi$.

\subsection{Implications for LyC Escape at High Redshift}
\label{sec:lycimplications}

While our analysis has focused primarily on the production and escape
of $\lya$, it has obvious implications for LyC production and escape
as well.  The small variation in $\langle \xi_{\rm ion}\rangle$ with
$\langle\wlya\rangle$ (Section~\ref{sec:primary}) implies a narrow
range of ionizing photon production efficiencies across the sample.
The variation in $\wlya$ is most directly explained by the covering
fraction of $\hi$ (Section~\ref{sec:primary}).  From a physical
standpoint, the low-column-density and ionized channels that
facilitate the escape of $\lya$ provide avenues for LyC leakage.  This
scenario is supported empirically by the strong connection between
$\wlyaem$ and the ionizing-to-non-ionizing flux density ratio measured
for $z\sim 3$ star-forming galaxies \citep{steidel18}, as well as
correlations between $\fesclya$ and $1-\fcovhi$ \citep{reddy16b,
  gazagnes20} (see also Section~\ref{sec:primary}) and between $f_{\rm
  esc}({\rm LyC})$ and $1-\fcovhi$ \citep{steidel18, gazagnes20} at
both low and high redshift.\footnote{Note that $\fcovhi$ derived in
  this work is based on the residual flux under the $\lyb$ line, which
  is optically thick for gas with column densities that are $\simeq 3$
  orders of magnitude lower than the column densities required for
  significant opacity ($\tau>1$) in the LyC; i.e., $\lognhi \simeq
  14.2$ versus $\lognhi \simeq 17.2$.}

The variation in $\fcovhi$ alone is sufficient to account for the
range of $\wlya$ measured in our sample (Section~\ref{sec:primary}).
However, there is evidence that the hardness of the ionizing radiation
field may play a more important role in the escape of $\lya$ and LyC
radiation at higher redshifts ($z\ga 4$) and for galaxies with fainter
continuum luminosities and stronger $\lya$ emission (e.g.,
\citealt{trainor16, maseda20}).  For instance, \citet{pahl20} report
higher $\civ$ P-Cygni emission and $\lya$ equivalent width for
$z\sim 5$ galaxies compared to lower-redshift galaxies at a fixed
interstellar absorption line equivalent width, suggestive of a harder
ionizing spectrum on average for the former.  Similarly, Atek et
al. (submitted) suggest an evolution of higher $\xi_{\rm ion}$ with
redshift for galaxies of similar mass based on a compilation of
$\xi_{\rm ion}$ measurements at different redshifts.  The strong
correlation between $\xi_{\rm ion}$ and the equivalent widths of
optical nebular emission lines (e.g., \citealt{chevallard18, tang19,
  reddy18b}; Atek et~al., submitted) and the evolution towards higher
equivalent widths with redshift at a fixed stellar mass (e.g.,
\citealt{fumagalli12, sobral14, khostovan16, faisst16, reddy18b})
imply an increase in the average hardness of the ionizing spectrum
with redshift.

This increase in $\xi_{\rm ion}$ with redshift (e.g., for galaxies of
a fixed mass) may also be accompanied by a decrease in the gas
covering fraction.  The well-studied size evolution of galaxies (e.g.,
\citealt{ribeiro16, allen17}) points to more compact sizes with
redshift at a fixed stellar mass.  This size evolution combined with
the increase in SFR with redshift at a fixed stellar mass (e.g.,
\citealt{noeske07, whitaker14, schreiber15}) together imply a more
compact configuration of star formation (i.e., higher $\Sigma_{\rm
  SFR}$) at higher redshifts that may favor the formation of
low-column-density and/or ionized channels through which $\lya$ and
LyC photons can escape.

Regardless of these evolutionary trends that may favor the escape of
ionizing radiation, it is worth noting that the SPSneb modeling
described in Section~\ref{sec:fuvfitting} implies that even {\em
  typical} star-forming galaxies at $z\sim 2$ are quite metal poor in
terms of their stellar abundances; $5-10\%$ of solar.  If the escape
of LyC radiation at higher redshifts ($z\ga 3$) is aided by an
increase in $\xi_{\rm ion}$, then it would ostensibly point to even
younger ($\la 10^7$\,yr) or more metal-deficient ($Z_\ast \la
0.05Z_\odot$) stellar populations at these redshifts.  However, for
the 100bin SPSneb model with an age of $10^7$\,yr, $\logxi = 25.58$
and $25.66$ for $Z=0.001$ and $Z=0.0001$ (i.e., $Z_\ast/Z_\odot =
0.07$ and $0.007$), respectively, implying a relatively modest change
of $\Delta\logxi \approx 0.08$\,dex.  The same numbers for the 300bin
SPSneb model are $\logxi = 25.65$ and $25.71$, again implying a modest
change of $\Delta\logxi \approx 0.06$.  

Thus, it appears that changes in stellar metallicity alone are
insufficient to cause significant increases in $\xi_{\rm ion}$
relative to the values found for typical star-forming galaxies at
$z\sim 2$.  The analysis described in Section~\ref{sec:sigmavslya}
suggests that $\Sigma_{\rm SFR}$ and, in particular, $\Sigma_{\rm
  sSFR}$, may have more influence on LyC escape than the actual
properties of the massive stars (e.g., stellar metallicity and age).
Alternatively, strong bursts of star formation can both temporarily
elevate $\xi_{\rm ion}$ (e.g., \citealt{emami19, emami20,
  nanayakkara20}) and reduce the covering fraction.  Additionally, a
harder ionizing spectrum than that predicted by the BPASS models at a
fixed metallicity may allow for significant production (and escape) of
LyC radiation without the need for invoking extremely metal-poor
stellar populations.  Further insight into the relevant factors for
LyC escape may be achieved by extending the joint FUV and optical
spectral modeling discussed here to galaxies with stronger $\lya$
emission and/or at higher redshifts where at least most of the salient
measurements are still possible, particularly at $2.7\la z\la 4.0$
where the IGM is still relatively transparent and the LyC region is
accessible using ground-based facilities.

\section{\bf CONCLUSIONS}
\label{sec:conclusions}

We used spectral fitting and photoionization modeling of typical
star-forming galaxies at redshifts $1.85\le z\le 3.49$ to deduce
several important characteristics of the massive stellar populations,
and neutral and ionized ISM, and explore how the emergent $\lya$ line
luminosity varies with these characteristics.  The sample consists of
136 galaxies with deep FUV and optical spectra obtained with the
Keck/LRIS and MOSFIRE spectrographs, respectively.  The galaxies were
binned according to $\wlya$, $\Sigma_{\rm SFR}$, and $\Sigma_{\rm
  sSFR}$, and composite FUV and optical spectra were constructed for
these bins.  Stellar population synthesis model fits to the composite
FUV spectra were used to infer stellar metallicity, age, and continuum
reddening of the massive stars.  Simultaneous fits to the interstellar
$\hi$ absorption lines including $\lya$ and $\lyb$ were used to infer
line-of-sight reddening, column density, and gas-covering fraction.
Photoionization modeling of the composite optical spectra of the same
sets of galaxies was used to infer the nebular reddening, gas-phase
oxygen abundance, and ionization parameter.  The joint FUV and optical
spectral modeling is also used to distinguish between single and
binary stellar evolution models.  Section~\ref{sec:fittingsummary}
summarizes most of the findings from the spectral fitting, including
confirmation of several previously found (anti-)correlations, such as
those between $\wlya$ and $\ebmvcont$, and between $\wlya$ and age.
Here, we summarize the key new results from our analysis.

\begin{itemize}

\item Based on a comparison of the inferred $\heii$ nebular emission
  from the FUV composite spectra and the predictions from
  photoionization modeling, we find that the galaxies in our sample
  are uniformly consistent with stellar population models that include
  the effects of stellar binarity, independent of $\wlya$
  (Section~\ref{sec:bpasstypemod}, Figure~\ref{fig:heii}).
  Furthermore, we find little variation in the stellar and nebular
  metallicity with $\wlya$ over the dynamic range probed by our
  sample, with $\langle Z_\ast\rangle \simeq 0.08Z_\odot$
  (Section~\ref{sec:wlyavssps}, Figure~\ref{fig:wlyavssps}) and
  $\langle Z_{\rm neb}\rangle \simeq 0.40Z_\odot$
  (Section~\ref{sec:wlyavsphot}, Figure~\ref{fig:wlyavsphot},
  Appendix~\ref{sec:directmethod}).  The offset in stellar and nebular
  metallicity implies that the stars and gas have ${\rm (O/Fe)} \simeq
  5\times {\rm (O/Fe)}_\odot$ irrespective of $\wlya$, consistent with
  primary enrichment from Type II (core-collapse) SNe
  (Section~\ref{sec:fittingsummary}).  This conclusion is corroborated
  by the spectrally-derived ages that vary from $\la 100$\,Myr for
  galaxies with net positive $\langle \wlya\rangle$ to $\simeq
  300$\,Myr for those with net negative $\langle\wlya\rangle$
  (Section~\ref{sec:wlyavssps}), implying that the galaxies are
  observed prior to the onset of significant Fe enrichment from Type
  Ia SNe.

\item The preference for binary stellar evolution models, the absence
  of any significant correlation between $\langle Z_{\ast}\rangle$ and
  $\langle \wlya\rangle$, and the modest anti-correlation between
  $\langle\log[{\rm Age}/{\rm Myr}]\rangle$ and $\langle \wlya\rangle$
  together imply a relatively narrow range of ionizing spectral shapes
  ($\langle \logxi\rangle \simeq 25.30-25.40$;
  Section~\ref{sec:primary}, Figure~\ref{fig:xiion}) that alone cannot
  account for the variation in $\wlya$ observed within the sample.  On
  the other hand, modeling of the Lyman series absorption lines and
  the depths of saturated low-ionization interstellar absorption lines
  suggests $\hi$ and metal-bearing gas covering fractions that are
  correlated with $\wlya$.  The covering fractions vary from
  $\langle\fcovhi\rangle \simeq 0.90$ and $\langle\fcovmetal\rangle
  \simeq 0.75$ for galaxies with the highest $\langle\wlya\rangle$, to
  $\langle\fcovhi\rangle \simeq 1.00$ and $\langle\fcovmetal\rangle
  \simeq 0.92$ for galaxies with the lowest $\langle\wlya\rangle$
  (Section~\ref{sec:wlyavsfcov}, Figures~\ref{fig:wlyavsfcov} and
  \ref{fig:fcovmetals}).

\item These inferred covering fractions are sufficient to account for
  the variation in $\wlya$ observed within the sample
  (Section~\ref{sec:primary}).  Thus, the gas covering fraction plays
  a more important role in modulating $\wlya$ than the ionizing
  stellar spectrum for galaxies in our sample.  Furthermore, we find
  that the fraction of $\lya$ photons that escapes along the line of
  sight, $\langle\fesclya\rangle$, correlates with
  $1-\langle\fcovhi\rangle$, suggesting that the majority of $\lya$
  photons in down-the-barrel observations of galaxies escapes through
  low-column-density or ionized channels in the ISM
  (Section~\ref{sec:primary}, Figure~\ref{fig:lyafracvsfcov}).  The
  dependence of $\wlya$ on $\fcovhi$ implies that $\lya$ visibility
  may be a highly stochastic function of viewing angle, consequently
  contributing to the scatter between $\wlya$ and several other galaxy
  properties, such as $\ebmvcont$, $\log U$, and $\Sigma_{\rm SFR}$
  (Section~\ref{sec:scatter}, Figure~\ref{fig:scatterdemo}).

\item The apparent large scatter in $\langle f_{\rm cov}\rangle$ at a
  fixed $\xi_{\rm ion}$ (Figure~\ref{fig:fcovvsxiion}) implies that
  there are factors other than the shape of ionizing spectrum that
  affects $f_{\rm cov}$.  In particular, we investigate the effects of
  compact star formation and galaxy potential on gas covering
  fraction.  We do not observe a clear monotonic relationship between
  $\langle\fcovhi\rangle$ and $\langle\Sigma_{\rm SFR}\rangle$, and we
  only find a significant difference in $\langle\fcovmetal\rangle$
  between galaxies in the lower and upper third of the $\Sigma_{\rm
    SFR}$ distribution (Section~\ref{sec:sigmavslya},
  Figure~\ref{fig:fcovvssigma}).  On the other hand, both
  $\langle\fcovhi\rangle$ and $\langle\fcovmetal\rangle$ are
  significantly lower in galaxies with higher $\Sigma_{\rm sSFR}$.
  Similarly, the difference in $\langle\fesclya\rangle$ for galaxies
  in the lower and upper thirds of the $\Sigma_{\rm sSFR}$
  distribution is larger than that of galaxies in the lower and upper
  thirds of the $\Sigma_{\rm SFR}$ distribution
  (Figure~\ref{fig:fesclyavssigma}) .  These results suggest that the
  galaxy potential may play an important role in the escape of $\lya$
  (and LyC) photons: compact star formation in a low potential may
  yield conditions (i.e., lower gas covering fractions) that aid the
  escape of these photons.  Furthermore, we suggest that the
  correlation between ionization parameter and $\Sigma_{\rm SFR}$ may
  be connected to the superlinear dependence of the latter on electron
  density, or interstellar pressure (Section~\ref{sec:sigmavsU},
  Figure~\ref{fig:logUvssigma}).

\end{itemize}

This paper presents a comprehensive analysis of the composite FUV and
optical spectra of a sample of typical star-forming galaxies at $z\sim
2$, focusing on how $\lya$ escape depends on key properties of the
massive stars, and neutral and ionized ISM.  To conclude, we suggest a
few investigations to further elucidate the mechanisms of $\lya$ and
LyC escape in high-redshift galaxies.  As noted in
Section~\ref{sec:ismfitting}, direct inferences of $\hi$ gas covering
fractions at high-redshift are limited to ensembles of galaxies given
the large line-of-sight opacity variations of the $\lya$ forest.
Nonetheless, several of the correlations presented here, and the
dependence of the scatter in these correlations on other galaxy
properties (e.g., such as those between $\wlya$, $Z_\ast$, and
spectrally-derived ages; between $\wlya$, $Z_{\rm neb}$, and $\log U$;
between $\wlya$ and $\xi_{\rm ion}$; between $\wlya$, and $\Sigma_{\rm
  SFR}$ and $\Sigma_{\rm sSFR}$), may be further investigated on an
individual galaxy basis provided sufficiently high S/N FUV and optical
spectra (e.g., \citealt{topping20b, du21}).  Additionally, several
ongoing ALMA programs to constrain molecular gas masses can be used to
probe gas surface densities, examine their connection to interstellar
pressure, and evaluate them in the context of the high electron
densities and high ionization parameters inferred for high-redshift
galaxies.  MOSDEF-LRIS, and the ancillary {\em HST} imaging, also
enable investigations of the correlation between gas covering fraction
and galaxy inclination, and the subsequent effect on $\lya$ and LyC
escape.  The joint FUV and optical spectral modeling presented here
can be extended to galaxies at slightly higher average redshifts
($2.7\la z\la 4.0$) for which the transparency of the intervening IGM
and the Earth's atmosphere allow the modeling of the higher-order
Lyman series lines and LyC emission to be probed with ground-based
observatories.  Upcoming observations with the {\em James Webb Space
  Telescope} will give access to the longer-wavelength optical lines
(e.g., $\ha$, $\nii$, $\sii$) for galaxies at these redshifts, aiding
in photoionization modeling of their ionized ISM.  Finally, ongoing
optical IFU observations with the Keck Cosmic Web Imager (KCWI) will
enable studies of the resonantly-scattered Ly$\alpha$ emission (and
LyC emission) in individual galaxies, likely revealing the diversity
of avenues through which $\lya$ and LyC emission escape galaxies.

\begin{acknowledgements}

We acknowledge support from NSF AAG grants AST1312780, 1312547,
1312764, and 1313171, grant AR13907 from the Space Telescope Science
Institute, and grant NNX16AF54G from the NASA ADAP program.  This work
made use of v2.2.1 of the Binary Population and Spectral Synthesis
(BPASS) models as described in \citet{eldridge17} and
\citet{stanway18}, and v17.02 of the Cloudy radiative transfer code
\citep{ferland17}.  We wish to extend special thanks to those of
Hawaiian ancestry on whose sacred mountain we are privileged to be
guests.  Without their generous hospitality, most of the observations
presented herein would not have been possible.

\end{acknowledgements}



\appendix

\section{Constraints on the Continuum Attenuation Curve}
\label{sec:sfrcomparison}

As noted in Section~\ref{sec:fuvfittingprocedure}, the composite FUV
spectra were initially fit with the SPSneb models assuming three dust
attenuation curves.  For the bulk of our analysis, however, we choose
to present the results with the SMC extinction curve, as this curve
yields SFRs from the spectral fitting, $\sfrsed$, most consistent with
those derived based on H$\alpha$, $\sfrha$.  The calculation of
$\sfrha$ for individual galaxies and composites (subsamples) are
presented in Section~\ref{sec:sigsfrcalc}.  For individual galaxies,
$\sfrsed$ was inferred from broadband SED modeling
(Section~\ref{sec:sedmodeling}).  For an ensemble of galaxies,
$\langle \sfrsed\rangle$ was determined by the normalization of the
SPSneb model that best fits the composite FUV spectrum
(Section~\ref{sec:fuvfittingprocedure}), and is similar to that
obtained by averaging the $\sfrsed$ of individual galaxies
contributing to the composite.  For the purposes of comparing
$\sfrsed$ and $\sfrha$, we highlight a few additional salient details
below.

Binary stellar evolution or a higher mass cutoff of the IMF results in
a larger ionizing photon luminosity, $\qh$, and hence larger $\lha$,
per unit SFR.  The factors used to convert the dust-corrected $\lha$
to $\sfrha$ are $[2.12,1.52,2.69,1.96]\times
10^{-42}$\,$M_\odot$\,yr$^{-1}$\,erg$^{-1}$\,s for the $Z_\ast =
0.001$ 100bin, 300bin, 100sin, and 300sin SPSneb models, respectively,
and are essentially constant as a function of age beyond $10^7$\,yr.
On the other hand, SFR(SED) is primarily determined by the
non-ionizing UV luminosity and therefore is not particularly sensitive
to binary stellar evolution or the high-mass cutoff of the IMF
\citep{theios19}.  Similarly, the variations in $\sfrha$ and $\sfrsed$
with $Z_\ast$ are generally small over the range of relevant
$Z_\ast$---e.g., $\approx 7\%$ variations in $\sfrha$ for the $Z_\ast
= 0.001$ versus $Z_\ast = 0.002$ models---compared to the variations
induced by changing the high-mass cutoff of the IMF or including the
effects of binary stellar evolution.  As noted in
Section~\ref{sec:sedmodeling}, we assumed the $Z_\ast = 0.001$ BPASS
models when fitting the broadband photometry of individual galaxies
while, for the galaxy ensembles, $Z_\ast$ is the best-fit value
obtained when fitting the SPSneb models to the composite FUV spectra
(Section~\ref{sec:fuvfittingprocedure}).

\begin{figure}
  \epsscale{1.15}
    \plotone{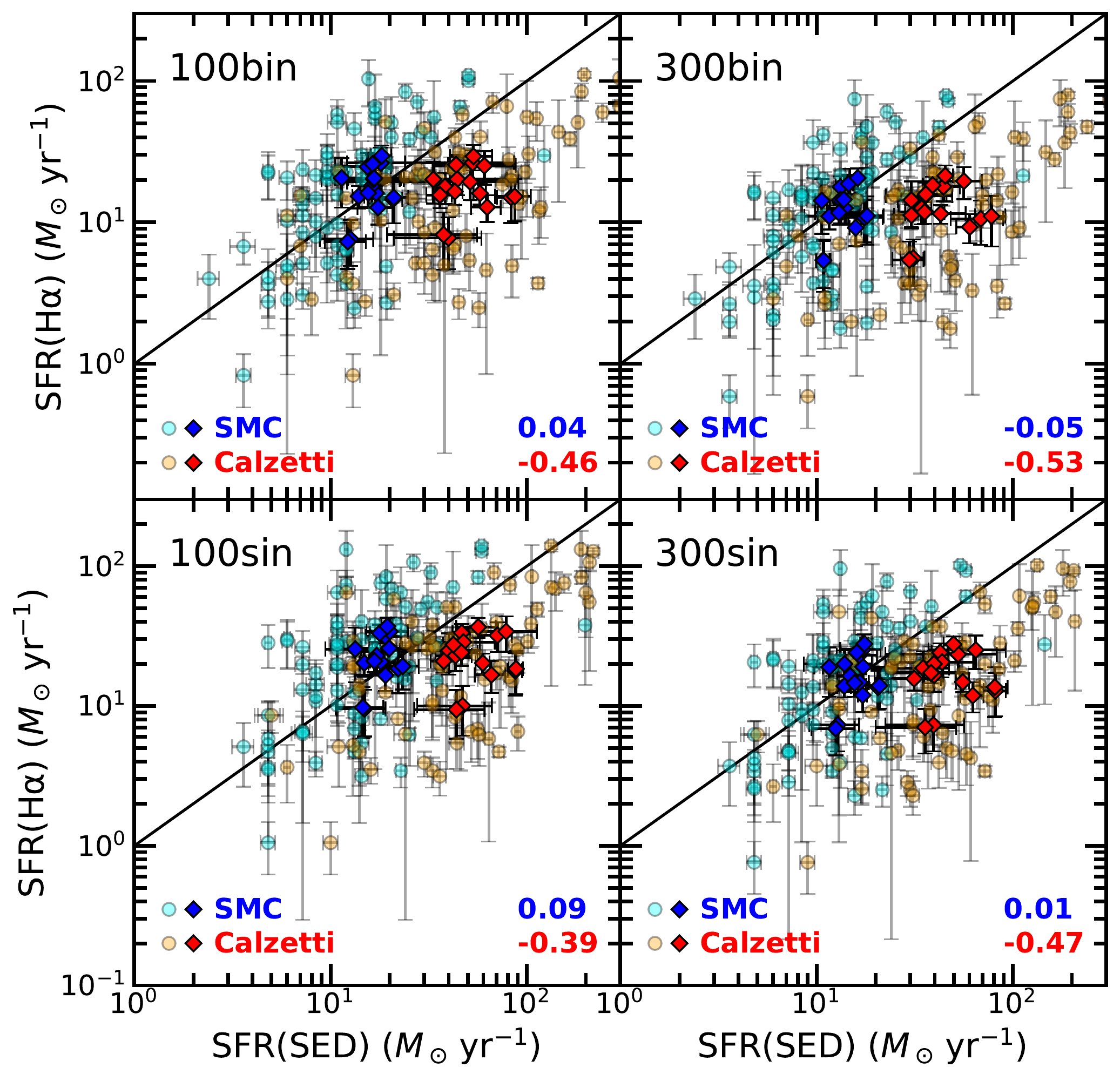}
    \caption{Comparison of $\sfrha$ and $\sfrsed$ for individual
      objects with significant detections of $\ha$ and $\hb$ (circles)
      and for the 16 composites with complete coverage of the optical
      nebular emission lines (diamonds).  SFRs derived assuming the
      SMC curve are denoted by the cyan and blue symbols; those
      derived assuming the \citet{calzetti00} curve are denoted by the
      orange and red symbols.  In all cases, $\sfrha$ is calculated by
      dust-correcting $\lha$ using the \citet{cardelli89} extinction
      curve.  The line indicates the one-to-one relation between
      $\sfrha$ and $\sfrsed$.  The panels show the comparisons for the
      four flavors of BPASS models discussed in
      Section~\ref{sec:sedmodeling}.  The average offsets
      (differences) in dex between $\log[\sfrha/M_\odot\,{\rm
          yr}^{-1}]$ and $\log[\sfrsed/M_\odot\,{\rm yr}^{-1}]$ for
      the 16 composites are indicated for the SMC and
      \citet{calzetti00} cases in the lower right corner of each
      panel.}
    \label{fig:sfrcomparison}
\end{figure}

Finally, we note that varying the dust attenuation curve can have a
large effect on $\sfrsed$, which can change by a factor of $\simeq
2-4\times$ depending on the colors of the object (or ensemble).  At a
fixed observed FUV color, curves with a steep dependence of
attenuation on wavelength yield lower reddening, and hence lower
$\sfrsed$, than shallower curves (e.g., \citealt{pettini98,
  reddy12b}).  Consequently, $\sfrsed$ computed with the SMC curve are
systematically lower than those computed with the \citet{calzetti00}
curve, while those computed with the \citet{reddy15} curve lie between
the SMC and \citet{calzetti00} determinations.

Figure~\ref{fig:sfrcomparison} shows comparisons between $\sfrha$ and
$\sfrsed$ for individual objects and composites of galaxies having
complete coverage of the optical nebular emission lines.  Though only
individual objects with significant $\ha$ and $\hb$ detections are
shown in the figure, for a fixed attenuation curve their
$\langle\sfrha\rangle$ and $\langle\sfrsed\rangle$ generally lie
within $20\%$ of the mean values obtained for the composites which
include galaxies irrespective of significant individual detections of
$\ha$ and $\hb$.

Having computed $\langle\sfrha\rangle$ and $\langle\sfrsed\rangle$
using the self-consistent modeling described above, we find that the
two agree within a factor of $\simeq 2$ only if we assume the SMC
extinction curve for the reddening of the stellar continuum.  In
contrast, the \citet{calzetti00} curve yields $\langle \sfrsed\rangle$
that is systematically larger than $\langle \sfrha\rangle$ by $0.4$ to
$0.5$\,dex in $\log[{\rm SFR}/M_\odot\,{\rm yr}^{-1}]$.  These
conclusions hold irrespective of the effects of binary stellar
evolution and variations in the high-mass cutoff of the IMF between
$100$ and $300$\,$M_\odot$, as indicated in the four panels of
Figure~\ref{fig:sfrcomparison}.\footnote{An upward correction to
  $\langle\sfrha\rangle$ to account for the fraction of escaping
  ionizing photons (as inferred from the clumpy ISM model;
  Sections~\ref{sec:ismfittingprocedure} and \ref{sec:wlyavsfcov})
  does not significantly affect the SFR comparisons discussed here.
  Such corrected $\langle\sfrha\rangle$ are considered in
  Section~\ref{sec:sfrdistribution}.}

The preference for the SMC extinction curve in describing the
reddening of the stellar continuum is not unique to this analysis.
Earlier studies suggested that young and low-mass galaxies at $z\ga 2$
may follow an attenuation curve that is steeper than the
commonly-assumed \citet{calzetti00} curve (e.g., \citealt{baker01,
  reddy06a, siana08, siana09, reddy10, reddy12a, kslee12, oesch13,
  debarros16}).  More recently, \citet{reddy18a} found that the
relationship between dust attenuation---parameterized by the ratio of
the infrared and unobscured FUV luminosities (i.e., IRX)---and FUV
slope, $\beta$, is best reproduced by the SMC extinction curve for
sub-solar metallicity stellar populations at $z\sim 2$ (see also
\citealt{theios19}).  ALMA dust continuum and $[\cii]$ surveys of
modestly-reddened $z\ga 2$ galaxies also point to a steep (SMC-like)
dust curve \citep{bouwens16b, fudamoto17, fudamoto20}.  Such a steep
curve is also favored by observations of local analogues of
high-redshift galaxies \citep{salim18}.

The requirement of an SMC-like dust curve for obtaining consistent
values of $\langle\sfrha\rangle$ and $\langle\sfrsed\rangle$ does not
preclude the possibility that some galaxies in our sample are better
described by shallower dust curves.  Specifically, \citet{shivaei20}
find that MOSDEF galaxies in the upper half of the stellar mass and
gas-phase abundance distributions have a dust curve similar in shape
to the \citet{calzetti00} curve, and shallower than the curve for
lower mass (and lower gas-phase abundance) MOSDEF galaxies.  This
conclusion corroborates previous findings of a mass-dependence of the
shape of the dust curve (e.g., \citealt{pannella09,
  reddy10,bouwens16b}).  In the present context, we simply note that
the comparison between $\langle\sfrha\rangle$ and $\langle
\sfrsed\rangle$ implies an attenuation curve that is steeper than the
\citet{calzetti00} curve for the subsamples formed by the parameters
listed in Table~\ref{tab:compositestats}.

\section{Direct Electron-Temperature-Based Metallicities}
\label{sec:directmethod}

The detection of the intercombination lines
$\interoiii$\,$\lambda\lambda 1660, 1666$ in the composite FUV spectra
provides an independent measurement of the oxygen abundance via the
``direct'' metallicity method.  The average
$\interoiii$\,$\lambda\lambda 1660, 1666$ luminosity was computed by
directly integrating over the line pair in the composite FUV spectrum
for the ALN subsample---the only subsample with complete coverage of
the optical nebular emission lines where $\interoiii$ is significantly
detected---after subtracting a locally-determined continuum.  The
$\interoiii$ luminosity was corrected for dust obscuration assuming
$\langle\ebmvneb\rangle$ and the \citet{cardelli89} extinction curve
(Table~\ref{tab:fuvlinemeasurements}).  The ratio of the
dust-corrected $\interoiii$\,$\lambda\lambda 1660, 1666$ luminosity to
that of $\oiii$\,$\lambda 5008$ yields an estimate of the electron
temperature of the $\oiii$-emitting region, $T_{\rm e}(\oiii) = 13400
\pm 1100$\,K (e.g., \citealt{villar04}).  Using the relationship
between the temperatures of the $\oiii$- and $\oii$-emitting regions
(i.e., the ``$T_2 - T_3$'' relation) given in \citet{campbell86}, and
the abundance relations provided in \citet{izotov06}, we derive an O
abundances of $12+\log({\rm O/H})_{\rm dir} \simeq 8.01 \pm 0.10$,
provided that O is predominantly in the singly- or doubly-ionized
stages.  Note that the abundances derived using the
collisionally-excited lines are most sensitive to the highest
temperature regions and will thus underestimate the abundances
relative to those derived from nebular recombination lines.  Based on
measurements of collisionally-excited and recombination lines in
star-forming knots of local galaxies \citep{esteban14} and $\hii$
regions \citep{blanc15}, the typical offset in abundances derived from
the two sets of lines is $\approx 0.24$\,dex (see also discussion in
\citealt{steidel16}).  Adding 0.24\, dex to the direct method
abundance yields $\langle 12+\log({\rm O/H})\rangle = 8.26\pm 0.10$,
or $\langle Z_{\rm neb}\rangle = 0.37\pm0.09$, in excellent agreement
with that obtained from the photoionization modeling.  Note that there
may be a potentially large (and unaccounted for) systematic error in
the extrapolation of the nebular dust attenuation curve to FUV
wavelengths and the resulting dust corrections to $\interoiii$.
  However, the agreement between the direct-method abundance and those
  obtained from the photoionization modeling suggests the dust
  corrections inferred from the \citet{cardelli89} curve are
  reasonable.

\section{$\ciii$\,$\lambda\lambda 1907, 1909$}
\label{sec:c3}

\begin{figure}
  \epsscale{1.15}
    \plotone{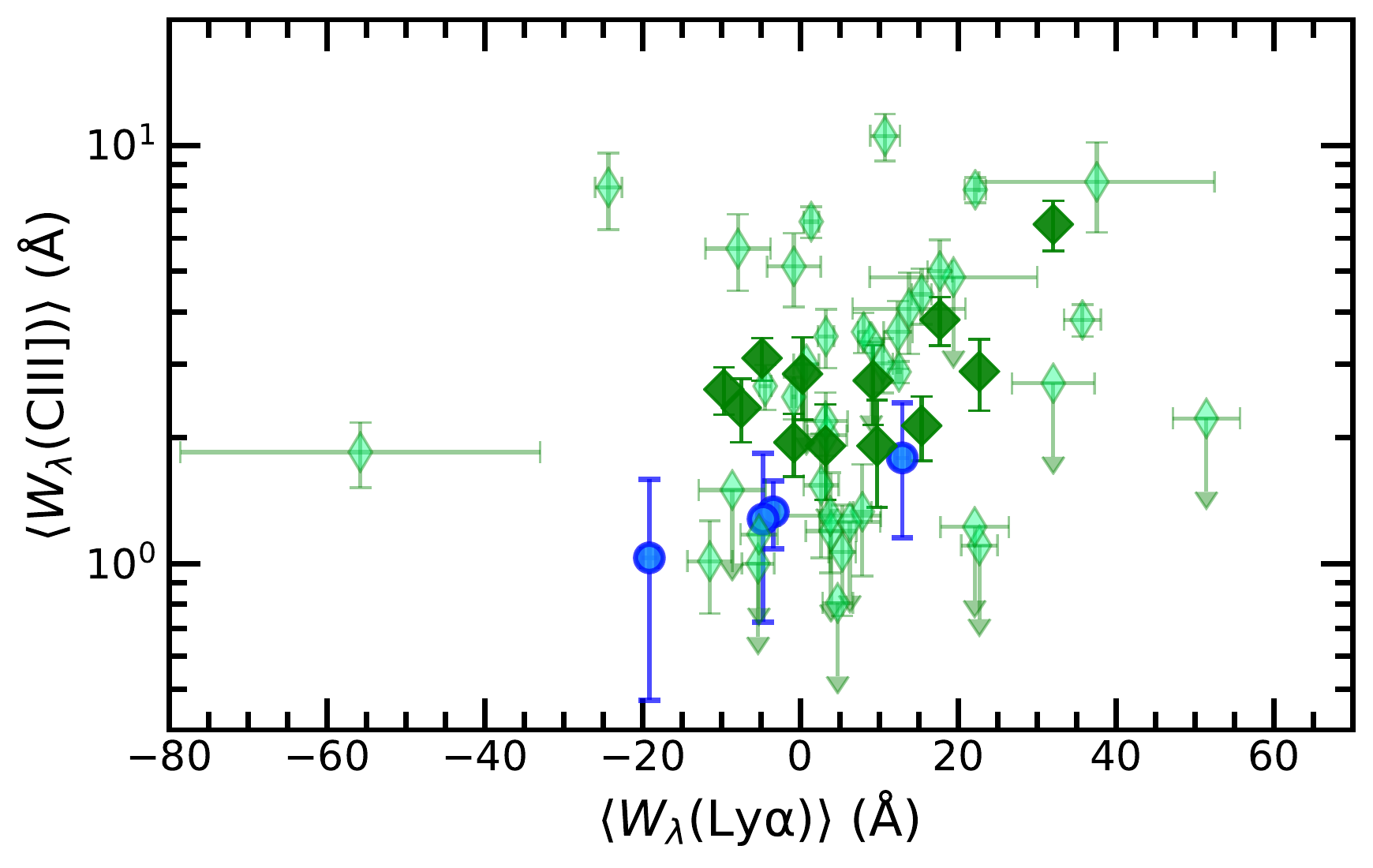}
    \caption{Variation of $\langle\wciii\rangle$ with
      $\langle\wlya\rangle$.  Values obtained for the ALN, WT1LN,
      WT2LN, and WT3LN subsamples which have complete coverage of the
      optical nebular emission lines are indicated by the blue filled
      circles.  For comparison, also shown are the individual
      measurements (small light green diamonds) of extreme
      emission-line galaxies (EELGs) from \citet{du20}, which lie at
      similar redshifts ($z\sim 2$) to MOSDEF-LRIS galaxies.  For
      clarity, one of the EELGs from \citet{du20} which has $\wlya =
      132\pm 11$\,\AA\, and $\wciii = 13.2\pm 0.8$\,\AA\, is not
      shown.  Composite measurements of the EELGs are indicated by the
      large dark green diamonds.}
    \label{fig:wciii}
\end{figure}

As noted in Section~\ref{sec:wlyavsphot}, additional evidence for the
higher ionization parameter of galaxies with stronger $\lya$ emission
comes from $\ciii$\,$\lambda\lambda 1907, 1909$, which is frequently
used as a probe of the ionizing radiation field (e.g.,
\citealt{garnett95, shapley03, erb10, stark15, vanzella16, berg16,
  senchyna17, maseda17, schaerer18, nakajima18a, hutchison19, du20,
  mainali20, feltre20, ravindranath20, tang21}).  The relationship
between $\wlya$ and $\wciii$ for subsamples ALN and WT with complete
coverage of the optical nebular emission lines---and hence complete
coverage of $\ciii$ in the composite FUV spectra given the redshift
distribution of the galaxies in these subsamples ($z<2.6$)---is shown
in Figure~\ref{fig:wciii}.  For context, the figure also includes
measurements for extreme-emission-line galaxies (EELGs) from
\citet{du20}, where $\wlya$ was measured using the same methodology
adopted here.  The uncertainties on $\wciii$ for the aforementioned
composites prevents us from independently confirming a trend between
$\langle \wciii\rangle$ and $\langle\wlya\rangle$.  However, these
measurements are consistent with those obtained from previous studies
(e.g., \citep{shapley03, stark14, stark15, nakajima18b,
  ravindranath20, feltre20}) that find a significant correlation
between $\wciii$ and $\wlya$.  

\begin{figure}
  \epsscale{1.15}
    \plotone{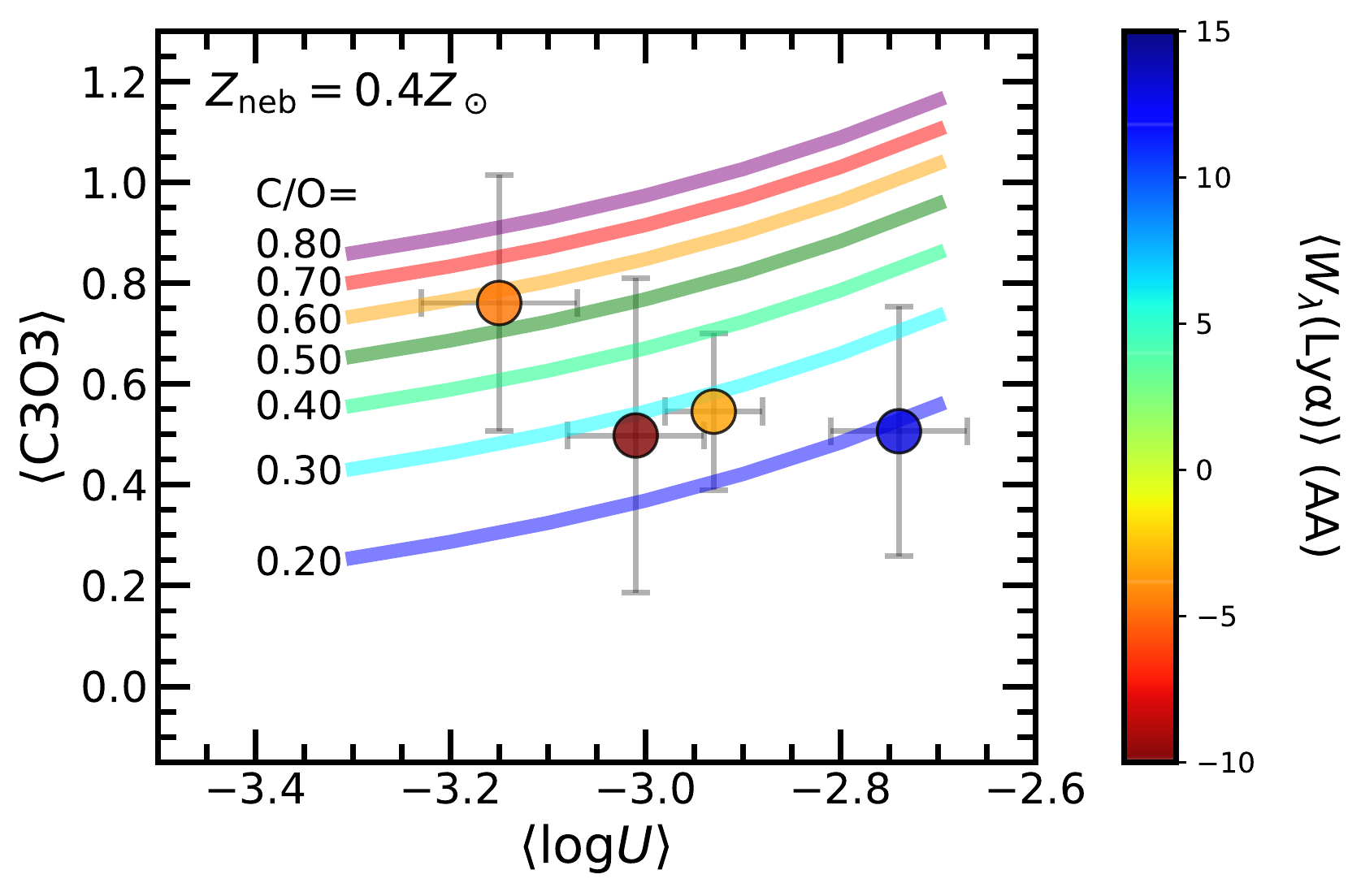}
    \caption{$\langle {\rm C3O3}\rangle$ versus $\langle\log U\rangle$
      for the ALN, WT1LN, WT2LN, and WT3LN subsamples which have
      complete coverage of the optical nebular emission lines,
      color-coded by $\langle \wlya\rangle$.  Also shown are the
      photoionization model predictions for the relationships between
      C3O3 and $\log U$ for $Z_{\rm neb} = 0.4Z_\odot$ and different
      values of ${\rm C/O}$.}
    \label{fig:c3o3vslogu}
\end{figure}

In addition to the shape of the ionizing spectrum, $\wciii$ is also
sensitive to the carbon abundance \citep{garnett95}.  To quantify the
extent to which variations in $\wciii$ may be driven by changes in C
abundance, we compared the C3O3 ratios (defined in
Table~4) measured for the subsamples with the
predictions of the best-fit photoionization models
(Section~\ref{sec:optfitting}) for a range of C/O.\footnote{As is the
  case with O abundances determined from collisionally-excited O lines
  (Appendix~\ref{sec:directmethod}), C abundances derived from
  $\ciii$\,$\lambda\lambda 1907,1909$ may underestimate those derived
  from the optical C recombination lines.  However, in the present
  context, we are interested in the {\em relative} abundances derived
  for the different subsamples and are therefore not concered with
  offsets in the absolute C abundance.}  Figure~\ref{fig:c3o3vslogu}
shows the predicted C3O3 as a function of $\log U$ and C/O for $Z_{\rm
  neb} = 0.4Z_\odot$, the typical nebular abundance indicated by the
photoionization modeling presented in Section~\ref{sec:optfitting}.
These predictions can be compared with the C3O3 calculated for each
subsample using the measurements in
Table~\ref{tab:fuvlinemeasurements}.  The measured C3O3 for subsamples
with complete coverage of the optical nebular emission lines are also
shown in Figure~\ref{fig:c3o3vslogu}, color-coded by the
$\langle\wlya\rangle$ for those subsamples.  In considering the sample
as a whole (i.e., subsample ALN), we find that the photoionization
models discussed in Section~\ref{sec:optfitting} reproduce the
measured C3O3 ratios if we assume $\log({\rm C/O}) = -0.46 \pm 0.13$,
a value similar to that of local dwarf galaxy $\hii$ regions
\citep{kobulnicky98, berg19} of the same O abundance of $12+\log({\rm
  O/H})\simeq 8.3$ ($Z_{\rm neb}\approx 0.4Z_\odot$).  The inferred
C/O in terms of the solar value ($\log({\rm C/O})_\odot = -0.26$;
\citealt{asplund09}) is $[{\rm C/O}] = -0.20 \pm 0.13$.  Note that
this is similar to the inferred value of $[{\rm N/O}] = \log({\rm
  N/O}) - \log({\rm N/O})_\odot = -0.34\pm 0.01$, where $\log({\rm
  N/O})=-1.20\pm 0.01$ (Section~\ref{sec:optfittingprocedure}) and
$\log({\rm N/O})_\odot = -0.86$ \citep{asplund09}, and suggests a
common nucleosynthetic origin for C and N.

The measurement uncertainties in C3O3 are not particularly
constraining insofar as the implied $\log({\rm C/O})$, and there does
not appear to be any significant trend between $\log({\rm C/O})$ and
$\langle\wlya\rangle$.  This result is perhaps not so surprising given
the lack of any significant trend in O abundance (or gas-phase
metallicity, $Z_{\rm neb}$) with $\langle\wlya\rangle$
(Section~\ref{sec:wlyavsphot}; Figure~\ref{fig:wlyavsphot}).
Consequently, the significant correlation between
$\langle\wciii\rangle$ and $\langle\wlya\rangle$
(Figure~\ref{fig:wciii}) is likely driven by changes in the ionizing
radiation field rather than by abundance variations (e.g., see also
\citealt{jaskot16, nakajima18a, ravindranath20}).

\section{Correlations between $\wlya$ and $R_{\rm e}$, SFR,
and $M_\ast$ for Individual Galaxies}
\label{sec:othercorrelations}

\begin{figure}
  \epsscale{1.15}
    \plotone{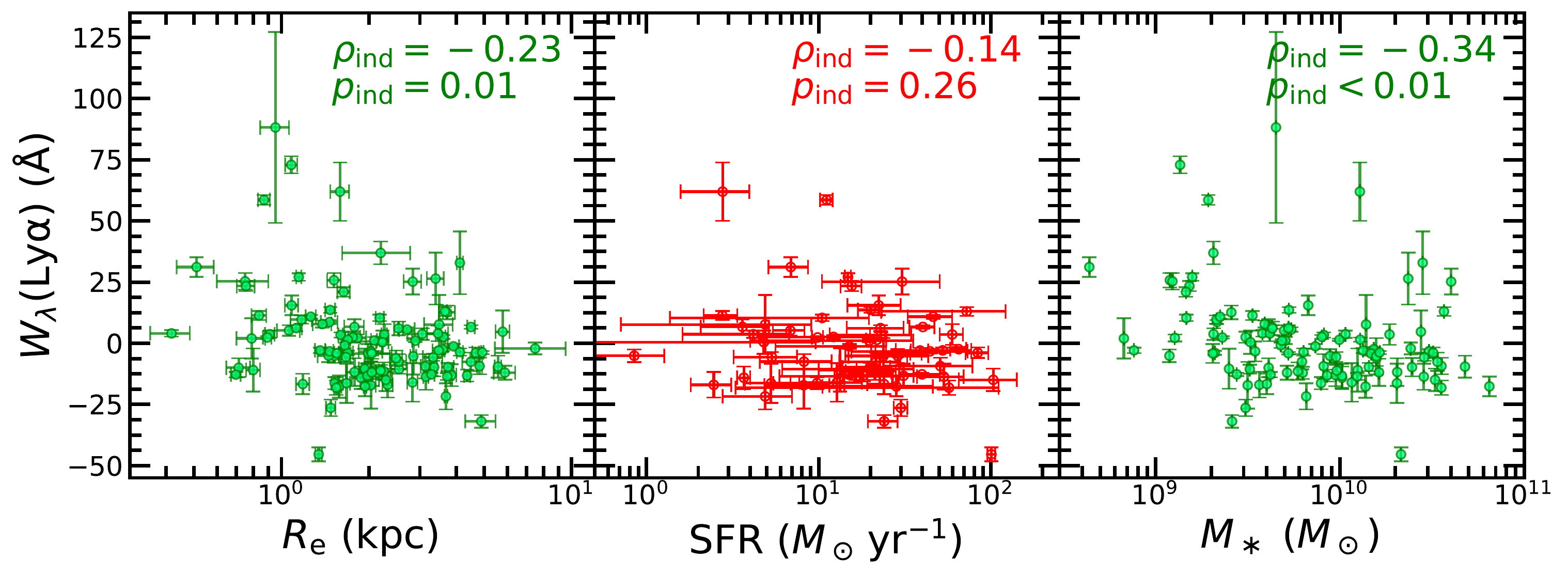}
    \caption{Variation in $\wlya$ with half-light radius $R_{\rm e}$
      (left), $\sfrha$ (middle), and $M_\ast$ (right) for individual
      galaxies.  $\sfrha$ is shown only for those galaxies where $\ha$
      and $\hb$ are both detected with $S/N>3$.  Each panel indicates
      the Spearman correlation coefficient and $p$-value for the
      displayed variables.}
    \label{fig:individualwlyavsother}
\end{figure}

Figure~\ref{fig:individualwlyavsother} summarizes the correlations
between $\wlya$ and $R_{\rm e}$, $\sfrha$, and $M_\ast$ for individual
galaxies in the sample.  The data indicate a significant
anti-correlation between $\wlya$ and $R_{\rm e}$, with a probability
of $p_{\rm ind} = 0.01$ that the two are uncorrelated based on a
Spearman rank correlation test.  The observation of compact ($\la
1$\,kpc) sizes for strong $\lya$ emitters has been noted in many
previous studies, and likely reflects the underlying correlations
between $\wlya$ and luminosity/mass, and between size and
luminosity/mass (e.g., \citealt{shibuya19} and references therein).

A Spearman test indicates a relatively high probability ($p_{\rm ind}=
0.26$) of a null correlation between $\wlya$ and $\sfrha$.  Finally,
the data indicate a significant anti-correlation between $\wlya$ and
$M_\ast$, with $p_{\rm ind}<0.01$ (right panel of
Figure~\ref{fig:individualwlyavsother}), a result that has been found
(or suggested) by a number of other studies (e.g., \citealt{gawiser06,
  pentericci07, finkelstein07, pentericci09, guaita11, jones12,
  hagen14, hathi16, du18, marchi19}).

The aforementioned correlations, particularly between $\wlya$ and both
size and stellar mass, have generally been interpreted to reflect the
less-evolved state of galaxies with stronger $\lya$ emission.  This
interpretation is supported by the apparent anti-correlation between
$\wlya$ and age, where the latter is inferred either from broadband
SED fitting or from FUV spectral fitting
(Section~\ref{sec:wlyavssps}).  However, the strong $\lya$ emission
observed in some evolved high-redshift galaxies implies that the
emergent $\lya$ emission is not solely related to galaxy youth and may
also depend on the covering fraction of gas or dust (e.g.,
\citealt{pentericci09, finkelstein09, kornei10}), an inference that is
supported by our findings (Sections~\ref{sec:physicalrole}).  Our
analysis also suggests that the galaxy potential may play an important
role in gas covering fraction (Section~\ref{sec:sfrdistribution})
which, consequently, gives rise to an anti-correlation between $\wlya$
and stellar mass (e.g., \citealt{kim20}).

\end{document}